\documentclass[
aps,
prd,
superscriptaddress,
nofootinbib,
onecolumn,
notitlepage,
longbibliography,
10pt,
floatfix
]{revtex4-2}

\usepackage{graphicx}

\usepackage{
	amsmath,
	amssymb,
	amsbsy,
	amstext,
	amsfonts,
	mathrsfs,
	latexsym,
	bm,
	physics
}

\usepackage{xcolor}

\usepackage{orcidlink}

\usepackage{subcaption}

\usepackage{hyperref}
\hypersetup{
	colorlinks=true,
	urlcolor=blue,
	anchorcolor=blue,
	citecolor=blue,
	linkcolor=blue,
	linktocpage=true
}
\begin{document}

\preprint{APS/123-QED}

\title{Gravitational Landscapes: black holes with linear equations of state in asymptotically safe gravity}

\author{Ramin Hassannejad}
\email{r.hassannejad@ut.ac.ir}
\affiliation{Department of Physics, University of Tehran, P.O. Box 14395-547, Tehran, Iran.}

\author{Fatimah Shojai}
\email{fshojai@ut.ac.ir}
\affiliation{Department of Physics, University of Tehran, P.O. Box 14395-547, Tehran, Iran.}

\author{Kazuharu Bamba}
\email{bamba@sss.fukushima-u.ac.jp}
\affiliation{Faculty of Symbiotic Systems Science, Fukushima University, Fukushima 960-1296, Japan.}

%\date{\today}
\begin{abstract}
We study black holes with a linear equation of state within the framework of asymptotically safe gravity. This study extends previous work on gravitational collapse in asymptotically safe gravity (that has been done for a dust fluid) by taking into account the pressure of stellar matter. We derive modified field equations containing the running gravitational coupling and the cosmological constant as functions of the energy density. The interior spacetime of the collapsing star is modeled by the Friedmann-Lemaître-Robertson-Walker metric, while the exterior is described by a static spherically symmetric spacetime. Different equations of state of matter are considered to investigate their impact on black hole structure and horizon formation. Our results illustrate that asymptotically safe gravity can introduce  non-singular black hole solutions under specific conditions.  These results provide new insights into black hole physics and the avoidance of singularities within the asymptotically safe gravity framework.

\end{abstract}
       
\maketitle
%\tableofcontents
\section{Introduction}
For nearly half a century, black holes (BHs) have been a focal point of theoretical physics, serving as a complex intersection of gravity, quantum mechanics, and cosmology \cite{Penrose:1964wq,Bekenstein:1973ur,Bekenstein:1974ax,Bekenstein:1972tm,Hawking:1975vcx,hawking1974black,Bardeen:1973gs,Hawking:1982dh,Hawking:1976ra,Almheiri:2020cfm,Ashtekar:2018cay,Ashtekar:1997yu,Ashtekar:2018lag,PhysRevD.62.043008}. BHs originating either from the gravitational collapse of massive stars \cite{oppenheimer1939continued, Datt1938,Frolov:1981mz,Shojai:2022pdq,PhysRevD.111.064069,Hassannejad:2023lrp,Lewandowski:2022zce,Bonanno:2017zen}, or from primordial fluctuations in
the early universe \cite{Zeldovich:1967lct, Carr:1974nx, Hawking:1971ei}, embody extreme physical conditions, where spacetime curvature becomes so intense that even light cannot escape \cite{Penrose:1964wq, Hawking1972}. According to GR, this collapse leads to the formation of a singularity, a region where physical quantities such as energy density and spacetime curvature diverge, indicating the breakdown of classical physics \cite{Penrose:1964wq, Hawking:1970zqf,Penrose:1969pc}. The failure of GR at singularities \cite{hawking1976breakdown} requires a quantum theory of gravity, since quantum mechanical effects are expected to dominate in such regimes \cite{hawking1974black, Bojowald:2005ah, Ashtekar:2018lag, PhysRevD.62.043008, reuter1998nonperturbative, Reuter:2001ag, bonanno2001}. Consequently, understanding the BH formation and the nature of the singularities is a fundamental challenge in gravity and a key step toward a unified quantum gravity framework \cite{Maldacena:1996ky, hawking1973, tHooft:1984kcu,Callan:1996ey,Maldacena:1997de,Horowitz:2003he,Maldacena:2013xja}. Traditional models of BHs, based on Einstein's field equations of GR, cannot fully address the problems posed by singularities. In classical GR, once the event horizon forms, the fate of the collapsing matter is to be crushed into an infinitely dense point \cite{Penrose:1964wq}. Although this result is theoretically consistent within the framework of Einstein's equations, it is physically unsatisfactory because it leads to violations of known physical laws, where infinities usually signal a limitation of the current theory \cite{ hawking1976breakdown, weinberg1979, thooft1974one,Bonanno:2022jjp}. Moreover, the singularity problem illustrates that GR, while successful in describing large-scale phenomena such as planetary motion and the expansion of the universe, requires modification at the quantum scale \cite{Dou:1997fg,Ashtekar:2008zu, Ashtekar:2018lag, Bojowald:2005ah}. The search for a quantum theory of gravity, that can resolve the issue of singularities and unify gravity with the other fundamental interactions, has led to various proposals, including string theory \cite{becker2006string}, loop quantum gravity \cite{Gambini:2011}, and AS gravity \cite{Platania:2018,Codello:2006in}. Of these, AS gravity is one of the most promising approaches to quantum gravity \cite{Fradkin:1978yf}. First introduced by Steven Weinberg in 1979 \cite{Weinberg:1980gg}, AS gravity suggests that gravity becomes ``safe'' at high energies due to the presence of a non-trivial ultraviolet (UV) fixed point \cite{Litim:2003vp, wetterich1993exact, goroff1986ultraviolet, reuter1998nonperturbative, Reuter:2001ag}. This fixed point ensures that the gravitational coupling constants, such as Newton’s constant, remain finite and well-behaved as the energy increases, thereby avoiding the divergences that typically plague quantum field theories at high energies \cite{weinberg1979, bonanno2001, Reuter:2001ag}. The theory is rooted in the renormalization group approach \cite{Polonyi:2001se,Litim:2003vp,Machado:2007ea,Lauscher:2002sq,Reuter:2001ag,Lauscher:2001ya,Souma:1999at,Knorr:2018kog, Falkenberg:1996bq}, which allows for the ``running” of the coupling constants with energy, meaning that the strength of gravity and the cosmological constant are functions of the energy density of the system \cite{bonanno2001, Bonanno:2023rzk}. Generally, the concept of asymptotic safety in gravity provides a potential solution to the singularity problem by suggesting that at extremely high densities, such as those found near the core of a BH, quantum gravitational effects modify the collapse dynamics and prevent the formation of a singularity \cite{Bosma:2019aiu,PhysRevD.62.043008,Platania:2019kyx,Platania:2025imw,Casadio:2010fw}. In addition to studies on BH solutions and massive objects in AS gravity, its cosmological implications have also been extensively investigated \cite{bonanno2001,Reuter:2005kb,Koch:2010nn,Bonanno:2007wg,Bonanno:2011yx,Zhumabek:2024tvp,Zholdasbek:2024pxi}. Notably, AS gravity has been applied to early cosmic inflation, as discussed in Ref.~\cite{Weinberg:2009wa,Bonanno:2015fga,Liu:2018hno,Silva:2024wit,Platania:2019qvo,Kofinas:2016lcz}. Other studies, such as AS gravity in the presence of a matter field, have been widely explored; see Ref.~\cite{Christiansen:2017cxa,Dona:2015tnf,Meibohm:2015twa,Dona:2013qba,Vacca:2010mj,Laporte:2021kyp}. 
\\

AS gravity predicts observable deviations from GR in astrophysical phenomena. Recent studies indicate that running couplings modify BH shadows in ways that are potentially detectable by the Event Horizon Telescope (EHT) \cite{Held:2019xde}, as well as the quasi-normal mode spectra in LIGO/Virgo ringdown observations \cite{Liu:2012ee}, and Hawking radiation spectra \cite{Bonanno:2025dry}. Despite current observational constraints, next-generation gravitational wave detectors may soon be able to probe these quantum gravitational effects. Furthermore, the construction of rotating BH solutions in AS gravity has made significant progress \cite{Kumar:2019ohr}, opening the door to more realistic astrophysical modeling.
Although this theory has been widely studied, many open questions remain. One of the most fascinating aspects is its application to the physics of gravitational collapse, which we address in the following sections.
\\

In this paper, we extend the study of BHs within the framework of AS gravity by incorporating the effects of pressure through an equation of state (EOS) of the form $ p = w \epsilon$, where $ p$ is the pressure, $ \epsilon$ denotes the energy density, and $ w$ is a parameter that determines the type of matter or energy under consideration. The value of $ w $ influences the properties of the collapsing matter, and thus, the structure and behaviour of the resulting BH. A wide range of EOS parameters are considered in this study, from ordinary matter ($w \ge 0$) to negative-pressure sectors of matter with  ($ -1\leq w <0$) \cite{Caldwell:2003vq,Semiz:2020lxj}. These different forms of matter significantly affect the nature of horizon formation, BH stability, and thermodynamic properties \cite{Caldwell:1999ew, Kiselev:2002dx, Semiz:2008rd}. The study of BHs with pressure is crucial because most astrophysical objects, such as stars, have significant internal pressure. 	The role of pressure and the EOS in gravitational collapse has already been studied extensively within classical general relativity. Analyses of pressure-bearing spherical collapse show that nonzero pressure does not uniquely determine the final state: BH or naked-singularity formation may depend on the initial density, pressure, and velocity profiles. In particular, Ref.~\cite{Mosani:2019wfu} showed that imposing a constant linear EOS on a general spherically symmetric perfect fluid places strong compatibility restrictions on the MS mass function, whereas homogeneous and self-similar collapse constitute symmetry-restricted sectors in which the relevant equations can be simplified. The FLRW interior adopted here belongs to this homogeneous sector. Related studies in Ref.~\cite{Joshi:2011rlc} further show that small pressure perturbations may change the final outcome of dust collapse and that pressure alone neither guarantees horizon formation nor removes singularities. These classical results provide the baseline needed to separate EOS effects from those induced by the RG-improved gravitational couplings. Previous studies of BHs in AS gravity have focused primarily on the collapse of pressureless matter (dust) \cite{Bonanno:2023rzk}, but this simplification omits essential physical properties that play a critical role in realistic gravitational collapse \cite{PhysRevD.111.064069,Shojai:2022pdq, Hassannejad:2023lrp}. 	More specifically, Ref.~\cite{Bonanno:2023rzk} considered AS collapse for pressureless dust and derived the corresponding logarithmic matter-gravity coupling, collapse potential, and exterior mass function. The present work generalizes that construction by adopting the linear EOS. On this basis, we derive the generalized Friedmann equation and collapse potential, and the MS mass. The result of Ref. ~\cite{Bonanno:2023rzk} is recovered in the limit $w=0$, providing a consistency check of the present formalism. By contrast, the $w\neq0$ sectors lead to new EOS-dependent masses, horizon structures, central regularity properties, and stability behavior. This EOS-dependent classification of regular and singular collapse outcomes is the main physical novelty of the
	present work; the junction conditions are used only as the matching tool required to connect the homogeneous
	interior to the static exterior geometry. The inclusion of pressure not only provides a more realistic model of BH formation but also allows a more comprehensive investigation into the role of different EOS parameters on the dynamics of the collapse. The present analysis focuses on representative EOS sectors in the range $-1\leq w\leq1$. Our analysis reveals that for  the solutions considered here, negative values of $w$ lead to static exterior geometries that are regular at $R=0$, whereas the non-negative-$w$ exterior metrics analyzed below possess curvature singularities in their $R\rightarrow0$ extensions. Whether a dynamical collapse trajectory actually reaches this region depends additionally on the effective potential, the Hamiltonian constraint, spatial curvature, and initial conditions.
This paper begins by deriving the modified field equations incorporating running gravitational coupling  and the cosmological constant, both of which vary with the energy density of the collapsing matter. We model the interior of the collapsing star using the Friedmann-Lemaître-Robertson-Walker (FLRW) metric, which is widely adopted in cosmology for describing homogeneous and isotropic spacetimes \cite{weinberg2008cosmology}. The exterior of the star, on the other hand, is described by a static, spherically symmetric solution, consistent with the traditional approach to modeling BH spacetimes. By analyzing a range of EOS parameters, we investigate how different types of matter affect BH structure, horizon formation, and the potential for non-singular solutions \cite{PhysRevD.62.043008, Carroll:2003st}. 
Furthermore, we reformulate the field equations in terms of kinetic and potential energy, where the potential term provides deeper insight into the system's behavior. To this end, we first construct the Lagrangian for the FLRW spacetime and perform a Legendre transformation to obtain the corresponding Hamiltonian. This allows us to express the field equations in a form that explicitly separates kinetic and potential contributions. The dynamical stability of the system is then examined by analyzing the first and second derivatives of the potential function.
\\
	A closely related investigation of perfect-fluid collapse in AS gravity was recently presented by Harada, Chen, and Mandal \cite{Harada:2025cwd}. Using a general parametrization of the density dependence of the effective gravitational coupling, they classified singularity resolution in terms of the ultraviolet suppression of gravity and the spatial curvature of the collapsing interior. The present work follows a different but related route: the density dependence of the gravitational coupling is derived from a specific proper-distance scale identification and is therefore tied directly to the fluid EOS. Our analysis focuses on the resulting EOS-dependent effective couplings, collapse potentials, exterior mass functions, horizon structures, central regularity properties, and dynamical stability.
\\

The paper is organized as follows. Section II introduces the modified field equations and the running gravitational and cosmological couplings. Section III discusses Oppenheimer-Snyder collapse, the junction conditions, and the EOS-dependent running coupling. Section IV derives the modified Friedmann equations and analyzes the collapse potentials and the stability of the reduced homogeneous dynamics. Section V presents the exterior MS mass functions and the corresponding static geometries for representative values of $w$. Section VI summarizes the main results, limitations, and directions for future work.
\\
Throughout this paper, the signature of the metric tensor is assumed to be $(-, +, +, +)$. 

In the numerical analysis, we use natural units $c=\hbar=1$. Unless stated otherwise in an individual caption, all dimensional quantities are expressed in reduced Planck units, with $M_{\rm Pl}=(8\pi G_0)^{-1/2}$ and $8\pi G_0=1$.
Accordingly, the plotted variables are
\begin{equation}
	\bar{k}=\frac{k}{M_{\rm Pl}}, \quad
	\bar{R}=M_{\rm Pl}R, \quad
	\bar{m}=\frac{m}{M_{\rm Pl}}, \quad
	\bar{\epsilon}=\frac{\epsilon}{M_{\rm Pl}^4}, \quad
	\bar{\rho}_e=\frac{\rho_e}{M_{\rm Pl}^4}, \quad
	\bar{p}_e=\frac{p_e}{M_{\rm Pl}^4}, \quad
	\bar{\Lambda}=\frac{\Lambda}{M_{\rm Pl}^2}.
\end{equation}
For readability, bars are omitted from the plot axes. The scale factor is normalized according to $a(\tau_0)=1$, so that $\epsilon_0=\epsilon(\tau_0)$. In figures comparing different values of $w$, we use the normalized density $x=\epsilon/\epsilon_0$ and the dimensionless parameter $\lambda=\frac{\varrho G_0}{r_b^3}$, for which $\alpha G_0\epsilon^{1/(1+w)}=\lambda x^{1/(1+w)}$.

%%%%%%%%%%%%%%%%%%%%%%%%%
%%%%%%%%%%%%%%%%%%%%%%%%%
%%%%%%%%%%%%%%%%%%%%%%%%%

\section{The gravitational field equation}
In this section, we review the modified gravitational field equations incorporating the effects of AS gravity, following the formalism developed in Refs.~\cite{Markov:1985py,fock1959theory}.
It has been shown that a certain type of modified gravity theory tends to a de Sitter universe at high energy density \cite{Markov:1985py}. This theory avoids the creation of singularities at high energy density limit. It is based on the idea that the gravitational coupling and the cosmological constant depend on the energy density $(\epsilon)$, represented as \( G(\epsilon) \) and \( \Lambda(\epsilon) \), respectively. The field equation describing this theory is as follows
\begin{eqnarray}\label{aa}
	G^{\mu\nu}=8\pi G(\epsilon)T^{\mu\nu}-\Lambda(\epsilon)g^{\mu\nu}=8\pi T_{\mathrm{e}}^{\mu\nu},
\end{eqnarray}
where $G^{\mu\nu}$ is the Einstein tensor, and $T^{\mu\nu}$
is the perfect fluid energy-momentum tensor,
\begin{eqnarray}\label{ddndn}
	T^{\mu\nu}=(\epsilon+p)u^{\mu}u^{\nu}+pg^{\mu\nu}.
\end{eqnarray}
We can write the effective energy-momentum tensor in a form analogous to a perfect fluid
\begin{eqnarray}\label{wswdjj}
	T^{\mu\nu}_{\mathrm{e}}=(\rho_{\mathrm{e}}+p_{\mathrm{e}})u^{\mu}u^{\nu}+p_{\mathrm{e}}g^{\mu\nu},
\end{eqnarray}
The functions $G(\epsilon)$ and $\Lambda(\epsilon)$ have the following properties: at high energy density regime, $G(\epsilon)\epsilon$ approaches zero and $\Lambda(\epsilon)$ approaches a constant value $\lambda$. However, these conditions are not mandatory, rather, they must be imposed in order to obtain a regular de Sitter spacetime at high energy density $\epsilon\gg 1$. For a general case in which the above conditions are not satisfied, the regularity of spacetime is not required \footnote{ In this situation, if $G(\epsilon)\epsilon$ does not vanish and $\Lambda(\epsilon)$ does not approach a constant in the limit $\epsilon \rightarrow \infty$, the resulting geometry is not constrained to approach the de Sitter configuration. As a consequence, curvature invariants may diverge or deviate from de Sitter asymptotics, and no mechanism enforces the avoidance of singular behavior in this regime. Therefore, such general choices of $G(\epsilon)$ and $\Lambda(\epsilon)$ correspond to models in which the high--energy limit of the theory is allowed to be irregular or singular, without imposing the emergence of a regular de Sitter core as a physical requirement.}
. 
In fact, this theory assumes that the fundamental coupling constants are dynamical and depend on the energy density. Their behavior at low and high energy densities alters the predictions of the field equations, and under certain conditions, it becomes possible to obtain regular spacetime.

It was later shown (see Ref.~\cite{Markov:1985py}) that the field equation \eqref{aa} can be derived from a general Lagrangian describing a gravitational field coupled to a hydrodynamic fluid, given by
\begin{eqnarray}\label{dn}
	S=\frac{1}{16\pi G_{0}}\int\sqrt{-g}d^{4}x\big[\mathcal{R}+2\chi(\epsilon)\mathcal{L}_{\epsilon}\big],
\end{eqnarray}
where $\mathcal{L}_{\epsilon}=-\epsilon$ is the matter Lagrangian, $\epsilon$ is the proper energy density, and  $\chi(\epsilon)$ represents the multiplicative gravity-matter coupling, which in the low energy limit tends to $\chi(\epsilon\to 0)\sim \Lambda_{0}/\epsilon+ 8\pi G_{0}$, where $\Lambda_{0}$ is the cosmological constant.\\ 
To derive the field equations with respect to $g_{\mu\nu}$, we apply the Fock method \cite{fock1959theory} (for more details, see Appendix~\ref{dkfwfu})
, where the density $\epsilon$ is expressed in the form
\begin{equation}\label{wedjnfrr}
	\epsilon=\rho\big(1+\Pi(\rho)\big),~~\Pi(\rho)=\int\frac{dp(\rho)}{\rho}-\frac{p(\rho)}{\rho},
\end{equation}
where $\Pi(\rho)$ is the internal energy per unit rest mass, and $\rho$ is the rest mass density which satisfies the continuity equation $\nabla_{\mu}(\rho u^{\mu})=0$. One can write a relation between the variation of the proper energy density $\epsilon$ and the rest mass energy density $\rho$, which is given by
\begin{equation}\label{dljcwdfwod}
	\delta \epsilon/\delta\rho=(p(\epsilon)+\epsilon)/\rho
\end{equation}
For more details, see Appendix~\ref{dkfwfu}.
Moreover, it has been shown \cite{Markov:1985py} that the variation of the rest mass density with respect to the metric yields
\begin{equation}\label{dcsdccc}
	\delta\rho/\delta g^{\mu\nu}=\rho(g_{\mu\nu}+u_{\mu}u_{\nu})/2
\end{equation}
see Appendix~\ref{dmdjdwfw} for a detailed discussion
. Substituting  Eq.~\eqref{dcsdccc} into Eq.~\eqref{dljcwdfwod}, we obtain an  equation that describes the variation of the proper energy density with respect to the metric in terms of the pressure and the other geometric elements. It is written as
\begin{equation}\label{efvergftrhgtr}
	\delta\epsilon=\frac{p(\epsilon)+\epsilon}{2}\big(g_{\mu\nu}+u_{\mu}u_{\nu}\big)\delta g^{\mu\nu},
\end{equation}
For the EOS $p(\epsilon) = -\epsilon$, the variation of the proper energy density vanishes, $\delta \epsilon = 0$, which shows that the proper energy density remains constant. Consequently, the term \(\chi(\epsilon)\epsilon\) in the Lagrangian \eqref{dn} is also a constant that effectively plays the role of the cosmological constant in the Einstein-Hilbert action. This EOS predicts a de Sitter or anti-de Sitter spacetime, depending on the sign of $\chi(\epsilon)$.   

The metric variation of the matter part of the Lagrangian \eqref{dn} gives
\begin{equation}\label{dcnnwdjbwej}
	\delta(2\sqrt{-g}\chi\epsilon)=\sqrt{-g}\Big(2\frac{\partial(\chi\epsilon)}{\partial\epsilon}\delta\epsilon-\chi\epsilon g_{\mu\nu}\delta g^{\mu\nu}\Big),
\end{equation}
Substituting the variation of the proper energy density \eqref{efvergftrhgtr} into Eq.~\eqref{dcnnwdjbwej} and  considering also the variation of the action \eqref{dn} leads to
\begin{equation}
	\delta S=\frac{1}{16\pi G_{0}}\int\sqrt{-g}d^{4}x\big[G_{\mu\nu}-\frac{\partial(\chi\epsilon)}{\partial\epsilon}\big(\epsilon+p(\epsilon)\big)(g_{\mu\nu}+u_{\mu}u_{\nu})+\chi\epsilon g_{\mu\nu}\big]\delta g^{\mu\nu}=0
\end{equation}
This yields the following field equations
\begin{equation}\label{gdg}
	G_{\mu\nu}= \frac{\partial(\chi\epsilon)}{\partial\epsilon}T_{\mu\nu}+\epsilon^2\frac{\partial\chi}{\partial\epsilon}g_{\mu\nu}.
\end{equation}
This is an interesting result. It is well-known that including a cosmological constant in the Einstein-Hilbert action leads to a field equation with an additional matter-like term proportional to the metric field, expressed as $ G_{\mu\nu} \sim G_{0} T_{\mu\nu} + \Lambda_{0} g_{\mu\nu}$. This is the most widely accepted field equation for gravity, but the cosmological constant is typically introduced into the Lagrangian by hand. In contrast,  there is no explicit term for the cosmological constant in the action \eqref{dn}. Instead, it contains the Ricci scalar and the matter terms, both of which must be included according to the principles of GR \cite{Einstein:1916vd}.

By comparing Eq.~\eqref{aa} and Eq.~\eqref{gdg}, the following relationships can be derived
\begin{equation}\label{xassk}
	8\pi G(\epsilon)=\frac{\partial(\chi\epsilon)}{\partial\epsilon}, \hspace{0.5cm} \Lambda(\epsilon)=-\epsilon^2\frac{\partial\chi}{\partial\epsilon},
\end{equation}
If the functional form of $ G(\epsilon)$ were known, the corresponding values of $\chi(\epsilon)$ and $ \Lambda(\epsilon)$ could be determined as functions of the proper energy density. 
Following Refs.~\cite{PhysRevD.62.043008,PhysRevD.62.043008} in the context of quantum Einstein gravity, we adopt the approximate running of $G$ derived from asymptotic safety in the appropriate limit (for more details, see Appendix \ref{Appendix A}),
\begin{equation}\label{dmndj}
	G(k)=\frac{G_{0}}{1+G_{0}k^{2}/g_*^{\mathrm{eff}}}.
\end{equation}
	Here, $k$ denotes the renormalization-group scale. We denote the fixed-point parameter by $g_*^{\text{eff}}$ in order to allow for possible matter-dressed shifts of the pure-gravity fixed point. In the numerical examples below, we use the pure-gravity Einstein--Hilbert estimate
	\begin{equation}\label{dcwkdc}
		g_*^{\mathrm{eff}}=\frac{570\pi}{833}.
	\end{equation}
	as a fiducial benchmark. Although quantum matter fluctuations in a coupled gravity-matter system can shift the fixed-point coordinates,
	critical exponents, and crossover scale, we do not regard the pure-gravity value as a universal prediction for 
	matter-dominated collapse. In the present phenomenological analysis, $g_*^{\mathrm{eff}}$ parameterizes the leading matter-dressed RG behavior while the numerical value in Eq. \eqref{dcwkdc} is retained only as a reference benchmark.\\
	
\begin{figure}[h]
	\centering
	\includegraphics[scale=0.5]{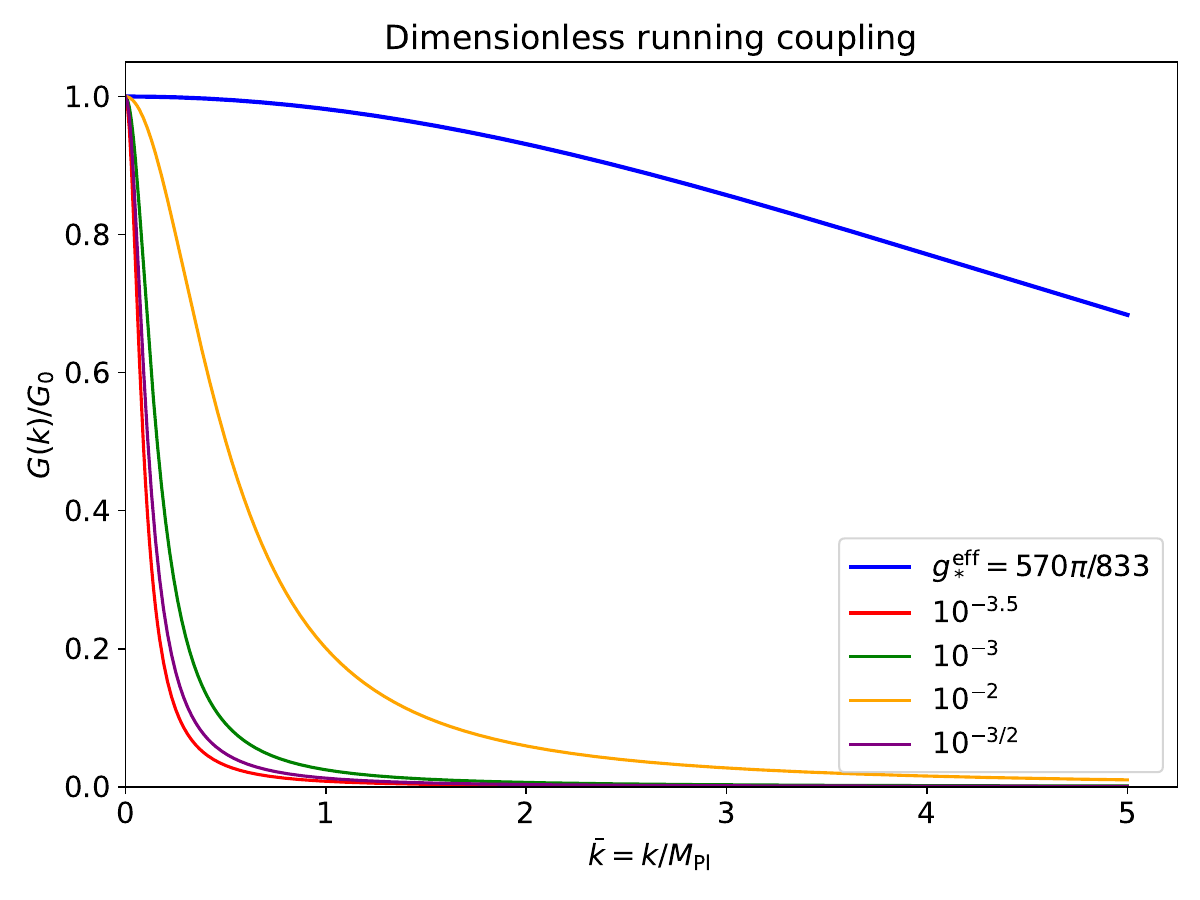}
	\caption{
		Dimensionless running coupling $G(k)/G_0$ versus $\bar{k}=k/M_{\rm Pl}$. The reference curve corresponds to $	g_*^{\mathrm{eff}}=570\pi/833$. The additional values $	g_*^{\mathrm{eff}}=\{10^{-3.5},10^{-3},10^{-2},10^{-3/2}\}$ are shown only to illustrate parameter sensitivity and are not identified with the physical fixed-point estimate. Reduced Planck units $8\pi G_0=1$ are used.}
	\label{fig1}
\end{figure}

	Figure~\ref{fig1} shows the variation of the running gravitational coupling with respect to the momentum scale according to Eq.~\eqref{dmndj}.
The momentum scale has been shown to depend on the distance as $ k(P) = \varpi / d(P)$, where $\varpi$ is a numerical constant, and $d(P)$ represents the proper distance (measured with respect to the classical Schwarzschild metric) from the point $P$ to the center of the BH along a curve $\mathcal{C}$ (see Ref.~\cite{PhysRevD.62.043008}). The proper distance is defined as  
$ d(P) = \int_{\mathcal{C}} \sqrt{|ds^2|}$.  
For a spherically symmetric spacetime, the proper distance depends only on the radial coordinate, i.e., $d(P) = d(R)$. In the UV limit, this distance has been calculated as:  
$ d(R) = 2/(3\sqrt{2G_0 m_0}) R^{3/2}$,  
where $m_0$ has the dimension of mass and ensures dimensional consistency \cite{PhysRevD.62.043008}.  
Using this result, the IR regulator scale becomes  
$k(R) = (3\beta/2) R^{-3/2}$,  
where $\beta = \varpi \sqrt{2G_0 m_0}$.  
Substituting this into Eq.~\eqref{dmndj}, the running gravitational coupling as a function of the radius is given by  
\begin{equation}\label{ajnsdiodod}
	G(R)=\frac{G_{0}}{1+\varrho G_{0}R^{-3}},	
\end{equation}
where $\varrho=9\beta^2/(4g_*^{\mathrm{eff}})$.

\begin{figure}[h]
	\centering
	\includegraphics[scale=0.43]{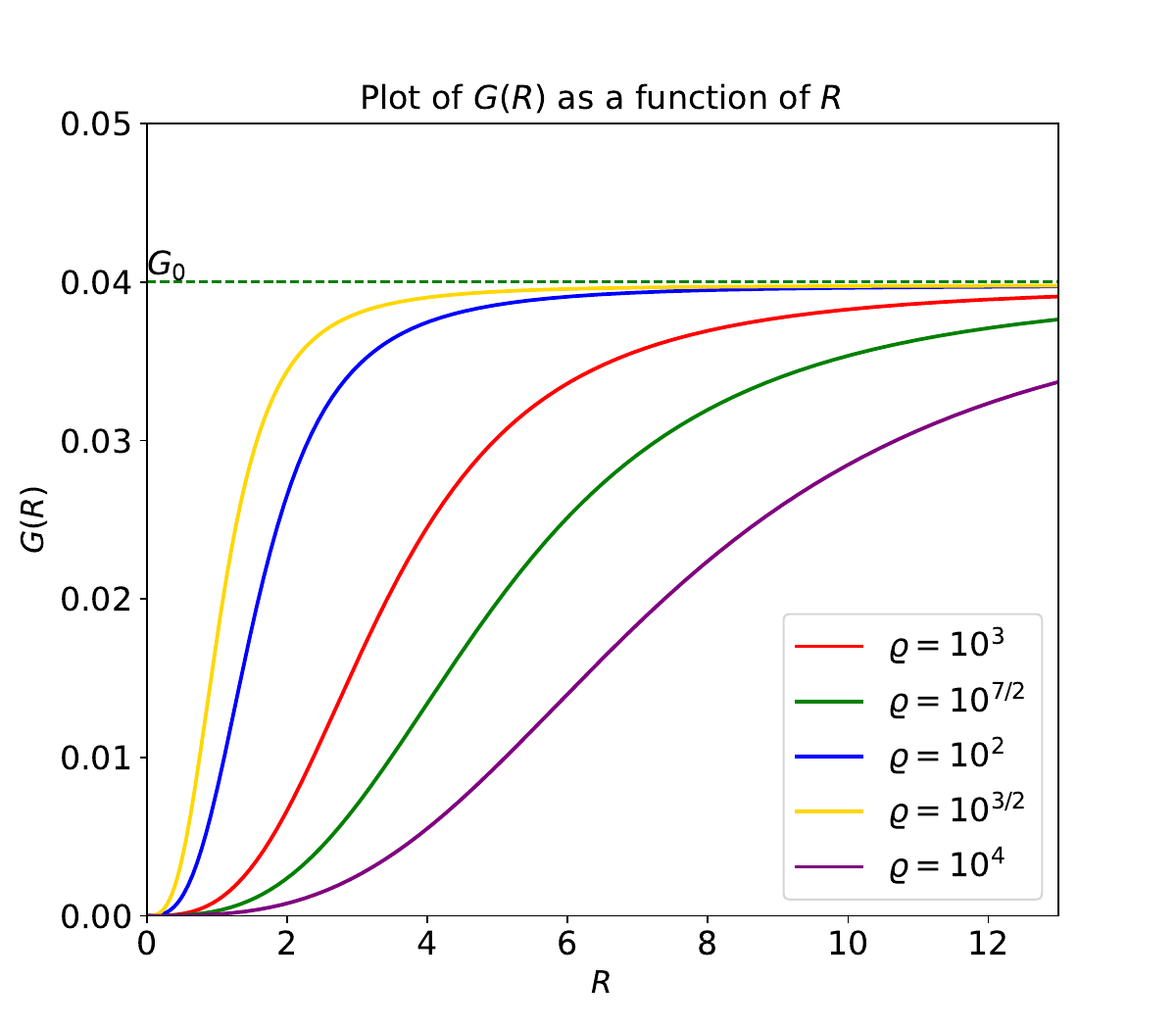}
	\caption{The behaviour of  \(G(R)\)  as a function of the radial distance \(R\) for different values of \(\varrho\), with \(8\pi G_{0} = 1\).  The running gravitational coupling decreases at small distances, while it approaches  $1/8\pi$ at sufficiently large distances.}
	\label{fig2}
\end{figure}

Figure \ref{fig2} shows the variation of the running gravitational coupling as a function of radial distance according to Eq.~\eqref{ajnsdiodod}.

	The cutoff identification is not unique. A commonly used covariant alternative is based on the Kretschmann scalar,
	\begin{equation}
		k_{K}(R)=\xi_{K}K^{1/4}(R).
	\end{equation}
	For the Schwarzschild background,
	\begin{equation}
		K_{\mathrm{Schw}}=\frac{48(G_0M)^2}{R^6},
	\end{equation}
	and therefore
	\begin{equation}
		k_{K}^2(R)=\xi_{K}^2\sqrt{48}\,G_0M R^{-3}.
	\end{equation}
	The ultraviolet proper-distance prescription gives
	\begin{equation}
		d_{\mathrm{Schw}}(R)\simeq\frac{2}{3\sqrt{2G_0M}}R^{3/2}, \quad
		k_d^2(R)=\frac{9}{2}\xi_d^2G_0M R^{-3}.
	\end{equation}
	Thus, both prescriptions yield the same leading ultraviolet scaling, $k^2\propto R^{-3}$, but with different normalization coefficients. The corresponding coupling may therefore be written as
	\begin{equation}
		G_s(R)=\frac{G_0}{1+\varrho_sG_0R^{-3}},
	\end{equation}
	where $\varrho_s$ depends on the cutoff prescription. In this Schwarzschild-based comparison, changing the cutoff scheme shifts the quantum crossover scale and hence modifies quantitative results such as horizon radii and Misner-Sharp mass profiles, while preserving the leading ultraviolet radial scaling.
\\

We remark that the cutoff identification $k \sim 1/d(R)$ employed here utilizes the proper distance computed with respect to the classical Schwarzschild background, following a standard IR-improvement scheme in the AS gravity literature \cite{PhysRevD.62.043008}. This choice avoids the implicit differential equations associated with self-consistent proper distance definitions and enables an analytical treatment of the matching conditions. While recent studies \cite{PhysRevD.109.024045} have correctly pointed out that self-consistent proper distance parametrizations require additional regularity conditions at the horizon to avoid curvature divergences, our prescription yields an explicit coupling $G(R)$ that remains finite and well-behaved for all $R > 0$. This ensures the mathematical consistency of the Israel junction conditions, which are applied at the stellar boundary and extended to characterize the horizon structure within the effective field theory description.
\\

	We emphasize that Eq.~\eqref{dmndj} is employed here as a minimal RG-improvement ansatz. The classical collapsing fluid affects the background geometry, the equation of state, and the cutoff identification, whereas quantum matter backreaction would additionally modify the beta functions governing $G(k)$ and $\Lambda(k)$.
\\	
Functional RG investigations of gravity-matter systems indicate that matter fields can appreciably shift the fixed-point coordinates and critical exponents. They also provide evidence that a positive non-Gaussian gravitational fixed point persists for broad classes of sufficiently weakly coupled matter sectors. Motivated by these results, we replace the pure-gravity fixed-point value with an effective parameter, $g_*^{\mathrm{eff}}$.
\\	
The subsequent results should therefore be applied to matter-dressed flows that preserve a positive fixed point and the leading ultraviolet behavior $G(k)\propto k^{-2}$. Matter backreaction may alter the numerical values of horizon radii, critical parameters, and Misner-Sharp mass profiles, and a sufficiently different coupled RG trajectory could also modify the qualitative behavior. However, a self-consistent finite-density matter-gravity flow on a dynamical collapse background lies beyond the scope of the present work.

\section{Oppenheimer-Snyder (OS) gravitational collapse}
Here, we study Oppenheimer-Snyder (OS) gravitational collapse in the context of a running gravitational coupling. Recently, the gravitational dust collapse in AS gravity has been studied in Ref.~\cite{Bonanno:2023rzk} with a running \( G(k) \) and \( \Lambda(k) \). The authors demonstrated that, while satisfying all energy conditions, the running cosmological constant leads to repulsive effects, thus avoiding the singularity. 
In this paper, we first extend the analysis in Ref.~\cite{Bonanno:2023rzk} to include the case in which the stellar perfect fluid matter possesses non-vanishing pressure. Second, we investigate various BH solutions along with FLRW spacetime for different EOS. To achieve this, we study the gravitational collapse within the framework of AS gravity. We take the interior of the star as a homogeneous and isotropic spacetime described by the FLRW metric
\begin{equation}\label{ddbcdb}
	ds^2_{-}=-d\tau^2+\frac{a^2(\tau)}{1-k_{c}r^2}dr^2+a^2(\tau)r^2d\Omega^2,
\end{equation}
where \(\tau\) and \(r\) denote the comoving time and radial coordinates, respectively. The spacetime outside the star is described by the static spherically symmetric metric in $(t,R,\theta,\phi)$ coordinates
\begin{equation}\label{dndn}
	ds^2_{+}=-f(R)dt^2+f^{-1}(R)dR^2+R^2d\Omega^2,
\end{equation}
where $f(R)=1-2m(R)/R$.
The two metrics inside and outside the star are smoothly joined at the hypersurface \(\Sigma\), which serves as the boundary of the star. Thus, the collapsing boundary of the star, as seen from the exterior, is described by the parametric equations $R = R_{b}(\tau)$ and $t = T(\tau)$, with the following induced metric
\begin{equation}\label{nxhhsh}
	ds^2_{\Sigma}=-\big(f(R_{b})\dot{T}^2-f^{-1}(R_{b})\dot{R}_{b}^2\big)d\tau^2+R_{b}^2d\Omega^2,
\end{equation}
where an overdot indicates differentiation with respect to $\tau$.
The hypersurface $\Sigma$ coincides with the surface of the collapsing star, which is located at $r_{b} = \text{constant}$ in the comoving coordinate system. Considering the interior spacetime, the induced metric on the star's comoving boundary $r_{b}$ is written as
\begin{equation}\label{djjjww}
	ds^2_{\Sigma}=-d\tau^2+a^2(\tau)r_{b}^2d\Omega^2.
\end{equation}
To ensure a smooth transition between the interior and exterior geometries, the metric and its first derivatives must remain continuous across the star's surface. The latter is equivalent to demanding that the extrinsic curvatures $ K_{\alpha \beta} $ of the inner and outer geometries match at the surface. These continuity conditions are known as the \textit{Israel junction conditions} \cite{Israel:1966rt}. The second fundamental forms of the interior and exterior metrics are given by
\begin{align}\label{dndnndo}
	&K^{-}_{\tau\tau}=0,\quad  \quad \quad  K^{-}_{\theta\theta}=r_{b}a(\tau)\sqrt{1-k_{c}r^{2}_{b}}
	\\
	& K^{+}_{\tau\tau}=-\frac{\dot{\Delta}(R_{b})}{\dot{R}_{b}}, \quad \quad \quad K^{+}_{\theta\theta}=R_{b}\Delta(R_{b}),
	\label{ddbdw}
\end{align}
where $\Delta(R_{b})=\sqrt{1-2m(R_{b})/R_{b}+\dot{R}^2_{b}}$. 
To achieve a smooth joining of the two geometries, the components of the metric and the second fundamental form must remain continuous across $\Sigma$. This leads to
\begin{align}\label{dgdgo}
	&R_{b}=a(\tau)r_{b},\\
	&\dot{T}=\frac{\Delta(R_b)}{f(R_{b})}\label{mjsjj},\\
	&\Delta(R_b)=\sqrt{1-k_{c} {r_b}^2}=\textit{constant},\\
	&H^{2}(\tau)+\frac{k_{c}}{a^2(\tau)}=\frac{2m(R_{b})}{R^3_{b}},
	\label{edbi4rbfr}
\end{align}
These dynamical equations will be used to investigate various properties associated with the BH spacetime in Eq.~\eqref{dndn}. For instance, by applying the Friedmann equations within a collapsing star, we can compute its Misner–Sharp (MS) mass as a function of the radius $R_{b}$ using Eq.~\eqref{edbi4rbfr}. We will extend this analysis in the following sections, however, it is first essential to establish some fundamental concepts. 

By differentiating Eq.~\eqref{gdg} and Eq.~\eqref{xassk} with respect to $\epsilon$, and then substituting the results into the Bianchi identity $\nabla_{\mu}T^{\mu\nu}_{\text{eff}} = 0$ the energy-momentum conservation law for $T^{\mu\nu}$ can be easily derived as follows
\begin{align}\label{nddbldkkd}
	\nabla_{\mu}T^{\mu\nu}=
	&-\frac{\partial\epsilon}{\partial x^{\mu}}\frac{G^{\prime}(\epsilon)}{G(\epsilon)}(\epsilon+p)(u^{\mu}u^{\nu}+g^{\mu\nu}),
\end{align}
Noting that the comoving energy density depends only on time, $\epsilon = \epsilon(\tau)$.  
Using the four-velocity in the comoving frame, $u^{\mu} = \delta^{\mu}_{0}$, and the FLRW metric \eqref{ddbcdb}, the energy-momentum conservation \eqref{nddbldkkd} simplifies to $\nabla_{\mu}T^{\mu\nu} = 0$, which reduces to $d\epsilon + 3(\epsilon + p(\epsilon))d\ln a = 0$\footnote{Some studies suggest that the cosmological constant can be neglected  $\Lambda(\epsilon) \simeq 0$ when studying local objects or systems, such as BHs \cite{Bonanno:2017pkg, PhysRevD.111.064069}. However, this assumption leads to an important consequence: the continuity equation becomes $\dot{\epsilon} + 3 H (\epsilon + p) = -\epsilon{\dot{G}(\epsilon)}/{G(\epsilon)}$.}, 
with the solution
\begin{equation}\label{ccncn}
	\epsilon=\epsilon_{0}a^{-3(1+w)},
\end{equation} for a linear EOS $p(\epsilon)=w \epsilon$.
\\

	Equation \eqref{ccncn} follows from the standard conservation law for a homogeneous perfect fluid with a linear EOS and is not, by itself, an AS correction. In a general inhomogeneous spherical collapse, imposing a constant linear EOS may introduce nontrivial compatibility restrictions on the MS mass function, as discussed in Ref. \cite{Mosani:2019wfu}. The FLRW assumption used here reduces the problem to the homogeneous symmetry-restricted sector, in which the fluid equations depend only on time and the generic compatibility obstruction is absent. This simplification permits an analytical treatment, but it also limits the direct applicability of the present conclusions to inhomogeneous collapse.
\\

Inside the star, the gravitational field equations \eqref{aa}, which include the effective cosmological fluid, reduce to the standard Friedmann equations
\begin{subequations}
	\begin{align}\label{dkaifnf}
		&H^2+\frac{k_c}{a^2}=\frac{8\pi}{3}\rho_{\mathrm{e}},\\
		&\dot{H}+H^2=-\frac{4\pi}{3}(\rho_{\mathrm{e}}+3p_{\mathrm{e}}),
		\label{ddndnkpskw}
	\end{align}
\end{subequations}
and the effective energy conservation law is satisfied:  
$
d\rho_{\mathrm{e}} + 3(\rho_{\mathrm{e}} + p_{\mathrm{e}})d\ln a = 0.
$ 
Comparing Eqs.~\eqref{edbi4rbfr} and \eqref{dkaifnf} at the star's surface, yields the effective energy density as a function of the star's radius
\begin{align}\label{dndndnsosw}
	\rho_{\mathrm{e}} = \frac{3}{4\pi}\frac{m(R_{b})}{R^{3}_{b}}, \quad\quad\quad
	p_{\mathrm{e}} = -\frac{m^{\prime}(R_{b})}{4\pi R_{b}^2},
\end{align}  
using the effective energy conservation law. For the exterior of the star, substituting Eq.~\eqref{dndn} into Einstein's equations \eqref{aa} gives the expressions for the density and the components of the anisotropic pressures as follows
\begin{equation}\label{ssbbjkdnwld}
	\rho_{\mathrm{\mathrm{out}}}=-p^{(R)}_{\mathrm{out}}=\frac{m^{\prime}(R)}{4\pi R^2},\quad p^{(\theta)}_{\mathrm{out}}=p^{(\phi)}_{\mathrm{out}}=-\frac{m^{\prime\prime}(R)}{8\pi R}.
\end{equation}
Furthermore, using Eq.~\eqref{dndndnsosw} and Eq.~\eqref{ssbbjkdnwld}, one can easily calculate the pure radial pressure at the star's surface as follows
\begin{equation}\label{djdnbdd}
	p^{(r)}_{\mathrm{star}} \equiv p_{\mathrm{e}}\big|_{R_{b}} - p^{(r)}_{\mathrm{out}}\big|_{R_{b}} = 0, \quad p^{\theta}_{\mathrm{star}}\big|_{R_{b}} = p^{\phi}_{\mathrm{star}}\big|_{R_{b}} \neq 0,
\end{equation}  
Equation \eqref{djdnbdd} should be interpreted as continuity of the radial normal stress across the stellar boundary rather than as the separate vanishing of either pressure. It is compatible with a smooth Darmois-Israel matching. Thus, Eq.~\eqref{djdnbdd} expresses the absence of a radial-pressure discontinuity and of a distributional surface stress tensor.
\\

The present analysis is restricted to a comoving, shell-free boundary satisfying Eq.~\eqref{djdnbdd}. 	
	Allowing a radial-pressure discontinuity introduces a surface energy density and surface pressure, modifies the relation between the interior and exterior mass functions, and shifts the horizon locations. For sufficiently small surface stresses, these changes are perturbative for ordinary simple horizons. Near degenerate horizons or critical configurations, however, even a weak surface layer may alter the number of horizons or their merger. We therefore regard the numerical horizon positions and critical parameters obtained here as specific to the shell-free matching prescription.
\\

We can determine the running gravitational coupling for a collapsing star with non-zero pressure by evaluating it at the star's surface, $ R = R_b = a(\tau) r_b$. By substituting this relation into the distance-dependent gravitational coupling $G(R)$, given in Eq.~\eqref{ajnsdiodod}, we can express the gravitational coupling on the surface and inside the star as a function of the scale factor, $G(a)$. Substituting Eq.~\eqref{ccncn} into $G(a)$, allows us to express the gravitational coupling as a function of the proper energy density, as follows
\begin{equation}\label{ggdhdh}
	G_{w}(\epsilon)=\frac{G_{0}}{1+\alpha G_{0}\epsilon ^{\frac{1}{1+w}}},
\end{equation}
where $\alpha=\varrho\epsilon_{0}^{-1/(1+w)}r_{b}^{-3}$ and $r_{b}$ is the comoving radius.\\
	In natural units, the dimensions of the quantities entering Eq.~\eqref{ggdhdh} are
	\begin{equation}
		[G_0]=M^{-2}, \quad [\epsilon]=M^4.
	\end{equation}
	Since the combination
	\begin{equation}
		\alpha G_0\epsilon^{1/(1+w)}
	\end{equation}
	must be dimensionless, the parameter $\alpha$ has the $w$-dependent mass dimension
	\begin{equation}
		[\alpha]=M^{2-\frac{4}{1+w}}, \quad w\neq-1.
	\end{equation}
	We therefore introduce the dimensionless parameter
	\begin{equation}
		\bar\alpha_w\equiv\alpha M_{\rm Pl}^{\frac{4}{1+w}-2}.
	\end{equation}
	All numerical values of $\alpha$ quoted in the figures refer to $\bar\alpha_w$. In particular,
	\begin{equation}
		\bar\alpha_0=\alpha M_{\rm Pl}^2, \quad
		\bar\alpha_{-1/2}=\alpha M_{\rm Pl}^6, \quad
		\bar\alpha_1=\alpha.
\end{equation}
\\
For $w=0$ and after redefining the parameters, the above expression is precisely the density-dependent coupling employed in the dust-collapse model of Ref.~\cite{Bonanno:2023rzk}.
\\
	The $w$-dependence in Eq. \eqref{ggdhdh} should be interpreted as the result of combining two distinct ingredients. The first is the classical EOS-dependent density scaling in Eq. \eqref{ccncn}, while the second is the RG-improved radial coupling obtained from the proper-distance cutoff identification. Consequently, the exponent $1/(1+w)$ in Eq. \eqref{ggdhdh} is neither a purely classical pressure effect nor an EOS-independent property of the RG flow. It arises when the distance-dependent coupling $G(R)$ is rewritten as a density-dependent coupling by means of the fluid conservation law.
\\
We now turn to the question of whether Eq.~\eqref{ggdhdh} is coordinate dependent. In the context of RG improvement, this issue has been the subject of sustained scrutiny and has been examined by several authors (see Refs.~\cite{Platania:2023srt,Held:2021vwd,Platania:2019kyx}). A fundamental requirement is that the identification of the renormalization group scale $k$ be rooted in coordinate-invariant quantities. Accordingly, $k$ is defined as a function of curvature invariants, most notably the Kretschmann scalar, such that $k = k(\mathcal{K})$. This construction ensures that the scale identification itself remains manifestly coordinate-independent \cite{Held:2021vwd}.

Kretschmann scalar is defined as the quadratic curvature invariant,
$
\mathcal{K}\equiv R_{\mu\nu\rho\sigma}R^{\mu\nu\rho\sigma}
$.
For FLRW spacetime, evaluating the contraction of the Riemann tensor yields
\begin{equation}\label{djdkfe}
	\mathcal{K} = 12(\dot H + H^2)^2 + 12\Big(H^2+\frac{k_c}{a^2}\Big)^2 .
\end{equation}
Substituting Eqs.~\eqref{dkaifnf} and \eqref{ddndnkpskw} into Eq.~\eqref{djdkfe}, the Kretschmann scalar is obtained as a function of the effective energy density and effective pressure
\begin{equation}\label{djdke}
	\mathcal{K} =\frac{64\pi^2}{3}\Big( 4\rho_{\mathrm{e}}^2 + (\rho_{\mathrm{e}}+3p_{\mathrm{e}})^2\Big)
\end{equation}
	The curvature-based prescription can also be implemented directly in the FLRW interior. For illustration, consider a spatially flat classical FLRW background with a linear equation of state, $p=w\epsilon$. The Kretschmann scalar is
	\begin{equation}
		K_{\mathrm{FLRW}}=\frac{64\pi^2G_0^2}{3}\left[(1+3w)^2+4\right]\epsilon^2.
	\end{equation}
	The identification $k_{K}=\xi_{K}K_{\mathrm{FLRW}}^{1/4}$ therefore gives
	\begin{equation}
		G_{K}(\epsilon)=\frac{G_0}{1+C_{K}(w)G_0^2\epsilon},
	\end{equation}
	where
	\begin{equation}
		C_{K}(w)=\frac{8\pi\xi_{K}^2}{\sqrt{3}\,g_*^{\mathrm{eff}}}\sqrt{(1+3w)^2+4}.
	\end{equation}
	This differs from the proper-distance result
	\begin{equation}
		G_d(\epsilon)=\frac{G_0}{1+\alpha_dG_0\epsilon^{1/(1+w)}}.
	\end{equation}
	For $w=0$, the two prescriptions have the same leading density dependence and differ mainly through the crossover coefficient. For $w\neq0$, the curvature prescription may also change the functional dependence on $\epsilon$, thereby modifying the shape of the effective density and Misner-Sharp mass profiles. This comparison illustrates explicitly that the coordinate invariance of a cutoff identification does not eliminate cutoff-scheme dependence. 
\\
Since $\epsilon$ is a scalar in the homogeneous FLRW interior, the coupling $G_{\rm w}(\epsilon)$ defines a scalar function on this
	symmetry-restricted background. If $K(\epsilon)$ is locally invertible, the same coupling can be formally rewritten as $G_{\rm w}(K)$.
	This observation, however, should not be interpreted as removing the cutoff-scheme dependence of the RG-improvement
	procedure. It only shows that, once the homogeneous background and the scale-identification prescription are fixed, the
	resulting coupling can be expressed in terms of scalar quantities.	
	The proper-distance prescription adopted in the main analysis
	is background-dependent but it is a standard RG-improvement
	scheme. The curvature-based prescription
	discussed above provides a covariant alternative
	and illustrates how different
	choices can modify the scalar
	crossover
	scale
	and, for $w\neq 0$, can even change the functional dependence of $G$ on the density. Therefore,
	precise
	horizon radii, critical values, and detailed  Misner-Sharp
	mass profiles should be regarded as scheme-dependent predictions of the chosen cutoff prescription.
\\

We derived the energy density-dependent gravitational coupling, Eq.~\eqref{ggdhdh}, using the Israel junction conditions \footnote{%
	The junction conditions employed here are the standard Israel conditions of GR \cite{Israel:1966rt}, namely continuity of the induced metric and the extrinsic curvature.  Should the underlying theory deviate from GR, or should the problem not be of the OS type (e.g., in the presence of a thin shell with non vanishing surface stress-energy tensor), then Eq.~\eqref{ggdhdh} would acquire additional energy-dependent terms. 
}. \\
	For a direct comparison with Ref.~\cite{Harada:2025cwd}, we note that the exponent $1/(1+w)$ is replaced by an arbitrary exponent in that work. We denote it by $\gamma$, in order to distinguish it from the coefficient $\alpha$ appearing in the present notation. Thus, in the present model, the ultraviolet density exponent is not independent but is fixed by the EOS parameter. For $0<w\leq1$, the mapping gives $1/2\leq\gamma<1$, which is the range in which Ref.~\cite{Harada:2025cwd} does not find complete singularity resolution. The dust case $w=0$ corresponds to the borderline value $\gamma=1$, while $-1<w<0$ gives $\gamma>1$, corresponding to their regular de Sitter-core regime.
\\
The running gravitational coupling  is now expressed as a function of the proper energy density. This dependence is sensitive to the choice of the EOS, with each EOS leading to a distinct functional form of  the running gravitational coupling. Such variations play a crucial role in understanding how gravitational interactions behave under different physical conditions, particularly in cosmological and high-energy regimes.

\begin{figure}[h]
	\centering
	\includegraphics[scale=0.43]{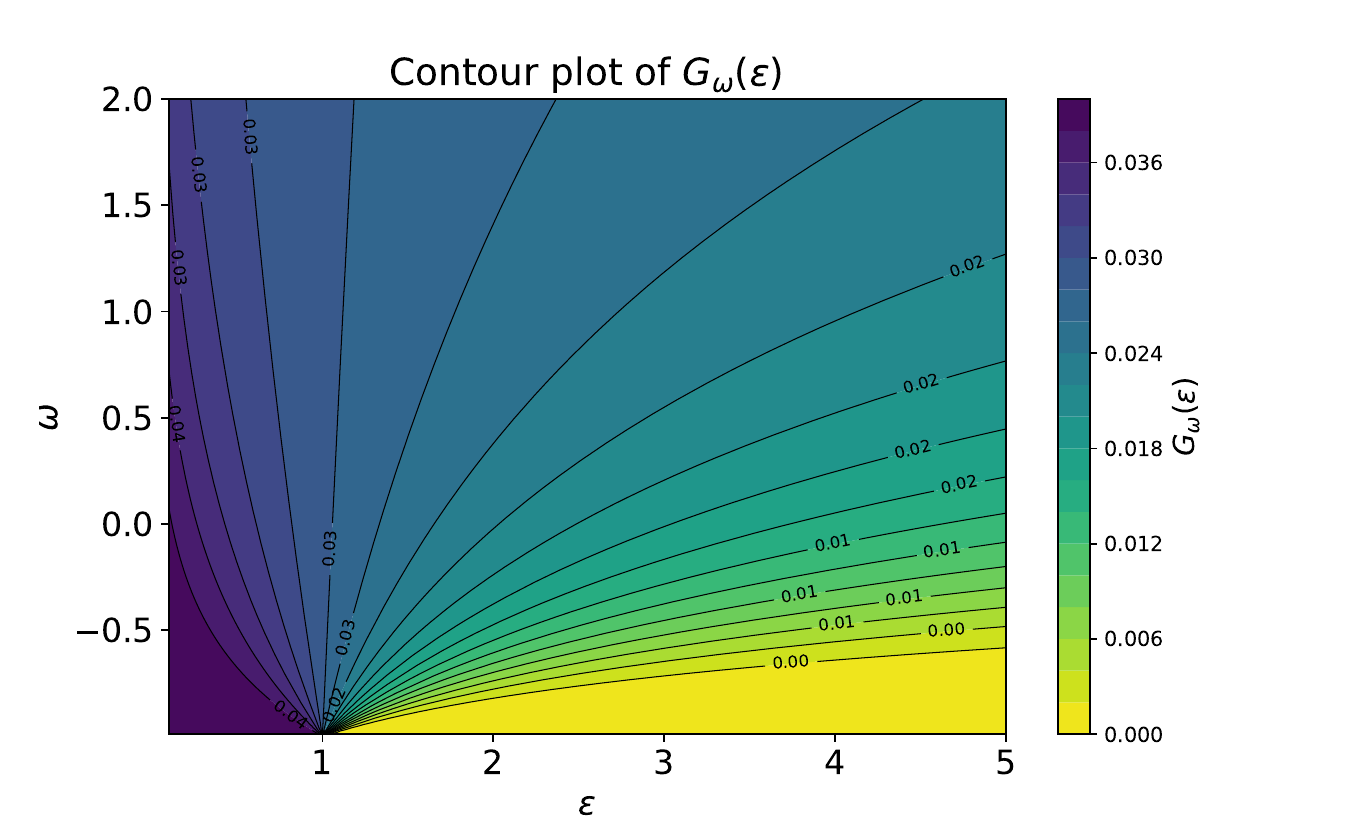}
	\caption{Contour plot of $G_{w}(\epsilon)$ as a function of the parameters $\epsilon$ and $w$. Lighter regions correspond to lower values of $G_{w}(\epsilon)$, whereas darker regions indicate higher values. In general $G_{w}(\epsilon)$ decreases with increasing $\epsilon$. The rate of this decrease is significantly affected  by $w$. The contour lines denote regions of constant $G_{w}(\epsilon)$, providing a detailed representation of the function's behavior in parameter space. The parameters are fixed to $8\pi G_{0}=1$ and $\bar{\alpha}=10$.}
	\label{contour_3}
\end{figure}

Figure \ref{contour_3} illustrates the variation of the running gravitational coupling \eqref{ggdhdh} as a function of the energy density $\epsilon$ and the EOS parameter $w$.

Substituting Eq.~\eqref{ggdhdh} into Eq.~\eqref{xassk} and integrating with respect to $\epsilon$  yields the
expressions for  $\chi(\epsilon)$ and $\Lambda(\epsilon)$ as follows
\begin{equation}\label{sjshsh}
	\chi_{w}(\epsilon)=\frac{\Lambda_{0}}{\epsilon}+8\pi G_{0} ~_2F_1(1,1+w;2+w;-\alpha G_{0}\epsilon^{\frac{1}{1+w}}),
\end{equation}
\begin{align}\label{nhhhh}
	\Lambda_{w}(\epsilon)=&\Lambda_{0}+8\pi G_{0}\epsilon~_2F_1(1,1+w;2+w;-\alpha G_{0}\epsilon^{\frac{1}{1+w}})\notag\\
	&-\frac{8\pi G_{0}\epsilon}{1+\alpha G_{0}\epsilon^\frac{1}{1+w}},
\end{align}
where both involve the hypergeometric function ${}_2F_1$. A detailed derivation is provided in Appendix \ref{dkdkjdkjx}. It should be noted that  $w+2 \notin -\mathbb{N}$, and $\Lambda_{0}$ is an integration constant that plays the role of the cosmological constant, corresponding to the zero point energy $\Lambda_{w}(\epsilon=0)=\Lambda_{0}$.
\\
	For $\Lambda_0=0$ and the normalization $8\pi G_0=1$, the $w=0$ limits of $\chi_w(\epsilon)$ and $\Lambda_w(\epsilon)$ reproduce the logarithmic expressions obtained in Ref.~\cite{Bonanno:2023rzk}. The general formulas derived here therefore constitute a genuine $w$-dependent extension of the dust construction.
\\
Furthermore the behavior of Eqs.~\eqref{sjshsh} and \eqref{nhhhh} can be readily analyzed in both the low- and high-energy-density regimes. In the low-energy-density limit, $\epsilon \ll (\alpha G_{0})^{-(1+w)}$, these expressions reduce to 
\begin{equation}\label{fvrfvrfvvvg}
	\chi(\epsilon)\simeq\frac{\Lambda_{0}}{\epsilon}+8\pi G_{0},\quad
	\Lambda(\epsilon)\simeq\Lambda_{0},
\end{equation}
Substituting the above results into Eq.~\eqref{gdg}, we obtain
\begin{equation}\label{fvfertgvgvg}
	G_{\mu\nu}\simeq 8\pi G_{0}T_{\mu\nu}-\Lambda_{0}g_{\mu\nu},
\end{equation}
which is Einstein's field equations for a perfect fluid in the presence of a cosmological constant.

We now have the explicit functions for the matter-gravity coupling, $\chi_{w}(\epsilon)$, and the running cosmological constant, $\Lambda_{w}(\epsilon)$. These functions provide the basic ingredients for studying the system through the action in Eq.~\eqref{dn}.

\begin{figure}[h]
	\centering
	\includegraphics[scale=0.44]{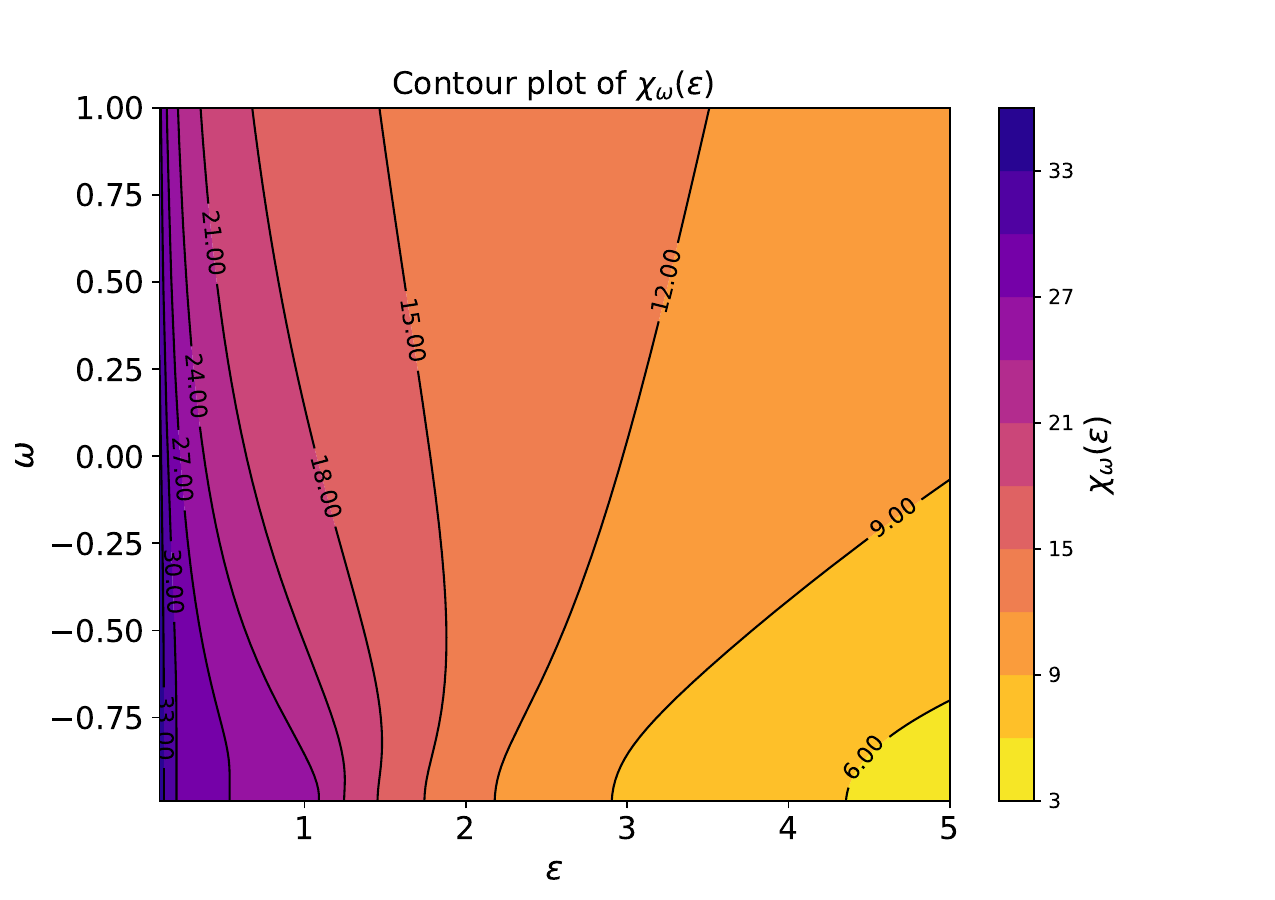}
	\caption{Contours of $\chi_{w}(\epsilon)$ as a function of $\epsilon$ and $w$. The color gradient represents the magnitude of $\chi_{w}$, with darker shades corresponding to higher values and lighter shades to lower values. Contour lines represent loci of constant  $\chi_{w}(\epsilon)$, revealing structural patterns and transitions in the parameter space. The system parameters are set to $8\pi G_{0}=\bar{\alpha}=\Lambda_{0}=1$
	}
	\label{contour_1}
\end{figure}

\begin{figure}[h]
	\centering
	\includegraphics[scale=0.43]{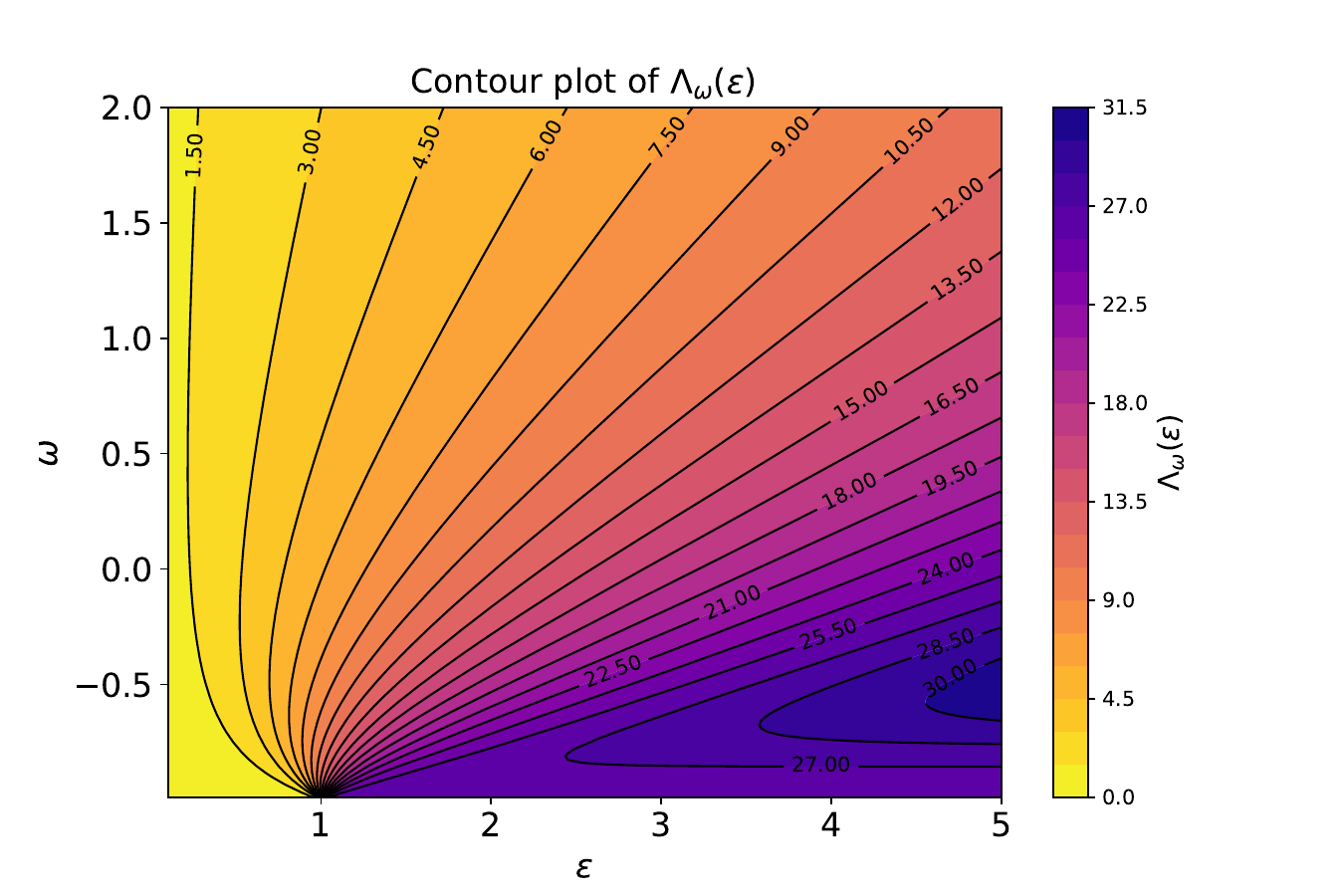}
	\caption{Contours of $\Lambda_{w}(\epsilon)$ as a function of the energy density $\epsilon$ and the parameter $w$. Contour lines indicate regions of constant  $\Lambda_{w}(\epsilon)$,  while the color gradient reflects its magnitude, with darker shades  corresponding to higher values. The parameters are fixed as $\Lambda_0=8\pi G_0 =\bar{\alpha} = 1$.
	}
	\label{contour_2}
\end{figure}

Figure \ref{contour_1} presents  the gravity-matter coupling, $\chi_{w}(\epsilon)$, as a function of the energy density $\epsilon$ and the EOS parameter $w$. Similarly, Fig.~\ref{contour_2} presents the behavior of the running cosmological constant, $\Lambda_{w}(\epsilon)$, with respect to the energy density $\epsilon$ and the parameter $w$.

In the following, assuming some specific values of $w$, we further analyze  Eqs.~\eqref{sjshsh} and \eqref{nhhhh}.
\subsubsection{$w=0$}
For dust matter, $w = 0$, Eqs.~\eqref{sjshsh} and \eqref{nhhhh} reduce to
\begin{align}\label{dhdhwdiu}
	&\chi(\epsilon)=\frac{\Lambda_{0}}{\epsilon}+\frac{8\pi}{\alpha}\frac{\ln(1+\alpha G_{0}\epsilon)}{\epsilon},\\
	&\Lambda(\epsilon)=\Lambda_{0}+\frac{8\pi}{\alpha}\ln(1+\alpha G_{0}\epsilon)-\frac{8\pi G_{0}\epsilon}{1+\alpha G_{0}\epsilon},
	\label{csdcwdcwekid}
\end{align}
These expressions were obtained using Eq.~\eqref{dnbkwdn}. By substituting Eqs.~\eqref{dhdhwdiu} and \eqref{csdcwdcwekid} into Eq.~\eqref{aa}, we derive the effective pressure $p_{\mathrm{e}}$ and the effective energy density $ \rho_{\mathrm{e}}$ as
\begin{align}\label{wdkjwfwrefb}
	&\rho_{\mathrm{e}}=\frac{\Lambda_{0}}{8\pi}+\frac{\ln(1+\alpha G_{0}\epsilon)}{\alpha},
	\\
	&
	p_{\mathrm{e}}=-\frac{\Lambda_{0}}{8\pi}-\frac{\ln(1+\alpha G_{0}\epsilon)}{\alpha}+\frac{G_{0}\epsilon}{1+\alpha G_{0}\epsilon},
	\label{dmndjdjd}	
\end{align}		
Combining these results leads to the following effective EOS
\begin{equation}\label{dmjdjdjhd}
	p_{\mathrm{e}}=-\rho_{\mathrm{e}}-\frac{e^{\alpha(\Lambda_{0}/8\pi-\rho_{\mathrm{e}})}-1}{\alpha},
\end{equation}
Expanding the above effective EOS around $\rho_{\mathrm{e}}= \Lambda_{0}/8\pi$ yields 
\begin{equation}\label{fgngbfd}
	p_{\mathrm{e}}\simeq -\frac{\Lambda_{0}}{8\pi}-\frac{\alpha \Lambda^{2}_{0}}{128\pi^2}+\frac{\alpha \Lambda_{0}}{8\pi}\rho_{\mathrm{e}}-\frac{\alpha}{2}\rho^2_{\mathrm{e}},
\end{equation}
On the right-hand side, a second-degree (quadratic)  negative pressure term appears, corresponding to a polytropic EOS with index $n = 1$. Such a negative EOS has been observed in gravitational collapse scenarios that result in the formation of a Hayward regular BH\footnote{In case, the stellar matter is modeled as a polytropic fluid with EOS 
	$
	p \propto \rho^{1+1/n},
	$
	where the polytropic index $n$ is unity yielding the specific form 
	$
	p = -\frac{8\pi}{3} l^2 \rho^2
	$ (see Ref.~\cite{Shojai:2022pdq}).}.

\begin{figure}[h]
	\centering
	\includegraphics[scale=0.43]{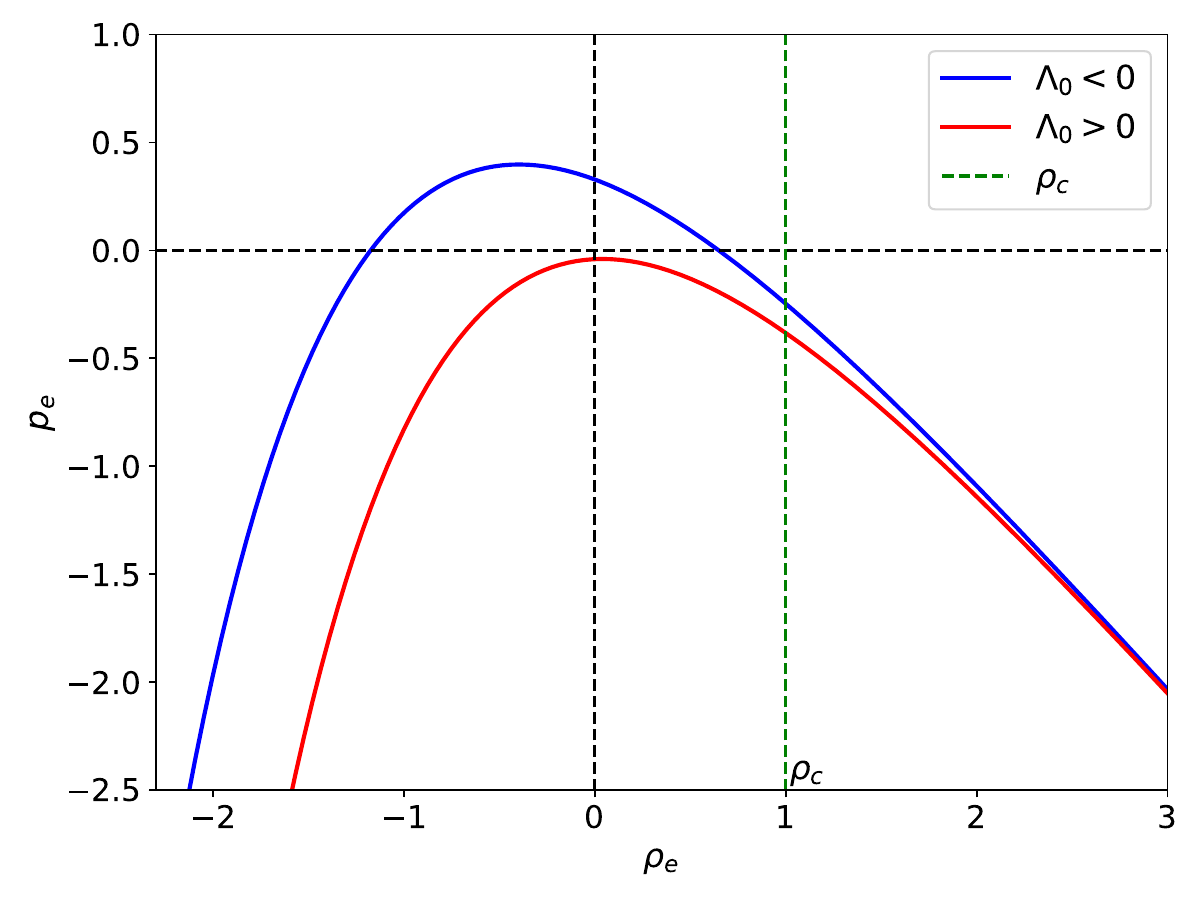}
	\caption{ Pressure $ p_{\mathrm{e}}$ as a function of the energy density $\rho_\mathrm{e}$ for two cases: $\Lambda_0 > 0$ (red curve) and $\Lambda_0 < 0$ (blue curve). The vertical dashed green line marks the critical energy density $ \rho_\mathrm{c}$. This plot depicts how the pressure varies with the energy density for different signs of the cosmological constant. The parameters are set as $\Lambda_{0}=1$ (positive value), $\Lambda_{0}=-10$ (negative value), and $\bar{\alpha}=1$.
	}
	\label{fgdfhg}
\end{figure}

At high effective energy densities, $\rho_{\mathrm{e}} \gg \Lambda_{0}/8\pi$, the effective EOS \eqref{dmjdjdjhd} approaches $p_{\mathrm{e}} \simeq -\rho_{e}$. Furthermore, for $\rho_{\mathrm{e}} = \Lambda_{0}/8\pi$, it  reduces exactly to $p_{\mathrm{e}} = -\rho_{\mathrm{e}}$. 
These results are noteworthy. At sufficiently high energy densities and for specific values of the effective energy density, we obtain a de Sitter (or anti-de Sitter) spacetime. Additionally, when the effective energy density vanishes, the effective pressure attains a constant value, expressed as $
p_{\mathrm{e}}=\big(1-e^{\alpha\Lambda_{0}/8\pi}\big)/\alpha
$, which is negative for a positive cosmological constant $\Lambda_{0}>0$ and positive for a negative cosmological constant  $\Lambda_{0}<0$. In the limit $\alpha \to 0$, this pressure converges to 
\begin{equation}
	p_{e}=\lim_{\alpha \to 0} \frac{1-e^{\alpha\Lambda_{0}/8\pi}}{\alpha}=-\frac{\Lambda_{0}}{8\pi}.
\end{equation}
The effective pressure described by Eq.~\eqref{dmjdjdjhd} varies between positive and negative values for different values of the effective energy density. The conditions that lead to a sign change of the effective pressure are as follows
\begin{itemize}
	\item $p_{\mathrm{e}} < 0 \;~~ \text{if} ~~ \;  e^{\alpha \left( \frac{\Lambda_{0}}{8\pi} - \rho_{\mathrm{e}} \right)} > 1 - \alpha \rho_{e}$,
	\item $p_{\mathrm{e}} > 0 ~~\; \text{if}~~ \; \{ e^{\alpha \left( \frac{\Lambda_{0}}{8\pi} - \rho_{\mathrm{e}} \right)} < 1 - \alpha \rho_{\mathrm{e}}, \; \rho_{\mathrm{e}} < \frac{1}{\alpha} \}$.
\end{itemize}
The critical effective energy density $\rho_{\mathrm{c}}=1/\alpha$ is of particular significance. When the effective energy density is below this critical threshold, the effective pressure can be positive. However, once $\rho_{\mathrm{e}}$ exceeds $\rho_{\mathrm{c}}$, the effective pressure becomes strictly negative . This negative pressure can play a fundamental role in facilitating the creation of non-singular spacetime at high energy densities.    
Fig.~\ref{fgdfhg} illustrates the general behavior of the effective pressure in Eq.~\eqref{dmjdjdjhd} as a function of the effective energy density.
\subsubsection{$w=-1$}
Among the various possible linear EOS, the case $w = -1$ has distinctive characteristics. In the limit $w \to -1$, we find
\begin{equation}\label{dwdlfjjw}
	\lim_{w \to -1^{+}} -\frac{8\pi G_0 \epsilon}{1 + \alpha G_0 \epsilon^{\frac{1}{1+w}}} =
	\begin{cases}
		0 & \text{if } \epsilon > 1, \\
		-8\pi G_0 \epsilon & \text{if } 0 < \epsilon < 1
	\end{cases}
\end{equation}
and 
\begin{equation}\label{dcjnwdjdwed}
	~_2F_1(1,1+w;2+w;-\alpha G_{0}\epsilon^{\frac{1}{1+w}})\Big|_{w\to -1}=1,
\end{equation}	
Substituting these equations into Eqs.~\eqref{sjshsh} and \eqref{nhhhh} leads to two different scenarios
\begin{itemize}
	\item $\epsilon >1$
	\begin{equation}\label{gthtr}
		\chi(\epsilon)=\frac{\Lambda_{0}}{\epsilon}+8\pi G_{0},\quad
		\Lambda(\epsilon)\simeq\Lambda_{0}+8\pi G_{0}\epsilon,
	\end{equation}
	Inserting these equations into Einstein's field equations \eqref{gdg} yields
	\begin{eqnarray}\label{wwjwhwh}
		G_{\mu\nu}\simeq-(\Lambda_{0}+16\pi G_{0}\epsilon_{0})g_{\mu\nu},
	\end{eqnarray}
	which is the Einstein  field equation with a cosmological constant.
	\item $0<\epsilon<1$
	\begin{equation}\label{gthtr}
		\chi(\epsilon)=\frac{\Lambda_{0}}{\epsilon}+8\pi G_{0},\quad
		\Lambda(\epsilon)\simeq\Lambda_{0},
	\end{equation}
	Similarly, substituting these equations into Einstein's field equations \eqref{gdg} yields
	\begin{eqnarray}\label{fvrgfbrfbvr}
		G_{\mu\nu}\simeq-(\Lambda_{0}+8\pi G_{0}\epsilon_{0})g_{\mu\nu},
	\end{eqnarray}
	This equation represents the Einstein field equations with a cosmological constant and an integration constant.
\end{itemize}
Indeed, when $w = -1$, the gravitational field equations given by Eq.~\eqref{aa} describe a de Sitter (or anti-de Sitter) spacetime with distinct constant values in the high and low energy density limits.
\subsubsection{$w=1$}
An important and well-known EOS is characterized by $w=1$ \cite{Mazur:2004fk}. For this EOS, Eqs.~\eqref{sjshsh} and \eqref{nhhhh} are simplified to
\begin{align}
	\label{rfergfer}
	\chi_{1}(\epsilon) &= \frac{\Lambda_{0}}{\epsilon} + \frac{16\pi}{\alpha\sqrt{\epsilon}} - \frac{16\pi\ln\big(1+\alpha G_{0}\sqrt{\epsilon}\big)}{\alpha^2 G_{0} \epsilon},  \\
	\Lambda_{1}(\epsilon) &= \Lambda_{0}+\frac{8\pi \sqrt{\epsilon}}{\alpha}\frac{2+\alpha G_{0} \sqrt{\epsilon}}{1+\alpha G_{0} \sqrt{\epsilon}} - \frac{16\pi\ln\big(1+\alpha G_{0}\sqrt{\epsilon}\big)}{\alpha^2 G_{0}}, 
	\label{rfergfergf} 
\end{align}
To derive the above equations, we used Eq.~\eqref{eferfrefr}.
Inserting Eqs.~\eqref{rfergfer} and \eqref{rfergfergf} into Eq.~\eqref{aa} yields the following expressions for the effective pressure and effective energy density 
\begin{align}\label{dkjwdfhbwfb}
	&\rho_{\mathrm{e}}=\frac{G_{0}\epsilon}{1+\alpha G_{0}\sqrt{\epsilon}}
	+\frac{\Lambda_{1}(\epsilon)}{8\pi},
	\\
	&p_{\mathrm{e}}=\frac{G_{0}\epsilon}{1+\alpha G_{0}\sqrt{\epsilon}}-\frac{\Lambda_{1}(\epsilon)}{8\pi}.
	\label{sfdgbrggr}
\end{align}	
In general, deriving an analytical expression for the effective EOS by combining Eqs.~\eqref{dkjwdfhbwfb} and \eqref{sfdgbrggr} is not possible. 
However, for small energy densities, $\epsilon \ll (\alpha G_{0})^{-2}$, the effective EOS is approximated by
\begin{equation}\label{dnwdfjbwdj}
	p_{\mathrm{e}}\simeq \rho_{\mathrm{e}}-\frac{\Lambda_{0}}{4\pi}-2\alpha \sqrt{G_{0}}\Big(\rho_{\mathrm{}e}-\frac{\Lambda_{0}}{8\pi}\Big)^{3/2},
\end{equation}
where the third term on the right-hand side represents the familiar polytropic EOS\footnote{The polytropic EOS,  $p= K \rho^\gamma$, where
	$\gamma = 1 +1/n$, (with $n$ as the polytropic index and $K$ as the proportionality constant) provides a simplified model to describe the thermodynamic behavior of self-gravitating fluids. By defining $\bar{\rho}_{\mathrm{e}}=\rho_{\mathrm{e}}-\Lambda_0/8\pi$, the effective EOS \eqref{dnwdfjbwdj}, becomes $p_{\mathrm{e}}\simeq -\Lambda_{0}/8\pi+\bar{\rho}_{\mathrm{e}}+K\bar{\rho}_{\mathrm{e}}^{3/2}$, where the third term corresponds to a polytropic EOS with a polytropic index $n=2$ and $K=-2\alpha \sqrt{G_{0}}$.}. 
For $\rho_{\mathrm{e}} = \Lambda_{0} / 8\pi$, Eq.~\eqref{dnwdfjbwdj} becomes exactly $p_{\mathrm{e}} = -\rho_{\mathrm{e}}$. 

\begin{figure}[h]
	\centering
	\includegraphics[scale=0.44]{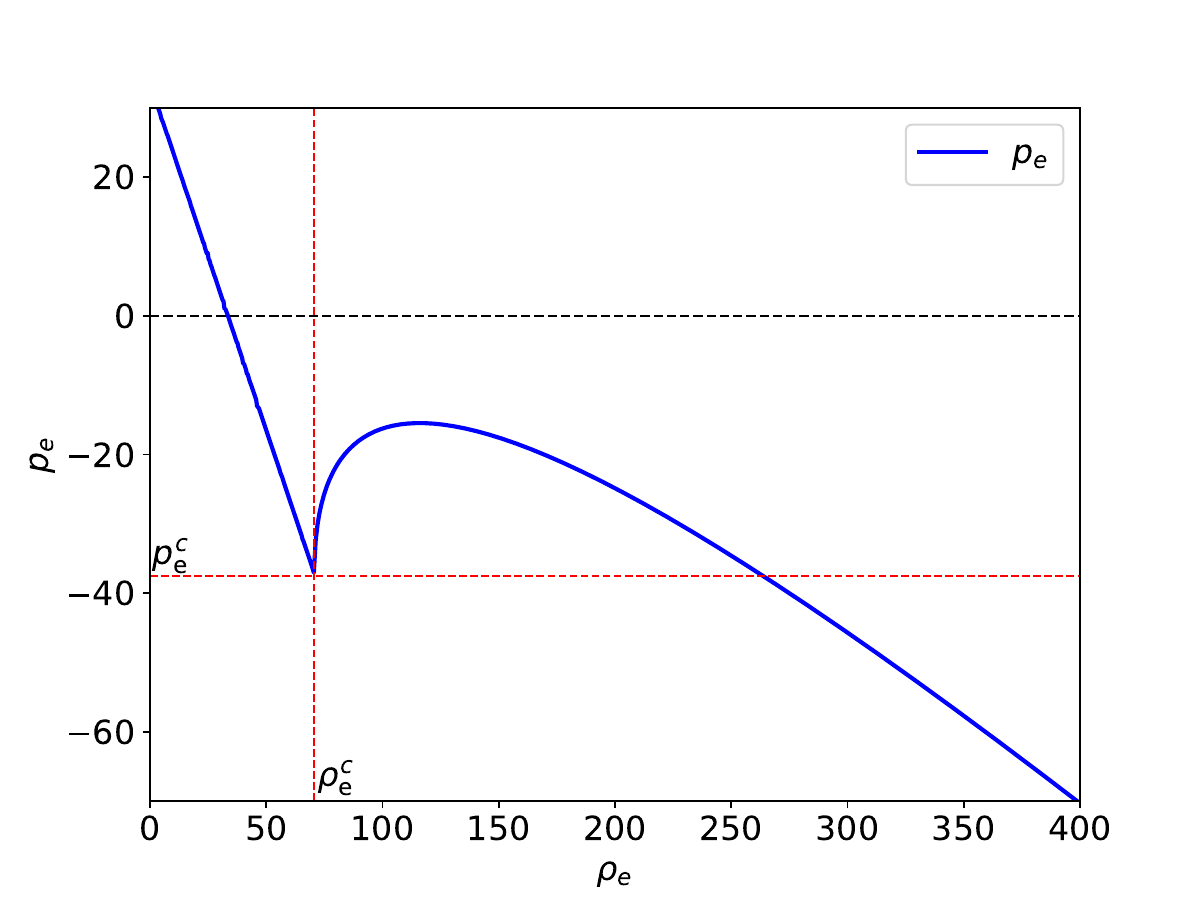}
	\caption{Effective pressure $p_\mathrm{e}$ as a function of the effective energy density $\rho_\mathrm{e} $ according to Eq.~\eqref{sdffdbggfggfrg}.  The parameters are set as $8\pi G_{0}=1$, and $\bar{\alpha}=\Lambda_{0}=1$.	}
	\label{fgdfghdgf}
\end{figure}

For high  energy densities, where $\epsilon \gg (\alpha G_{0})^{-2}$, the EOS obtained from Eqs.~\eqref{dkjwdfhbwfb} and \eqref{sfdgbrggr} 
\begin{equation}\label{sdffdbggfggfrg}
	(p_{\mathrm{e}}+\rho_{\mathrm{e}})e^{-\alpha^2 G_{0}(3p_{\mathrm{e}}+\rho_{\mathrm{e}})/4}\simeq \frac{2}{\alpha^2 G_{0}}e^{\alpha^2 G_{0}\Lambda_{0}/16\pi}.
\end{equation}
Figure~\ref{fgdfghdgf} illustrates the general relationship between the effective pressure, given by Eq.~\eqref{sdffdbggfggfrg} and the effective energy density.
This figure demonstrates that the pressure in the effective EOS \eqref{sdffdbggfggfrg} is positive in the low effective-energy density limit and becomes negative at high effective energy densities.  This negative pressure may slow stellar collapse or avoid
the formation of a singularity in the final stage of gravitational collapse \cite{Shojai:2022pdq}.\\

Figure \ref{fgdfghdgf} illustrates a discontinuity in the pressure equation as a function of the effective energy density, as described by Eq.~\eqref{sdffdbggfggfrg}. To investigate this behavior in greater detail, we differentiate both sides of Eq.~\eqref{sdffdbggfggfrg} with respect to $\rho_{\mathrm{e}}$, yielding
\begin{equation}\label{dwkjdjk}
	\frac{dp_{\mathrm{e}}}{d\rho_{\mathrm{e}}}= -\frac{4-\alpha^2 G_{0}(p_{\mathrm{e}}+\rho_{\mathrm{e}})}{4-3\alpha^2 G_{0}(p_{\mathrm{e}}+\rho_{\mathrm{e}})},
\end{equation}
A slope discontinuity occurs when the denominator vanishes, leading to
\begin{equation}
	p^{c}_{\mathrm{e}}=-\rho^{c}_{\mathrm{e}}+\frac{4}{3\alpha^2 G_{0}},
\end{equation}
where $p^{c}_{\mathrm{e}}$ denotes the critical effective pressure, and $\rho^{c}_{\mathrm{e}}$ represents the critical effective energy density.
\subsubsection{$w=\frac{1}{2}$}
Another important case is $w=1/2$ for which Eqs.~\eqref{sjshsh} and \eqref{nhhhh} are simplified as
\begin{equation}\label{dbfgbgbf}
	\chi_{1/2}(\epsilon)=\frac{\Lambda_{0}}{\epsilon}+\frac{24\pi}{\alpha \epsilon}\bigg(\epsilon^{1/3}-\frac{\arctan\big(\sqrt{\alpha G_{0}}\epsilon^{1/3}\big)}{\sqrt{\alpha G_{0}}}\bigg),
\end{equation}
\begin{align}
	\label{rfdfwdfwdf}
	\Lambda_{1/2}(\epsilon)=&\Lambda_{0}+\frac{24\pi}{\alpha }\bigg(\epsilon^{1/3}-\frac{\arctan\big(\sqrt{\alpha G_{0}}\epsilon^{1/3}\big)}{\sqrt{\alpha G_{0}}}\bigg)\notag\\
	&-\frac{8\pi G_{0}\epsilon}{1+\alpha G_{0}\epsilon^\frac{2}{3}},
\end{align}
where we applied Eq.~\eqref{rtgtrgtr}.  Substituting Eqs.~\eqref{dbfgbgbf} and \eqref{rfdfwdfwdf} into Eq.~\eqref{aa} gives the following expressions for the effective pressure and effective energy density
\begin{align}\label{gergre}
	&\rho_{\mathrm{e}}=\frac{G_{0}\epsilon}{1+\alpha G_{0}\epsilon^{2/3}}
	+\frac{\Lambda_{1/2}(\epsilon)}{8\pi},
	\\
	&p_{\mathrm{e}}=\frac{G_{0}\epsilon}{2(1+\alpha G_{0}\epsilon^{2/3})}-\frac{\Lambda_{1/2}(\epsilon)}{8\pi},
	\label{fgbrgftr}
\end{align}	
It is not possible to combine these equations to obtain an analytic effective EOS. However, they can be analyzed in the low- and high-energy density limits. In the small energy density limit $ \epsilon \ll (\alpha G_{0})^{-3/2}$, we can approximate $\arctan(x)$ as 
$\arctan(x) \simeq x -x^3/3 $. Combining Eqs.~\eqref{gergre} and \eqref{fgbrgftr} yields the following EOS
\begin{equation}\label{tgrewgwre}
	p_{\mathrm{e}}\simeq \frac{\rho_{\mathrm{e}}}{2}-\frac{3\Lambda_{0}}{16\pi}-\frac{3}{2}\alpha G^{1/3}_{0}\left(\rho_{\mathrm{e}}-\frac{\Lambda_{0}}{8\pi}\right)^{5/3},
\end{equation}	
The third term on the right-hand side takes the familiar form of a polytropic EOS\footnote{Defining $\bar{\rho}_{\mathrm{e}} = \rho_{\mathrm{e}} - \Lambda_0 / 8\pi$, the effective EOS in Eq.~\eqref{tgrewgwre} takes the form $p_{\mathrm{e}} \simeq -\Lambda_0 / 8\pi + \bar{\rho}_{\mathrm{e}} / 2 + C \bar{\rho}_{\mathrm{e}}^{5/3}$, where the third term represents a polytropic EOS with a polytropic index $n = 3/2$ and $C = -3\alpha G_0^{1/3} / 2$.}, which has been observed in the gravitational collapse of an old star into the Bardeen spacetime (see Ref.~\cite{Shojai:2022pdq})\footnote{For the Bardeen BH, the stellar matter is modeled as a polytropic fluid governed by the EOS $p=-(4\pi/3m)^{2/3}g^2 \rho^{5/3}$.}, and in the low energy density limit of scale-dependent gravity (see Ref.~\cite{PhysRevD.111.064069})\footnote{For a BH in scale-dependent gravity, the stellar matter is modeled as a polytropic fluid, governed by the EOS in the low-energy density limit $\tilde{p}\simeq-K\tilde{\omega}\tilde{\rho}^{5/3}$.}. 
\\
At the high energy density limit, $\epsilon \gg (\alpha G_{0})^{-3/2}$, Eqs.~\eqref{gergre} and \eqref{fgbrgftr} can be combined to give the following effective EOS
\begin{equation}\label{wdkwdwkjbwj}
	p_{\mathrm{e}}\simeq-\frac{1}{2}\rho_{\mathrm{e}}+\frac{1}{4}\bigg(\frac{3\pi}{\alpha\sqrt{\alpha G_{0}}}-\frac{\Lambda_{0}}{4\pi}\bigg),
\end{equation}		
At high effective energy density, this EOS approaches $p_{\mathrm{e}}\simeq -\rho_{\mathrm{e}}/2$. A similar EOS at high energy densities has been found in the gravitational collapse of a massive object described by the 4D Einstein-Gauss-Bonnet (4D-EGB) theory of gravity\footnote{Gravitational collapse in 4D-EGB \cite{Hassannejad:2023lrp}, predicts the stellar
	EOS as
	\begin{align}\label{SSlp}
		\tilde p^{\text{eff}}=-\frac{128 \pi^2 \tilde \alpha  (\tilde \rho^{\text{eff}})^2}{3} (1+ \frac{256 \pi^2 \tilde \alpha}{3}\tilde \rho^{\text{eff}})^{-1},
	\end{align}
	For sufficiently large energy density, Eq.~\eqref{SSlp} reduces to the linear EOS $\tilde p^{\text{eff}}=-\tilde \rho^{\text{eff}}/2$
	.}.
\subsubsection{$w=-\frac{1}{2}$}
Another notable case is $w = -1/2$, where Eqs.~\eqref{sjshsh} and \eqref{nhhhh} reduce to
\begin{equation}\label{fvedfwdf}
	\chi_{(-1/2)}(\epsilon)=\frac{\Lambda_{0}}{\epsilon}+8\pi \sqrt{\frac{G_{0}}{\alpha}}\frac{\arctan\big(\sqrt{\alpha G_{0}}\epsilon\big)}{\epsilon},
\end{equation}
\begin{align}\label{fefvfevfe}
	\Lambda_{(-1/2)}(\epsilon)=&\Lambda_{0}+8\pi \sqrt{\frac{G_{0}}{\alpha}}\arctan\big(\sqrt{\alpha G_{0}}\epsilon\big)\notag\\
	&-\frac{8\pi G_{0}\epsilon}{1+\alpha G_{0}\epsilon^{2}},
\end{align}
To obtain the above equations, we employed Eq.~\eqref{dndndnd}. 
Substituting Eqs.~\eqref{fvedfwdf} and \eqref{fefvfevfe} into Eq.~\eqref{aa}, we obtain the following expressions for the effective pressure and energy density
\begin{align}\label{hdisdcb}
	&\rho_{\mathrm{e}}=\frac{G_{0}\epsilon}{1+\alpha G_{0}\epsilon^{2}}
	+\frac{\Lambda_{(-1/2)}(\epsilon)}{8\pi},
	\\
	&p_{\mathrm{e}}=-\frac{G_{0}\epsilon}{2(1+\alpha G_{0}\epsilon^{2})}-\frac{\Lambda_{(-1/2)}(\epsilon)}{8\pi},
	\label{fvfevfev}
\end{align}	
At the low energy density limit, $ \epsilon \ll (\alpha G_{0})^{-2}$, the effective EOS can be expressed analytically by combining Eqs.~\eqref{hdisdcb} and \eqref{fvfevfev}, yielding
\begin{equation}\label{wdevffev}
	p_{\mathrm{e}}\simeq-\frac{1}{2}\rho_{\mathrm{e}}-\frac{\Lambda_{0}}{16\pi}-\frac{\alpha}{2G_{0}}\left(\rho_{\mathrm{e}}-\frac{\Lambda_{0}}{8\pi}\right)^3,
\end{equation}
where the third term on the right-hand side corresponds to a polytropic EOS \footnote{By introducing $\bar{\rho}_{\mathrm{e}} = \rho_{\mathrm{e}} - \Lambda_0 / 8\pi$, the effective EOS in Eq.~\eqref{wdevffev} can be expressed as $p_{\mathrm{e}} \simeq -\Lambda_0 / 8\pi - \bar{\rho}_{\mathrm{e}} / 2 + K \bar{\rho}_{\mathrm{e}}^{3}$. Here, the third term corresponds to a polytropic EOS with a polytropic index $n = 1/2$ and $K = -\alpha / (2G_{0})$.}.

For the high-energy density limit, $\epsilon \gg (\alpha G_{0})^{-2}$, Eqs.~\eqref{hdisdcb} and \eqref{fvfevfev} can be combined to yield the effective EOS,
\begin{equation}\label{rgbgrs}
	p_{\mathrm{e}}\simeq-\frac{3}{2}\rho_{\mathrm{e}}+\frac{\Lambda_{0}}{16\pi}+\frac{\pi}{4}\sqrt{\frac{G_{0}}{\alpha}},
\end{equation}
\subsubsection{$|w| \gg 1$}
In the limit $|w| \gg 1$,  Eqs.~\eqref{sjshsh} and \eqref{nhhhh} reduce to
\begin{equation}\label{fgbfgbfdsg}
	\chi_{w}(\epsilon)=\frac{\Lambda_{0}}{\epsilon}+\frac{8\pi G_{0}}{1+\alpha G_{0}\epsilon^{\frac{1}{w}}}, \quad \quad \Lambda_{w}\simeq \Lambda_{0},
\end{equation}
To obtain the above equation, we used Eq.~\eqref{dckjwdwk}. The first important result of Eq.~\eqref{fgbfgbfdsg} is that the running cosmological constant simply becomes constant.  Substituting these expressions  into Eq.~\eqref{aa}, yields
\begin{align}\label{wefwefwe}
	&\rho_{\mathrm{e}}=\frac{\Lambda_{0}}{8\pi},
	\\
	&p_{\mathrm{e}}=\frac{G_{0}w\epsilon}{1+\alpha G_{0}\epsilon^{\frac{1}{w}}}-\frac{\Lambda_{0}}{8\pi},
	\label{wdfwrefwre}
\end{align}	
The effective pressure \eqref{wdfwrefwre} in the limit $w \gg 1$ can be simplified to
\begin{equation}
	p_{\mathrm{e}}\simeq\frac{G_{0}w \epsilon}{1 + \alpha G_{0}} \left(1 - \frac{\alpha G_{0} \ln(\epsilon)}{(1 + \alpha G_{0})w }\right) + O\left(\frac{1}{w^2}\right),
\end{equation}
As $w$ approaches infinity, this pressure converges to  $p_{\mathrm{e}}\simeq G_{\infty}w\epsilon$, where 
\begin{align}\label{feffce}
	G_{\infty}=\frac{G_{0}}{1+\alpha G_{0}},
\end{align}
This can also be seen from Eq.~\eqref{ggdhdh}. It shows that for a finite energy density, the running gravitational coupling approaches a finite constant  value $G_{\infty}$, when the pressure diverges ($w \to \pm \infty$). However, $G_{\infty}$ is smaller than the Newtonian gravitational constant $G_0$. The Newtonian gravitational constant is measured in the weak gravity regime, while Eq.~\eqref{feffce} predicts a reduced value in the high-pressure limit (similar to conditions at the center of gravitationally collapsing compact objects or in the early universe).
\\
	To interpret the preceding results, it is important to separate classical EOS effects from those generated by the RG improvement. In classical GR, pressure and the value of the EOS parameter can change the collapse rate, the evolution of the MS mass, and the timing of trapped-surface formation. In inhomogeneous collapse, the initial density, pressure, and velocity profiles may also change the final state between a BH and a naked singularity. Therefore, the $w$-dependence of the collapse potential should not, by itself, be regarded as a quantum-gravity effect. In the present construction, the specifically RG-induced contributions arise from the ultraviolet decrease of the gravitational coupling and the effective cosmological term. The regular central behavior obtained for the negative-$w$ cases considered here is therefore not caused by negative pressure alone, but by the interplay between the EOS-dependent density scaling and the induced cosmological contribution. The present analysis is also restricted to homogeneous collapse and does not address the known sensitivity of inhomogeneous models to their initial profiles.
%%%%%%%%%%%%%%%%%%%%%%%%%
%%%%%%%%%%%%%%%%%%%%%%%%%
%%%%%%%%%%%%%%%%%%%%%%%%%
\section{Friedmann Equations}
The study of FLRW spacetime dynamics is of great importance for any gravitational theory. According to the cosmological principle it can describe the dynamics of our universe  and can serve as a reliable model for the interior of gravitationally collapsed objects. In this section, we investigate FLRW solutions of the gravitational field equations \eqref{gdg}. 
Substituting the FLRW metric \eqref{ddbcdb}, the running coupling \eqref{ggdhdh}, and the density-
dependent cosmological term  \eqref{nhhhh} into the field equations \eqref{gdg} we obtain the modified
Friedmann equation
\begin{equation}\label{dnjd}
	H^{2}+\frac{k_c}{a^2}=\frac{8\pi G_{0}\epsilon}{3}~_2F_{1}\Big(1,1+w;2+w;-\alpha G_{0} \epsilon^{\frac{1}{1+w}}\Big),
\end{equation}
where we assumed $\Lambda_{0}=0$. Eq.~\eqref{dnjd} reduces to the standard Friedmann equation $H^{2}+k_c/a^2=8\pi G_{0}\epsilon/3$ in the limit $\epsilon\ll (\alpha G_{0})^{-(1+w)}$. In the following analysis, we will examine the Friedmann equation for various values of the EOS parameter.
\begin{itemize}
	\item{ 
		For a dust fluid, $w=0$, the Friedmann equation \eqref{dnjd} can be written according to Eq.~\eqref{dnbkwdn} as follows
		\begin{equation}\label{ddnknd}
			H^{2}+\frac{k_c}{a^2}=\frac{8\pi }{3\alpha}\ln\big(1+\alpha G_{0}\epsilon\big),
		\end{equation}
		where it tends to 
		$
		H^{2}+k_c/a^2\simeq\big(8\pi G_{0}\epsilon/3\big)\big(1-\epsilon/\epsilon_{cr}\big)
		$
		using the definition of $\epsilon_{cr}=2/(\alpha G_{0})$ for sufficiently small energy density.
		Eq.~\eqref{ddnknd} is similar to the modified Friedmann equation in loop quantum cosmology \cite{Ashtekar:2008zu}, and the FLRW solution of the 4D-EGB theory of gravity\footnote{The 4D-EGB gravity theory proposed in Ref.~\cite{Glavan:2019inb} makes some significant predictions in both BH physics and cosmology \cite{Fernandes:2022zrq}. Its Friedmann equations at small energy densities reduce to $H^2+\frac{k_c}{a^2}\simeq \frac{8\pi G_{0}\epsilon}{3}\big(1-\frac{64\pi\alpha G_{0}\epsilon}{3}\big)$.}.}
	\item{For the EOS parameter $w=-1/2$, using Eq.~\eqref{dndndnd}, the Friedmann equation \eqref{dnjd} becomes as follows
		\begin{equation}\label{fdveferef}
			H^{2}+\frac{k_c}{a^2}=\frac{8\pi }{3}\sqrt{\frac{G_{0}}{\alpha}}\arctan(\sqrt{G_{0}\alpha}\epsilon),
		\end{equation}
		%We know that $\lim_{\epsilon \to \infty} \arctan(\sqrt{G_{o}\alpha}\epsilon) = \pi/2$, so
		At high energy density, Eq.~\eqref{fdveferef} becomes 
		$H^{2}+k_c/a^2=\Lambda/3$, where $\Lambda=8\pi\sqrt{G_{0}/\alpha}$. This is the Friedmann equation in the presence of a cosmological constant. 
		In the low energy density limit $\epsilon \ll 1/\sqrt{G_{0}\alpha}$, Eq.~\eqref{fdveferef} tends to $H^{2}+k_c/a^2\simeq\big(8\pi G_{0}\epsilon/3\big)\big(1-\epsilon^2/\epsilon_{cr}\big)
		$, where $\epsilon_{cr} = 3/(\alpha G_{0})$.}
	%=3/G_{0}\alpha$.
	\item{For the EOS parameter $w=1/2$, the Friedmann equation \eqref{dnjd} takes the form 
		\begin{equation}\label{gfbeffbvfd}
			H^{2}+\frac{k_c}{a^2}=\frac{8\pi \epsilon^{\frac{1}{3}}}{\alpha}\bigg(1-\frac{\arctan(\sqrt{\alpha G_{0}}\epsilon^{\frac{1}{3}})}{\sqrt{\alpha G_{0}}\epsilon^{\frac{1}{3}}}\bigg),
		\end{equation}
		At high energy density, Eq.~\eqref{gfbeffbvfd} becomes $H^{2}+k_c/a^2 \simeq (8\pi \epsilon^{1/3})/\alpha$. Conversely, in the low energy density limit, $\epsilon \ll (G_{0}\alpha)^{-3/2}$, 
		it reduces to $H^{2}+k_c/a^2\simeq\big(8\pi G_{0}\epsilon/3\big)\big(1-\epsilon^{2/3}/\epsilon_{cr}\big)
		$, where $\epsilon_{cr}=5/(3\alpha G_{0})$.}
	\item{For the case of EOS $w=1$, the Friedmann equation \eqref{dnjd} is simplified to
		\begin{equation}\label{defbdfbdsffbg}
			H^{2}+\frac{k_c}{a^2}=\frac{16\pi \sqrt{\epsilon}}{3\alpha}\bigg(1-\frac{\ln(1+\alpha G_{0}\sqrt{\epsilon})}{\alpha G_{0}\sqrt{\epsilon}}\bigg),
		\end{equation}
		At high energy density, the above equation 
		%\eqref{gfbeffbvfd}
		is simplified to  
		$
		H^{2} + k_c/a^2 \simeq (16\pi\sqrt{\epsilon})/(3\alpha)
		$, while in the low energy density limit, $\epsilon \ll (G_{0} \alpha)^{-2}$, 
		%equation \eqref{fdveferef} 
		it takes the form  
		$
		H^{2} + k_c/a^2 \simeq (8\pi G_{0} \epsilon/3) \left( 1 - \sqrt{\epsilon}/\epsilon_{cr} \right),
		$
		where $\epsilon_{cr} = 3/(2\alpha G_{0})$.}
	\item{For the case $w=-1$, the Friedmann equation \eqref{dnjd} takes the following simple form
		\begin{equation}\label{mhchgchgkcgh}
			H^{2}+\frac{k_c}{a^2}=\frac{8\pi G_{0}\epsilon}{3},
		\end{equation}
		which gives  $\epsilon(a)|_{w=-1}=\epsilon_{0}$ according to Eq.~\eqref{ccncn}. Thus, Eq. \eqref{mhchgchgkcgh} describes the vacuum FLRW space time in the presence of the cosmological constant with $\Lambda=8\pi G_{0} \epsilon_{0}$.} 
	\item{Another significant case is when $|w| \gg 1$ which simplifies Eq.~\eqref{dnjd}  to the following form
		\begin{align}\label{fdwefwdf}
			H^{2}+\frac{k_c}{a^2}
			&\simeq \frac{8\pi G_{0}\epsilon}{3(1+\alpha G_{0} \epsilon^{\frac{1}{w}})}
			\notag\\
			&\simeq \frac{8\pi G_{\infty}\epsilon}{3} \left(1 - \frac{\alpha G_{\infty} \ln(\epsilon)}{w }\right),
		\end{align}
		In the limit $w \to \pm\infty$,  Eq.~\eqref{fdwefwdf} tends to the standard Friedmann equation
		\begin{align}\label{sdfddffevfe}
			H^{2}+\frac{k_c}{a^2}\simeq \frac{8\pi G_{\infty}\epsilon}{3},
		\end{align}
		with a modified gravitational coupling constant.}
\end{itemize}
\subsection{Stability} 
The modified equations outlined above can be studied extensively in cosmological contexts,  especially in the early universe with extremely high energy density. While the study of these equations is beyond the scope of this paper, further calculations could provide deeper insights. It is instructive to express the field equations in terms of kinetic and potential energy \cite{Markov:1985py}, where the potential term allows for a more detailed analysis of the system. The action given in  Eq.~\eqref{dn} for the FLRW geometry can be written as $ S=\int \mathcal{L} d\tau
$, in which the Lagrangian is defined as follows 
\begin{equation}\label{dkwedwed}
	\mathcal{L}=	\frac{-3V_{3}}{8\pi G_{0}}\Big(a\dot{a}^2-k_c a+\frac{a^3}{3}\chi(\epsilon)\epsilon\Big),
\end{equation}	
where $V_{3}$ is the volume of a three-dimensional space\footnote{In a 4-dimensional spacetime with three spatial dimensions, the volume of a three-dimensional space is given by $V_{3}=4\pi\int_{0}^{r_{b}}r^2/\sqrt{1-k_c r^2}dr$.}. Defining the generalized momentum as $p=\partial{\mathcal{L}}/\partial{\dot{a}}=	{-3V_{3}a\dot{a}}/{4\pi G_{0}}$ and constructing the Hamiltonian by performing a Legendre transformation on the Lagrangian: $\mathcal{H}=p\dot{a}-\mathcal{L}$, we obtain 
\begin{equation}
	\mathcal{H}=-\frac{2\pi G_{0}}{3V_{3}}\frac{p^2}{a} -\frac{3 V_{3}a}{8\pi G_{0}} \big(V(a)+k_c \big),
\end{equation}
We know that the $G_{00}$ component of the Einstein equations reduces to the condition on the Hamiltonian $\mathcal{H}=0$. Thus, one can write
\begin{equation}\label{fdvefvfrbgrbgfrb}
	\dot{a}^2+V(a)=-k_c,
\end{equation}
This shows that our problem is equivalent to that of a particle with energy $-k$ moving in the potential $V(a)$.
The integral to be solved is 
\begin{equation}\label{wdfwdfnw}
	V(a)=-\frac{a^2}{3}\int_0^{\epsilon(a)} \frac{8\pi G_0}{1 + \alpha G_{0} \epsilon^{\frac{1}{1 + w}}} \, d\epsilon,
\end{equation}
The solution of this integral involves the Hypergeometric function and can be expressed as
\begin{align}\label{djjdjjreb}
	V(a) = -\frac{\Lambda_{0}}{3}a^2-&\frac{8 \pi G_0\epsilon_{0}}{3} a^{-3w-1}\notag\\
	&\times \, {}_2F_1\left(1, 1+w; 2+w; -\frac{\alpha G_0 \epsilon_{0}^{\frac{1}{1+w}}}{a^3}\right),
\end{align}
where the condition $w > -1$ holds and $\alpha, G_0 \in \mathbb{R}$. Building upon Eq.~\eqref{jbbkk}, we derive the asymptotic behavior of $V(a)$ as $a \to 0$ and $a \to \infty$. A systematic analysis  yields the following result  
\begin{equation}\label{wdfwdfwd}
	\begin{cases}
		\lim_{a\to 0} V(a)\simeq- \frac{\Lambda_{0}}{3}a^2, & \text{if } w<-\frac{1}{3}, \\
		\lim_{a\to \infty} V(a)\simeq- \frac{\Lambda_{0}}{3}a^2, & \text{if } w>-\frac{1}{3},
	\end{cases}
\end{equation}  
This result demonstrates that for a fluid with an EOS parameter satisfying $w < -\frac{1}{3}$, the function $V(a)$ exhibits quadratic dependence at small scale factor. Conversely, for $w > -\frac{1}{3}$, the same quadratic form arises at large values. The presence of symmetry in these asymptotic behaviors highlights the significant impact of $w$ on the evolution of $V(a)$ and suggests a direct connection between the dynamical properties of the system and the underlying EOS.

Substituting Eq.~\eqref{wdfwdfwd} into Eq.~\eqref{fdvefvfrbgrbgfrb}, yields the following result
\begin{equation}\label{feefgergf}
	\left(\frac{\dot{a}}{a}\right)^2+\frac{k_c}{a^2}\simeq \frac{\Lambda_{0}}{3},
\end{equation}  
This equation represents the Friedmann equation in the presence of a cosmological constant $\Lambda_0$. The resulting expression corresponds to the vacuum solution of Einstein's field equations for the FLRW metric, in which the energy density is dominated by the cosmological constant. In this scenario, the expansion of the universe is governed solely by $\Lambda_0$ and the curvature term $k/a^2$. For a spatially flat spacetime with $k=0$, the scale factor in Eq.~\eqref{feefgergf} evolves as $a(t)\propto e^{\pm\sqrt{\Lambda_{0}/3}\tau}$. Therefore, depending on the values of $w$ at small and large distances, the scale factor exhibits exponential evolution. For the collapse process we have $\dot{a}<0$, this implies $R_{b}(\tau)=R_{b}(\tau_{0})e^{-\sqrt{\Lambda_{0}/3}\tau}$. This describes a rapid contraction at small radii for an EOS with $w<-1/3$, a behavior we refer to as "deflation" analogous to early cosmic inflation. Deflation has been observed in the gravitational collapse of an old star into a regular spacetime\footnote{It has been shown that the gravitational collapse of a star into Hayward spacetime predicts an exponential decay for the star's radius at small distances, $ \tilde{R} \propto e^{-\tilde{\tau}/\tilde{l}}$ (see Ref.~\cite{Shojai:2022pdq}). For Bardeen spacetime, the radius follows $\tilde{R} \propto e^{-\tilde{\tau}/\tilde{g}^{3/2}}$ (see Ref.~\cite{Shojai:2022pdq}), and for a spherically symmetric regular spacetime in scale-dependent gravity, it is given by $ \tilde{R} \propto e^{-\tilde{\tau}/\tilde{A}}$ (see Ref.~\cite{PhysRevD.111.064069}).}.

In the following we derive the potential function \eqref{djjdjjreb} for specific values of $w$, and analyze its stability. Subsequently, we consider the case of a zero cosmological constant, $\Lambda_{0}=0$.
\subsubsection{$w=0$}
\begin{figure}[htb]
	\centering
	\begin{subfigure}[b]{0.45\textwidth}
		\includegraphics[width=\textwidth]{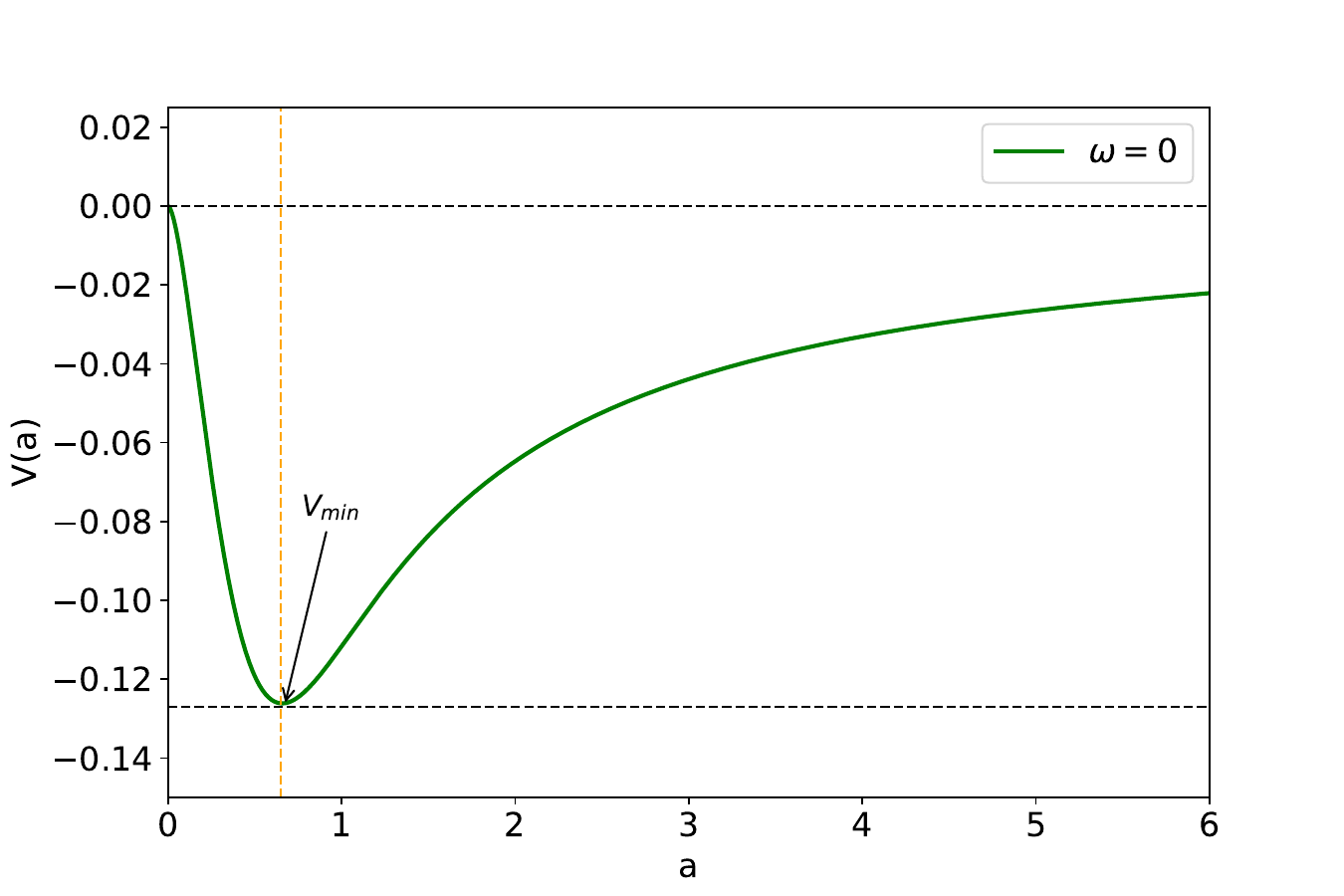}
		\caption{}
		\label{figfff}
	\end{subfigure}
	\hfill
	\begin{subfigure}[b]{0.45\textwidth}
		\includegraphics[width=\textwidth]{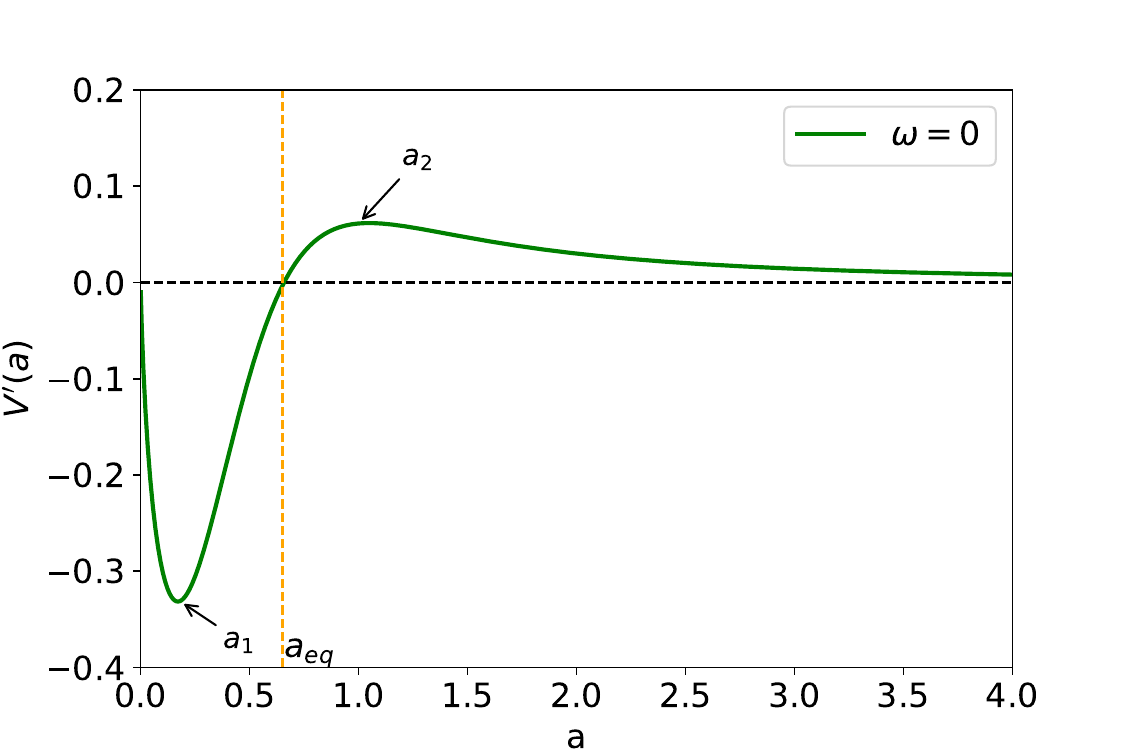}
		\caption{}
		\label{figvvv}
	\end{subfigure}
	\hfill
	\begin{subfigure}[b]{0.45\textwidth}
		\includegraphics[width=\textwidth]{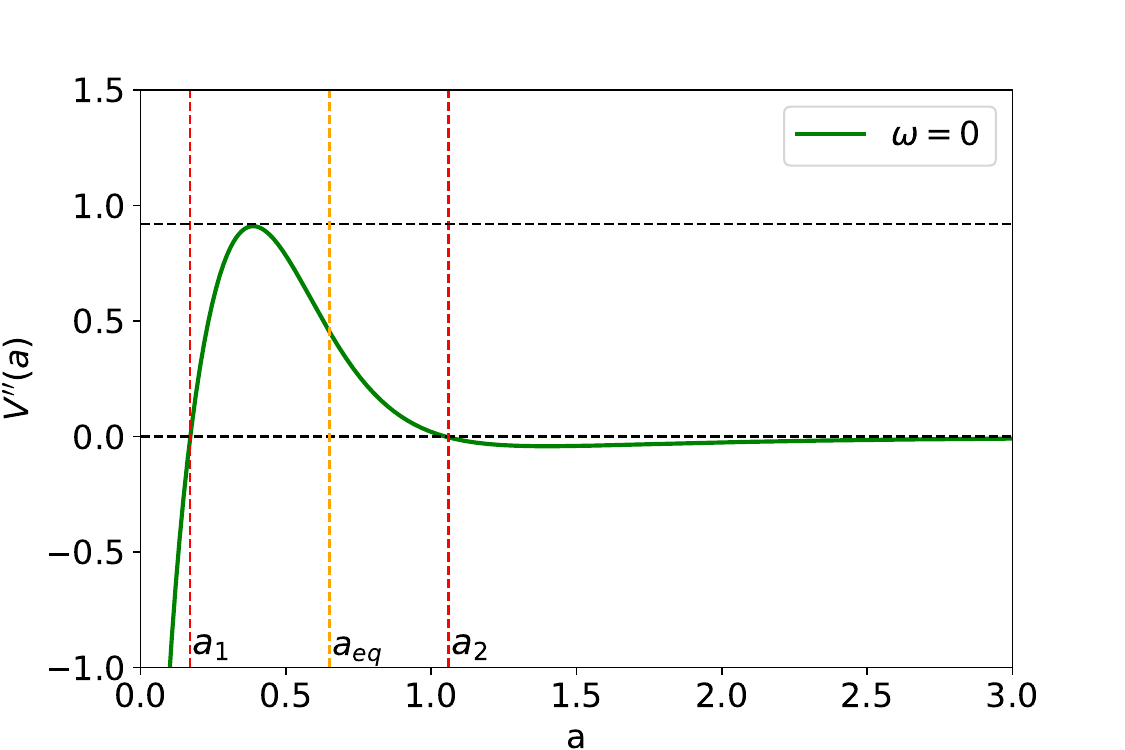}
		\caption{}
		\label{fiffhve}
	\end{subfigure}
	\caption{The potential $V$, its first derivative $dV/d\bar a$, and its second derivative $d^2V/d\bar a^2$, corresponding to Eqs.~\eqref{dwodweif}, \eqref{wdcwdbwed}, and \eqref{dwlwdwdn}, as functions of the dimensionless scale factor $\bar a=M_{\rm Pl}a$ for $w=0$. All quantities are expressed in reduced Planck units, $M_{\rm Pl}=1$, equivalently, $8\pi G_0=1$. The dimensionless parameter values are $\bar\epsilon_0=\frac{5}{4\pi}$, $\bar\alpha_0=8\pi$, $G_0=\frac{1}{8\pi}$. Bars are omitted from the plotted axes for readability.}
	\label{efgregfref}
\end{figure}

If the system behaves like dust,$w = 0$, the potential function approaches
\begin{equation}\label{dwodweif}
	V(a)=- \frac{8\pi a^2}{3\alpha}\ln\Big(1+\frac{\alpha G_{0}\epsilon_{0}}{a^3}\Big),
\end{equation}
At large distances where $a(\tau)\gg (\alpha G_{0}\epsilon_{o})^{1/3}$, the potential function approaches $V(a)\approx- G_{0}\epsilon_{0}/3a$ which corresponds to the Newtonian gravitational potential. Conversely, for small values of the scale factor, the potential function in Eq.~\eqref{dwodweif} takes the form $V(a)\simeq (8\pi a^2/\alpha) \ln a$. 
Assuming that the classical concept of the Newtonian force holds true in the small scale factor regime\footnote{Clearly, the accuracy of the classical notion of gravity, expressed as the gradient of the gravitational potential, $ F(a) = -\partial V(a)/\partial a $, has not been empirically tested at small distances. Here, we use it to evaluate and compare the given potential function providing a deeper insight into its evolution.
}, the force function can be expressed as $F(a)\simeq -(16\pi a/\alpha)\ln a$, indicating a repulsive force.
However, this is not the complete picture. We can also express the force function corresponding to the potential derived in Eq.~\eqref{dwodweif}. Its behavior is illustrated in Fig.~\ref{figvvv}. It is evident that at small distances ($a < a_{\mathrm{eq}}$), the force is repulsive, while at large distances ($a > a_{\mathrm{eq}}$), it becomes attractive.

The stability of a potential function describes how a system behaves when subjected to small perturbations around its equilibrium points. To determine the stability of a potential function $V(a)$, it is essential to analyze the first and second derivatives of the potential. Equilibrium points, denoted as $a_{\mathrm{eq}}$, are determined by solving the equation $V'(a) = 0$. These points represent locations where the force, or equivalently the gradient of the potential, is zero.

To assess the stability of equilibrium points further, we evaluate the second derivative of the potential, $V^{\prime\prime}(a)$. The sign of $V^{\prime\prime}(a)$ at the equilibrium points provides insight into the stability of the system. A positive value of  $V^{\prime\prime}(a)$ at an equilibrium point indicates stability, meaning the system will return to equilibrium after a small perturbation. Conversely, a negative value indicates instability, where perturbations drive the system away from equilibrium.

The first and second derivatives of the potential function $V(a)$ are given by
\begin{equation}\label{wdcwdbwed}
	V'(a) = \frac{8\pi G_0 \epsilon_0 a}{a^3+\alpha G_0 \epsilon_0} - \frac{16\pi a \ln\left(1 + \frac{\alpha G_0 \epsilon_0}{a^3}\right)}{3 \alpha},
\end{equation}
and
\begin{equation}\label{dwlwdwdn}
	V''(a) = \frac{24\pi\alpha G_0^2 \epsilon_0^2}{ \left(a^3+ \alpha G_0 \epsilon_0\right)^2} - \frac{16\pi\ln\left(1 + \frac{\alpha G_0 \epsilon_0}{a^3}\right)}{3 \alpha}.
\end{equation}
respectively.

Dynamical stability often requires that the potential has local minima $V_{min}$, where particles can settle. At these minima, a small displacement causes a restoring force that drives the particle back to the equilibrium point.
For a potential $V(a)$ at an equilibrium point $a_{\mathrm{eq}}$, stability generally requires that $ V^{\prime\prime}(a_{\mathrm{eq}}) > 0$. For the potential function given by Eq.~\eqref{dwodweif}, there is a minimum value $ V_{min} $ located at $a_{\mathrm{eq}}$, as shown in Fig.~\ref{figfff}. This figure illustrates the behavior of the potential function as a function of the scale factor. The potential is negative for all values of the scale factor, starting at zero, rapidly decreasing to a minimum and then slowly increasing while asymptotically approaching zero at infinity. This behavior resembles the Newtonian gravitational potential $V(a) \sim -G_0/a$, which also tends to zero at large distances but diverges to negative infinity at small distances. Consequently, the potential function given by Eq.~\eqref{dwodweif} corrects the Newtonian gravitational potential behavior at short distances.

The second derivative of the potential function \eqref{dwlwdwdn} is shown in Fig.~\ref{fiffhve}. It demonstrates that for $a_{1} < a < a_{2}$, the second derivative is positive. This
indicates that the potential function can be stable within this range. The equilibrium point $a_{\mathrm{eq}}$ lies within this interval, meaning that the system has a dynamically stable potential given by Eq.~\eqref{dwodweif}, with a minimum value at $ a_{\mathrm{eq}}$. The stability discussed here is a local property of the homogeneous minisuperspace potential. It should not be
	confused with curvature regularity of the static exterior extension.\\

We can also examine the behavior of the scale factor near the local minimum $a_{\text{eq}}$. Let
\begin{equation}
	y = a - a_{\text{eq}}, \qquad b = V''(a_{\text{eq}}) > 0.
\end{equation}
Expanding the potential around the minimum gives
\begin{equation}
	V(a) \simeq V(a_{\text{eq}}) + \frac{b}{2} y^2.
\end{equation}
Using the Hamiltonian equation $\dot a^2 + V(a) = -k_c$, we obtain
\begin{equation}
	\dot y^2 = \gamma^2 - \frac{b}{2} y^2,
\end{equation}
where
\begin{equation}
	\gamma^2 = -k_c - V(a_{\text{eq}}).
\end{equation}
Thus, for $\gamma^2 > 0$, the small-amplitude solution is
\begin{equation}
	a(\tau) = a_{\text{eq}} + \gamma \sqrt{\frac{2}{b}} \sin\left[\sqrt{\frac{b}{2}} (\tau - \tau_0)\right].
\end{equation}
Equivalently, the stellar surface radius is
\begin{equation}\label{jwdwew}
	R_b(\tau) = r_b a_{\text{eq}} + r_b \gamma \sqrt{\frac{2}{b}} \sin\left[\sqrt{\frac{b}{2}} (\tau - \tau_0)\right].
\end{equation}

A strictly static equilibrium corresponds to the special case $\gamma = 0$, or $V(a_{\text{eq}}) = -k_c$. The oscillatory solution above describes nearby initial data with a small but nonzero amplitude around the local minimum.

\begin{figure}[h]
	\centering
	\includegraphics[scale=0.44]{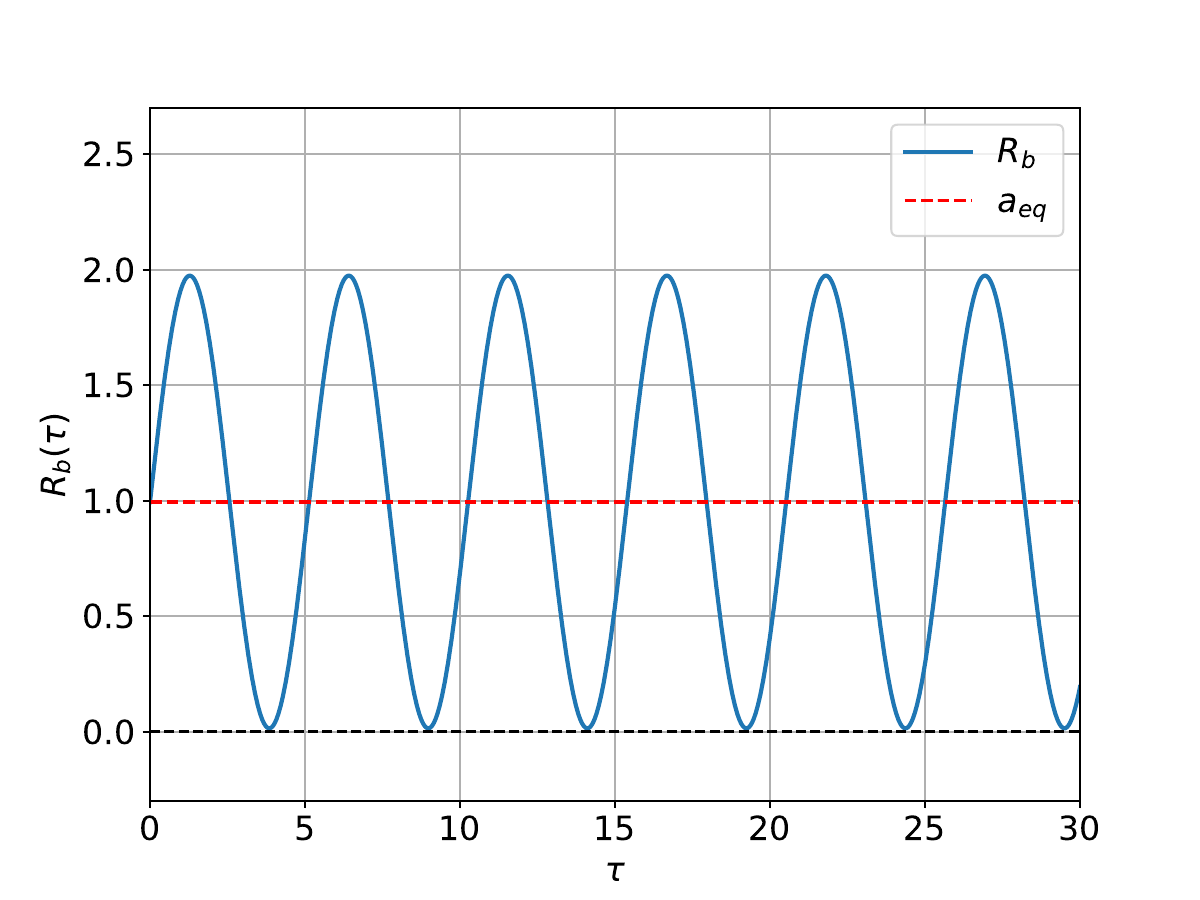}
	\caption{ Normalized stellar radius $R_b(\tau)/r_b$ obtained from Eq.~\eqref{jwdwew}, for $\gamma=1.2$, $b=3$, $a_{\rm eq}=1$, and $\tau_0=0$. The dashed line denotes the equilibrium value $a_{\rm eq}$.}
	\label{eferfr3f}
\end{figure}

For a short time period $\tau\ll 1$, Eq.~\eqref{jwdwew} can be approximated as $R_{b}(\tau)\simeq r_{b}a_{\mathrm{eq}}+r_{b}\gamma \tau$. The behavior of Eq.~\eqref{jwdwew} is shown in Fig.~\ref{eferfr3f}. It oscillates around the equilibrium point $a_{\mathrm{eq}}$. The oscillation of a star's surface radius around the equilibrium point is an intriguing subject in both theoretical and observational physics.  	 

\subsubsection{$w=-1/2$}
For a system composed of exotic matter with $w=-1/2$, the potential function takes the form
\begin{equation}\label{wdkwddbowfd}
	V(a)=-\frac{8\pi}{3}\sqrt{\frac{G_{0}}{\alpha}}a^2\arctan\Big(\frac{\epsilon_{0}\sqrt{\alpha G_{0}}}{a^{3/2}}\Big),
\end{equation}
At small distances, the potential function \eqref{wdkwddbowfd} simplifies to
$ V(a) = -C a^2 / 3 $, where $ C = 4\pi^2\sqrt{G_0/\alpha} $. This corresponds to the potential function of de Sitter spacetime. Applying the classical definition of force to this potential gives
$ F(a) \sim C a $, indicating a repulsive force. 
At large distances, the potential function approaches $ V(a) \sim -(8\pi G_{0}\epsilon_{0}/3)\sqrt{a} $, which also leads to a repulsive force given by $ F(a) \sim (4\pi G_{0}\epsilon_{0}/3)/\sqrt{a} $. The potential in Eq.~\eqref{wdkwddbowfd} exhibits an intriguing property: its derivative, interpreted as a classical force, remains repulsive and exhibits distinct behaviors at different distances. At large distances, the force decreases inversely with the square root of the distance, while at small distances it varies linearly with distance. 

To analyze the dynamics and stability of the potential in Eq.~\eqref{wdkwddbowfd}, we must compute both its first and second derivatives, as previously stated. These can be computed as
\begin{align}\label{aawdbwdbw}
	V^{\prime}(a)=&-\frac{16\pi a}{3}\sqrt{\frac{G_{0}}{\alpha}}\arctan\bigg(\frac{\epsilon_{0}\sqrt{G_{0}\alpha}}{a^{3/2}}\bigg)\notag\\
	&+\frac{4\pi \epsilon_{0} G_{0}}{a^3+\alpha G_{0}\epsilon^{2}_{0}}a^{5/2},
\end{align}	
and
\begin{align}\label{dffefefre}
	V^{\prime\prime}(a)=&-\frac{16\pi}{3}\sqrt{\frac{G_{0}}{\alpha}}\arctan\bigg(\frac{\epsilon_{0}\sqrt{G_{0}\alpha}}{a^{3/2}}\bigg)\notag\\
	&+\frac{18\pi \epsilon_{0} G_{0}}{a^3+\alpha G_{0}\epsilon^{2}_{0}}a^{3/2}-\frac{12 \pi G_{0}\epsilon_{0} a^{9/2}}{(a^3+\alpha G_{0}\epsilon_{0}^2)^2},
\end{align}

\begin{figure}[htb]
	\centering
	\begin{subfigure}[b]{0.48\textwidth}
		\includegraphics[width=\textwidth]{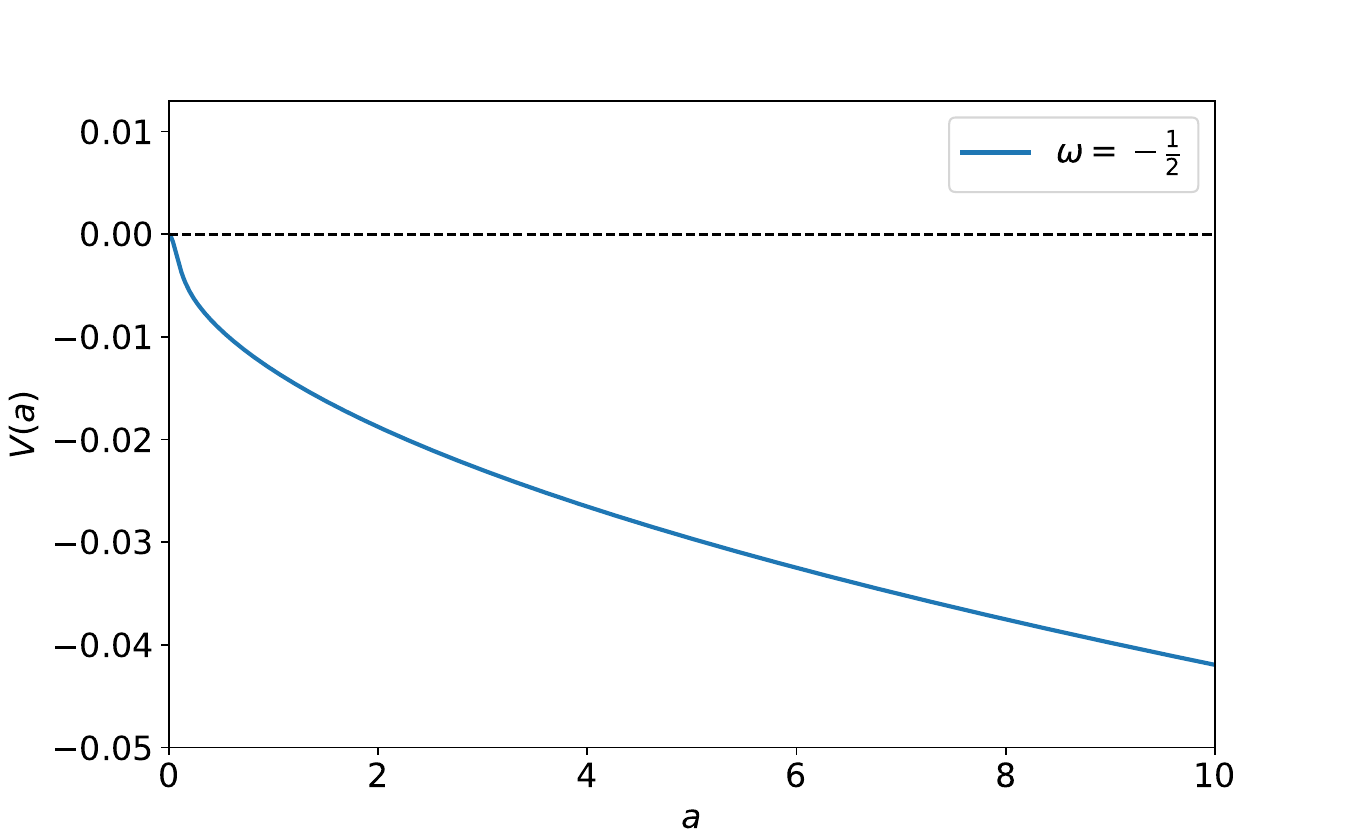}
		\caption{}
		\label{efgwethb}
	\end{subfigure}
	\hfill
	\begin{subfigure}[b]{0.48\textwidth}
		\includegraphics[width=\textwidth]{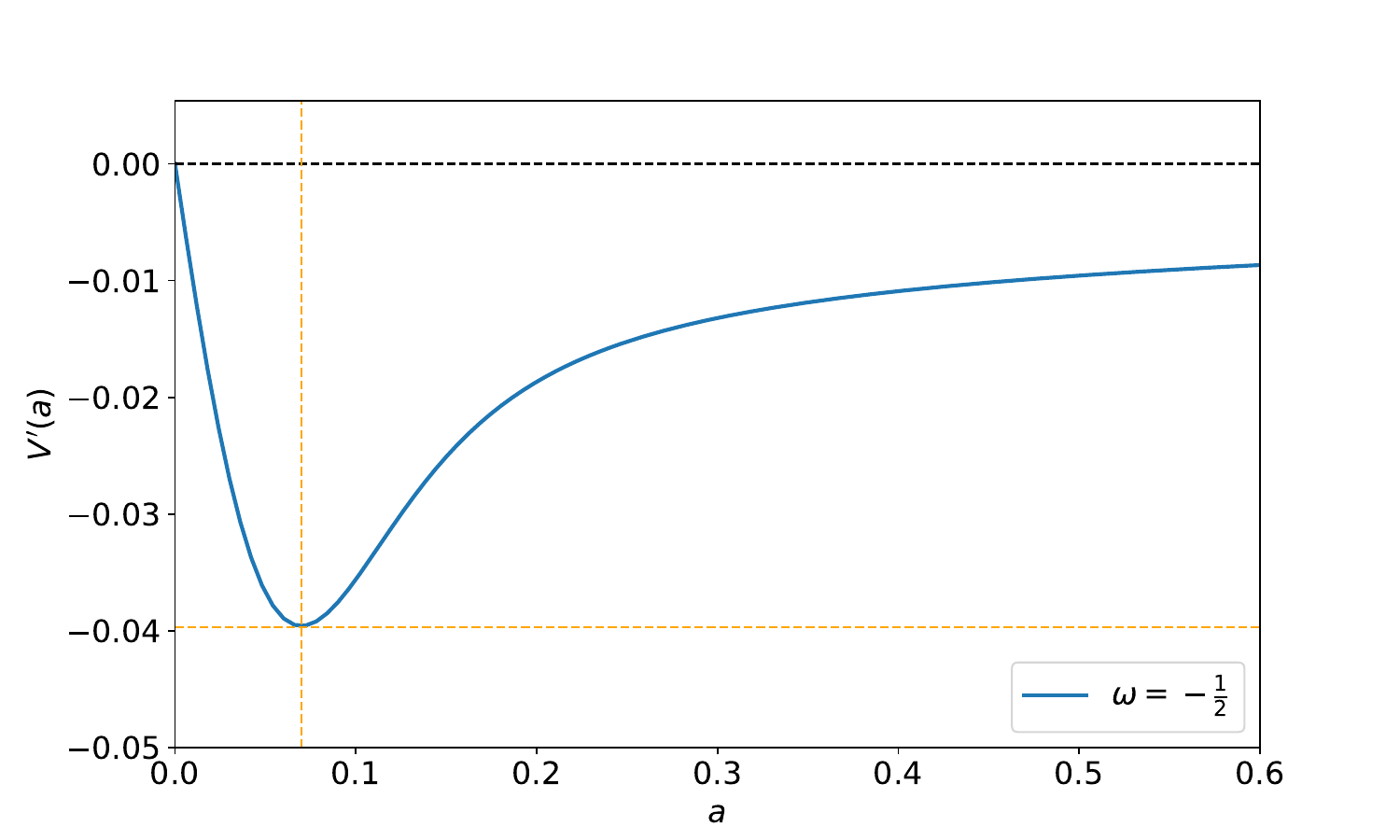}
		\caption{}
		\label{efrwegre}
	\end{subfigure}
	\hfill
	\begin{subfigure}[b]{0.48\textwidth}
		\includegraphics[width=\textwidth]{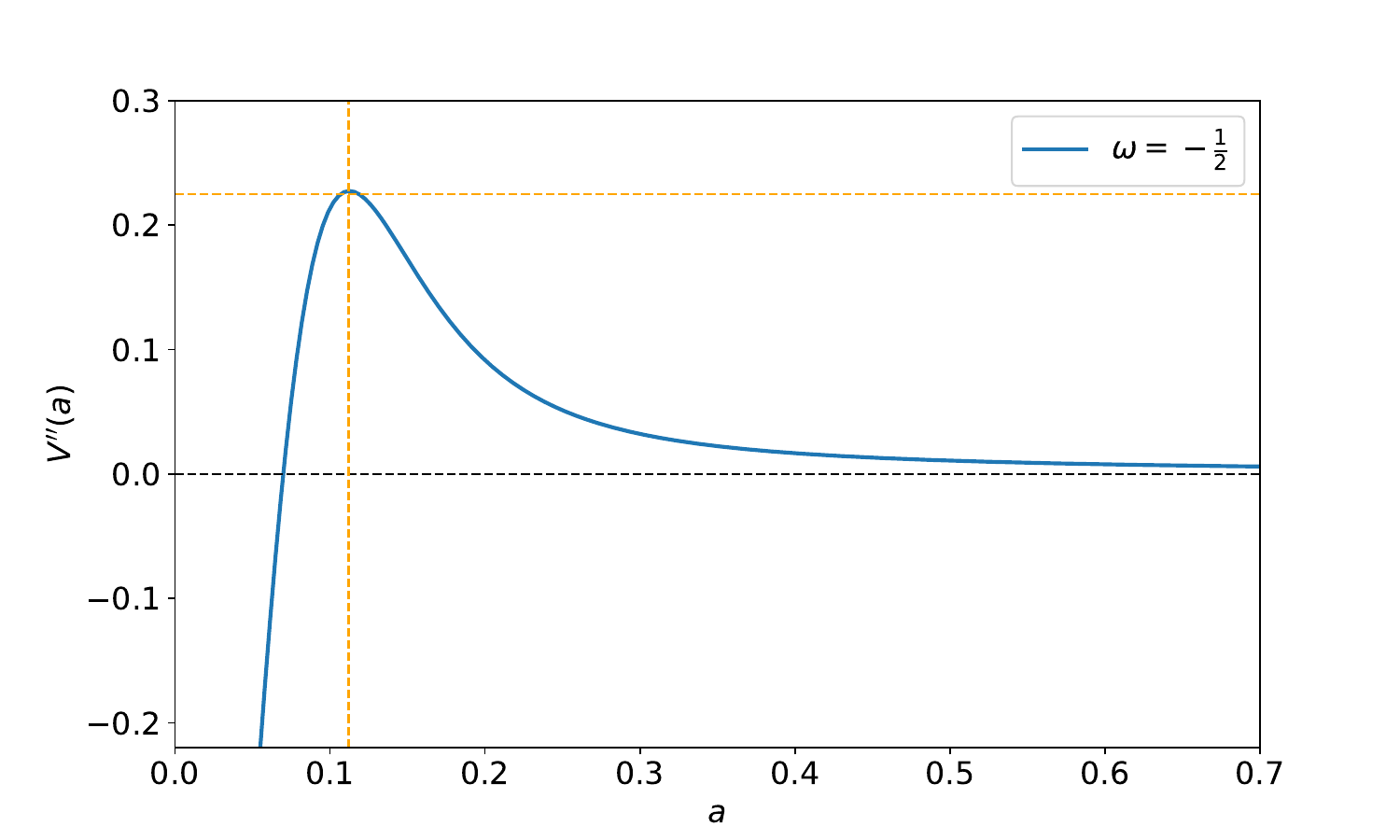}
		\caption{}
		\label{efrgegfr}
	\end{subfigure}
	\caption{The potential $V$, its first derivative $dV/d\bar a$, and its second derivative $d^2V/d\bar a^2$, corresponding to   Eqs.~\eqref{wdkwddbowfd},  \eqref{aawdbwdbw}, and \eqref{dffefefre} as functions of the dimensionless scale factor $\bar a=M_{\rm Pl}a$ for $w=-1/2$. All quantities are expressed in reduced Planck units, $M_{\rm Pl}=1$, equivalently, $8\pi G_0=1$. The dimensionless parameter values are $\bar\alpha_{-1/2}=8\pi$, $\bar\epsilon_0=\frac{1}{8\pi}$, $G_0=\frac{1}{8\pi}$. Here, $\bar\alpha_{-1/2}=\alpha M_{\rm Pl}^6$. Bars are omitted from the plotted axes for readability.}
	\label{fgeffgfeg}
\end{figure}

The analysis of Eqs.~\eqref{wdkwddbowfd}--\eqref{dffefefre} illustrated in Fig.~\eqref{fgeffgfeg}, supports a qualitatively different collapse dynamics from the dust case. Although the small-$a$ asymptotic form of the potential is de Sitter-like, the potential does not provide a finite-radius stable minimum for the homogeneous scale factor. The regularity of the corresponding exterior solution discussed below is therefore not a consequence of local potential stability, but follows from the small-$R$ behavior of the Misner-Sharp mass and the resulting de Sitter-like core.

The potential function defined by Eq.~\eqref{wdkwddbowfd} does not exhibit a local minimum, as shown in Fig.~\ref{efgwethb}. It decreases monotonically with a decreasing scale factor, implying that no stable equilibrium exists. However, at $a = 0$, both the potential function and its derivative, as described by Eq.~\eqref{aawdbwdbw}, reach zero, as shown in Fig.~\ref{efrwegre}. Nevertheless, the second derivative, presented in Eq.~\eqref{dffefefre} and depicted in Fig.~\ref{efrgegfr}, is negative, indicating that the potential is unstable at $a = 0$. Thus, the potential lacks a minimum value and is unstable, unlike the potential 
function for $w = 0$.

\subsubsection{$w=\frac{1}{2}$}
For a system consisting of an EOS with $w=1/2$, the potential function \eqref{wdfwdfnw} takes the form
\begin{equation}\label{fdvfegref}
	V(a)=\frac{8\pi a^2}{\alpha\sqrt{\alpha G_{0}}}\arctan\bigg(\frac{\epsilon_{0}^{1/3}\sqrt{\alpha G_{0}}}{a^{3/2}}\bigg)-\frac{8\pi \epsilon_{0}^{1/3}}{\alpha}a^{1/2},
\end{equation}
At large distances $ a \to \infty $, the potential function behaves asymptotically as $
V(a) \simeq -8\pi \epsilon_{0} G_{0}/3a^{5/2}.
$ 
This result indicates that the potential approaches zero from the negative side, decaying  inversely as a power law with respect to $ a $. The associated force function is given by $
F(a)\simeq -20\pi \epsilon_{0} G_{0}/3 a^{7/2}.
$ Since this force is negative, it acts as an attractive force, that pulls objects toward smaller values of $ a $. The power-law dependence ($ a^{-7/2} $) implies that the attractive force rapidly weakens as $ a $ increases, leading to a diminishing effect at very large distances.  
In contrast, at small distances, $a\to0$, the potential takes a different asymptotic form:  
$
V(a) \simeq -8\pi \epsilon_{0}^{1/3}\sqrt{a}/\alpha .
$
Unlike the large-distance case, this expression exhibits a square-root dependence on $ a $, indicating a much slower variation of the potential near $ a = 0 $. The corresponding force function, obtained by differentiating the potential, is $
F(a) \simeq 4\pi \epsilon_{0}^{1/3}/\alpha \sqrt{a}.
$
This force is  positive, meaning it remains repulsive. However, in this regime, the force diverges as $ a \to 0 $. This implies that as the system approaches very small distances, the repulsive interaction becomes extremely strong. This suggests a dominant influence of the potential at small scales, which could prevent the system from reaching $ a = 0 $ due to an infinitely increasing repulsive force.

\begin{figure}[htb]
	\centering
	\begin{subfigure}[b]{0.48\textwidth}
		\includegraphics[width=\textwidth]{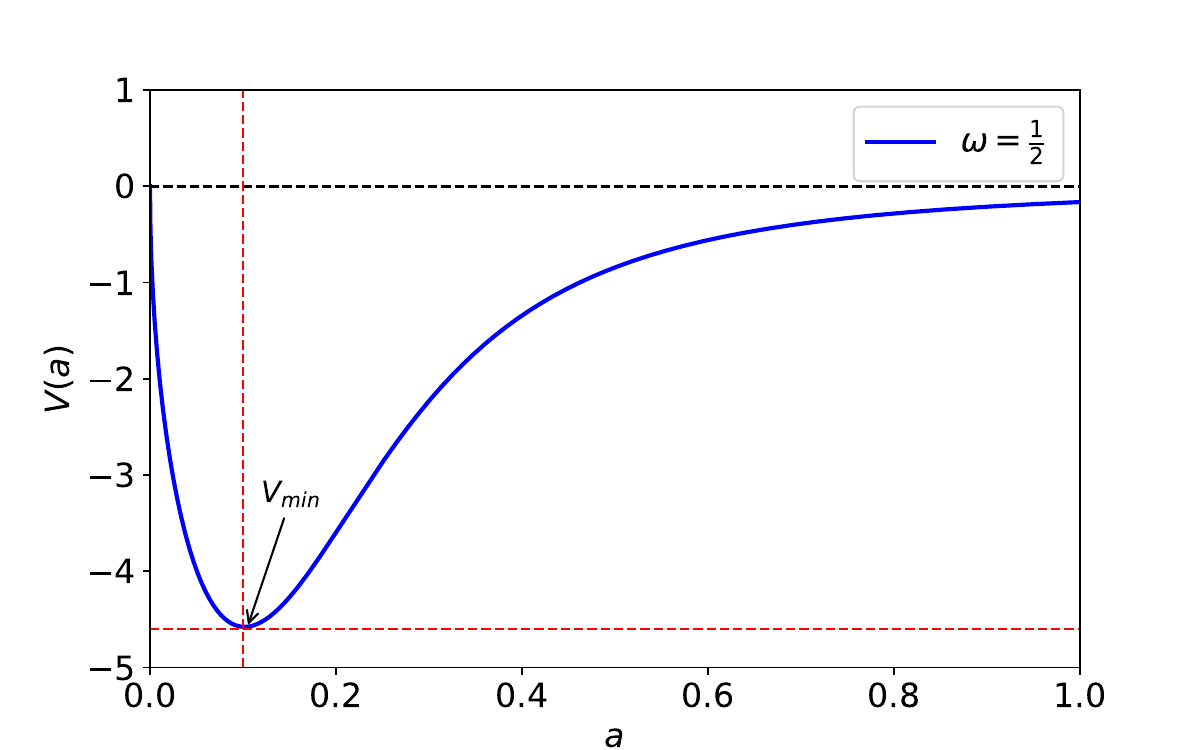}
		\caption{}
		\label{ffvfev}
	\end{subfigure}
	\hfill
	\begin{subfigure}[b]{0.48\textwidth}
		\includegraphics[width=\textwidth]{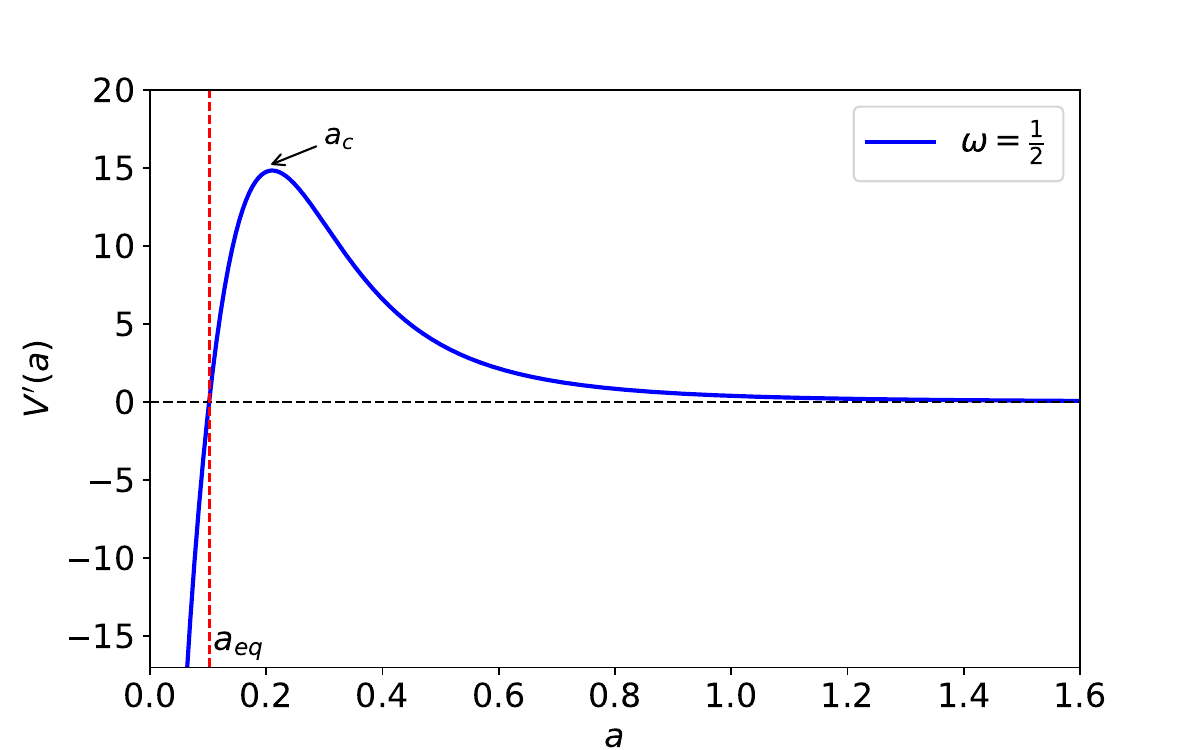}
		\caption{}
		\label{sdcwdcdsc}
	\end{subfigure}
	\hfill
	\begin{subfigure}[b]{0.48\textwidth}
		\includegraphics[width=\textwidth]{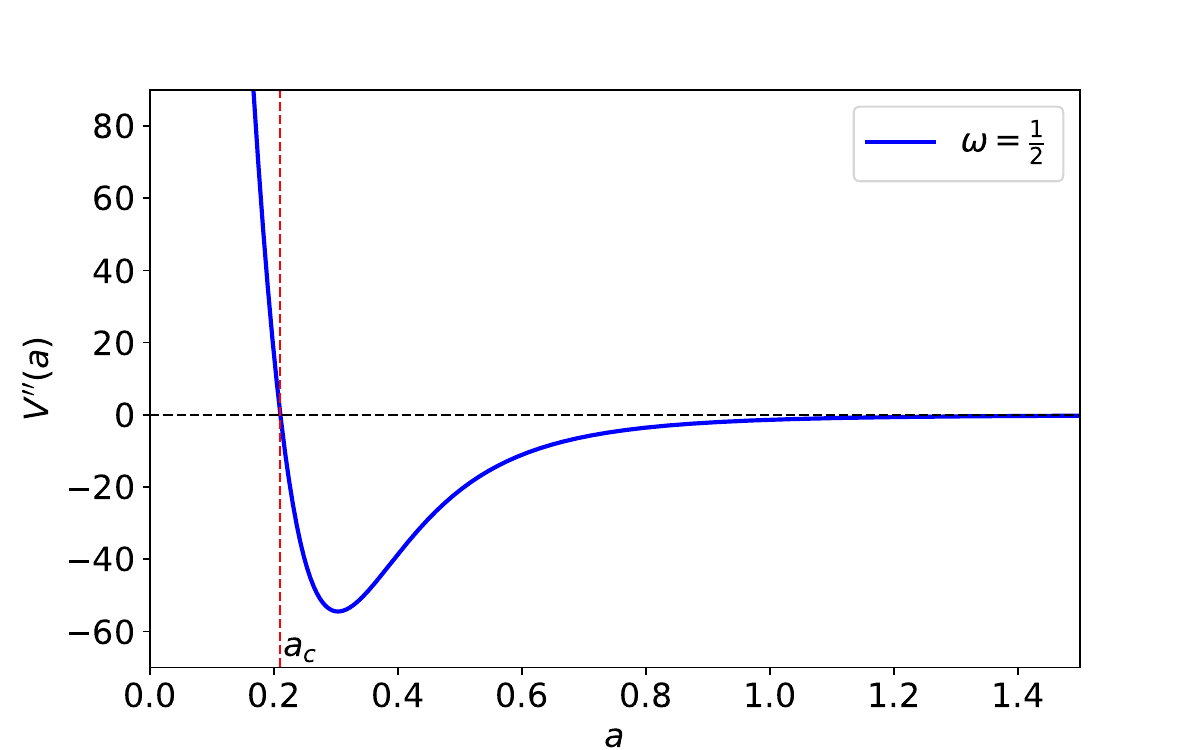}
		\caption{}
		\label{cdcc}
	\end{subfigure}
	\caption{The potential function and its first and second derivatives, Eqs.~\eqref{fdvfegref}, \eqref{skjlskjs}, and \eqref{dajaj} as functions of the scale factor. The parameters are set as $ \bar{\alpha}=1, \epsilon_{0}=0.5, G_{0}=1/8\pi$.}
	\label{ffeve}
\end{figure}

To examine the dynamics and stability of the potential \eqref{fdvfegref}, we must calculate its first and second derivatives, as mentioned earlier. These derivatives can be expressed as follows
\begin{align}\label{skjlskjs}
	V^{\prime}(a)=&-\frac{4\pi \epsilon_{0}^{1/3}}{ \alpha\sqrt{a}}-\frac{12\pi \epsilon_{0}^{1/3}}{\alpha}\frac{a^{5/2}}{a^3+\alpha G_{0}\epsilon_{0}^{2/3}}\notag\\
	& + \frac{16\pi a}{\alpha \sqrt{\alpha G_{0} }} \, \arctan\left(\frac{\sqrt{\alpha G_{0} } \epsilon_{0}^{1/3}}{a^{3/2}}\right),
\end{align}

\begin{align}\label{dajaj}
	V^{\prime\prime}(a)=&\frac{-16\pi a^6 \epsilon_{0}^{1/3} - 50\pi a^3 \alpha G_{0}  \epsilon_{0} + 2\pi \alpha^2 G_{0}^2  \epsilon_{0}^{5/3}}{ a^{3/2} \alpha (a^3 + \alpha G_{0}  \epsilon_{0}^{2/3})^2} 
	\notag\\
	&+ \frac{16\pi G_{0} \arctan\left(\frac{\sqrt{\alpha G_{0} } \epsilon_{0}^{1/3}}{a^{3/2}}\right)}{(\alpha G_{0} )^{3/2}}.
\end{align}
The potential function \eqref{fdvfegref} has a minimum value, $V_{\mathrm{min}}$, at $ a_{\mathrm{eq}}$, as shown in Fig.~\ref{ffvfev}. This figure illustrates how the potential varies with the scale factor. The second derivative of the potential, Eq.~\eqref{dajaj}, is plotted in Fig.~\ref{cdcc}. This plot shows that for $a < a_c$, the second derivative is positive, indicating that the potential is stable in this region. Since the equilibrium point $a_{\mathrm{eq}}$ lies within this range, the system remains dynamically stable.
\\
The stability discussed here, such as the case of $w=0$, is a local minisuperspace property of the homogeneous scale-factor dynamics. A static equilibrium at $a=a_{\rm eq}$ requires not only $V'(a_{\rm eq})=0$, but also the Hamiltonian condition $V(a_{\rm eq})=-k$. The inequality $V''(a_{\rm eq})>0$ implies local stability against homogeneous radial perturbations. The existence of a local minimum does not by itself imply that every collapse trajectory settles at $a_{\rm eq}$; the realized motion also depends on the initial conditions.

\subsubsection{$w=1$}
For a system containing  matter with $w=1$,
the potential function \eqref{djjdjjreb} is given by
\begin{equation}\label{wfwdfdwf}
	V(a)=\frac{16\pi a^2}{3\alpha^2 G_{0}}\ln\bigg(1+\frac{\alpha G_{0}\sqrt{\epsilon_{0}}}{a^3}\bigg)-\frac{16\pi \sqrt{\epsilon_{0}}}{3\alpha a},
\end{equation}
At large distances, the potential function $V(a)$ asymptotically approaches  
$V(a) \simeq -8\pi G_0 \epsilon_0 / (3\alpha a^4).$
The corresponding force function is given by  
$F(a) \simeq -32\pi G_0 \epsilon_0 / (3\alpha a^5).$
This force is attractive, as indicated by its negative sign, and it exhibits an inverse power-law dependence on the scale factor. The steep falloff with increasing $a$ suggests that the interaction weakens significantly at large distances.  
Conversely, in the limit of small distances, $a \to 0$, the potential is dominated by the term  
$V(a) \simeq -16\pi \sqrt{\epsilon_0} / (3\alpha a).$
The force associated with this potential is  
$F(a) \simeq -16\pi \sqrt{\epsilon_0} / (3\alpha a^2).$
This force diverges as $a \to 0$, implying an increasingly strong attractive interaction at small distances. The $a^{-2}$ dependence suggests a resemblance to gravitational or Coulomb-like behavior, where the force becomes significantly stronger as the separation decreases. This divergence highlights the presence of a dominant short-range interaction that governs the system at sufficiently small scales.

\begin{figure}[htb]
	\centering
	\begin{subfigure}[b]{0.48\textwidth}
		\includegraphics[width=\textwidth]{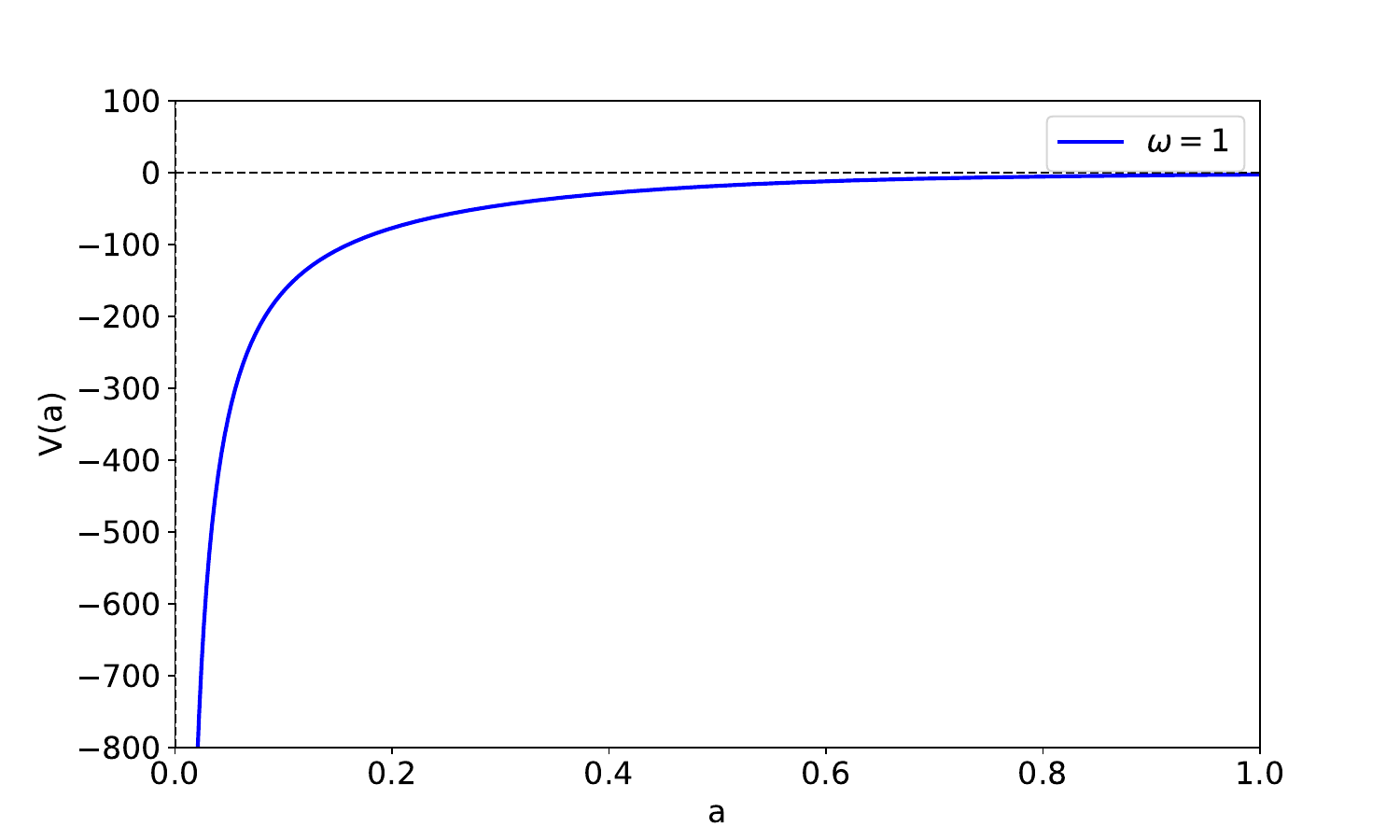}
		\caption{}
		\label{jhhvugvuh}
	\end{subfigure}
	\hfill
	\begin{subfigure}[b]{0.48\textwidth}
		\includegraphics[width=\textwidth]{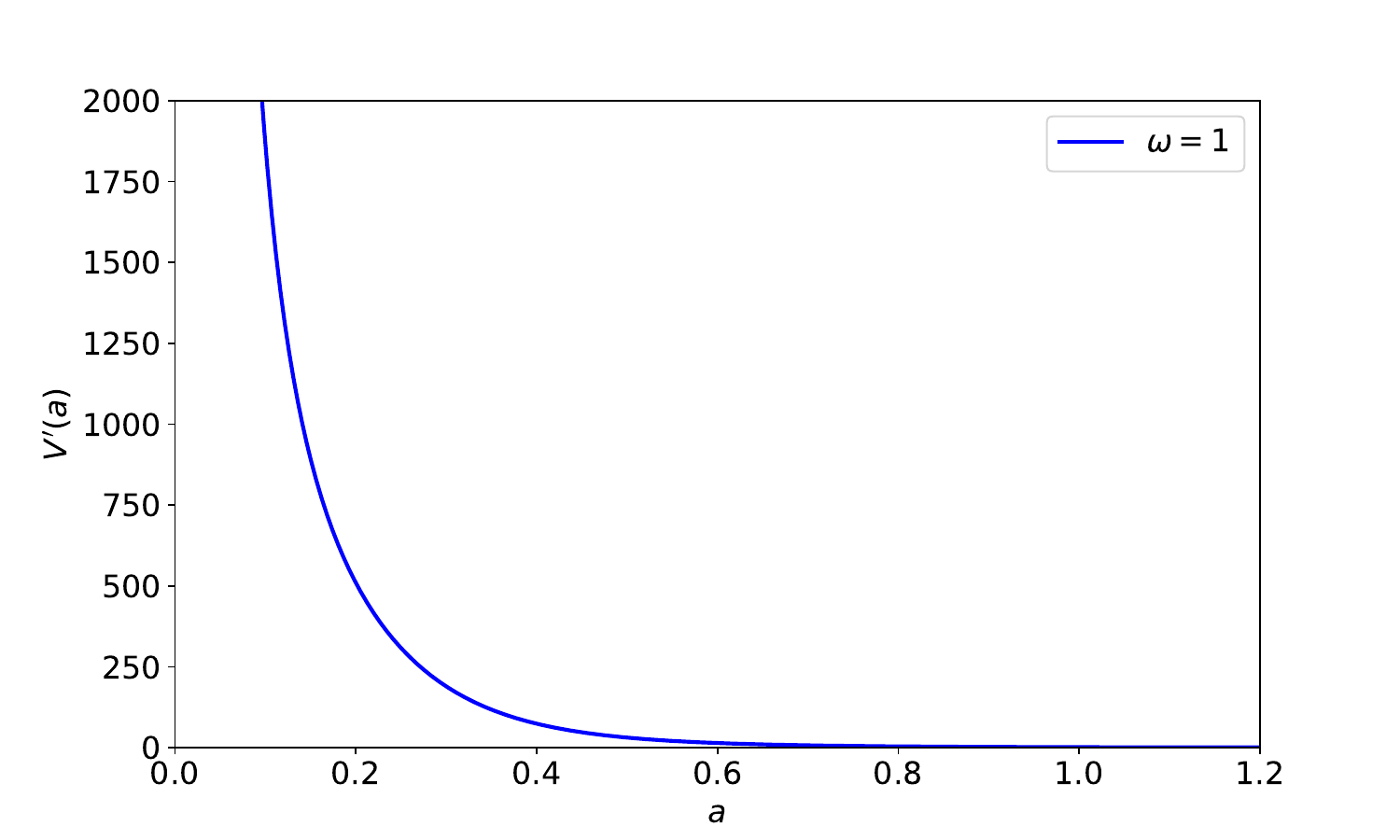}
		\caption{}
		\label{jhvugvyu}
	\end{subfigure}
	\hfill
	\begin{subfigure}[b]{0.48\textwidth}
		\includegraphics[width=\textwidth]{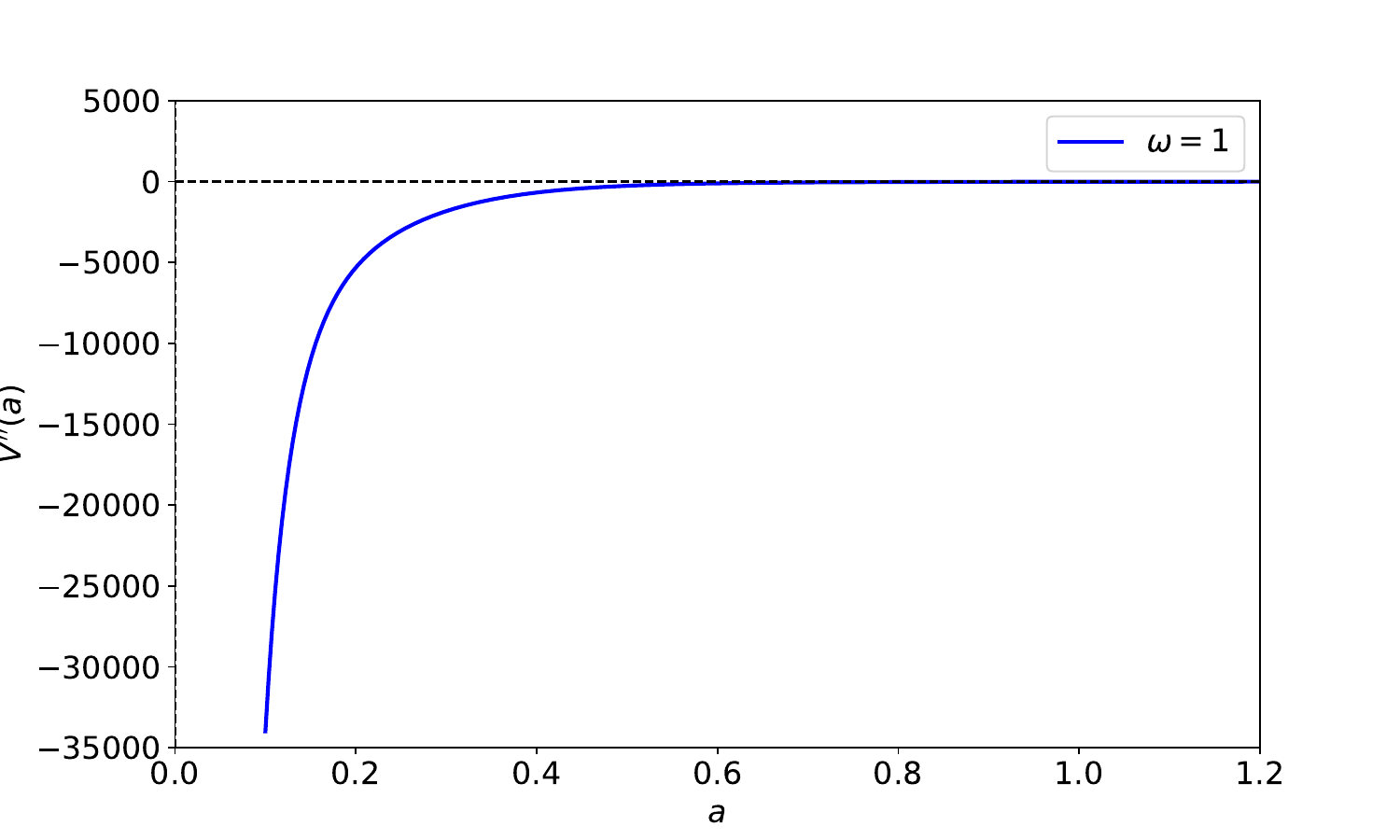}
		\caption{}
		\label{hbiuuvu}
	\end{subfigure}
	\caption{The potential function and its first and second derivatives, Eqs.~ \eqref{wfwdfdwf}, \eqref{dfcdcwdc}, and \eqref{fvfvf}, as functions of the scale factor. The parameters are set as $ \epsilon_{0}=\bar{\alpha}=1, G_{0}=1/8\pi$.}
	\label{xsdcdc}
\end{figure}

The dynamics and stability of the potential in Eq.~\eqref{wfwdfdwf} require computing its first and second derivatives. As previously stated, the first and second derivatives are given by
\begin{align}\label{dfcdcwdc}
	V^{\prime}(a)=&\frac{32\pi a}{3\alpha^2 G_{0}}
	\ln \bigg( 1+\frac{\alpha G_{0}\sqrt{\epsilon_{0}}}{a^3}\bigg)+\frac{16\pi \sqrt{\epsilon_{0}}}{3\alpha a^2}\notag\\
	&-\frac{16\pi \sqrt{\epsilon_{0}}}{\alpha }\frac{a}{a^3+\alpha G_{0}\sqrt{\epsilon_{0}}},
\end{align}
and
\begin{align}\label{fvfvf}
	V^{\prime\prime}(a)=&-\frac{32\pi\sqrt{\epsilon_{0}}}{3\alpha a^3}-\frac{48\pi G_{0}\epsilon_{0}}{(a^3+\alpha G_{0}\sqrt{\epsilon_{0}})^2}\notag\\
	&+\frac{32\pi}{3\alpha^2G_{0}}\ln\bigg(1+\frac{G_{0}\alpha \sqrt{\epsilon_{0}}}{a^3}\bigg),
\end{align}
For stability, an equilibrium point $a_{\mathrm{eq}}$ must satisfy $V^{\prime}(a_{\mathrm{eq}}) = 0$. However, Figs.~\ref{jhhvugvuh} and  \ref{jhvugvyu} demonstrate that $V^{\prime}(a_{\mathrm{eq}}) \neq 0$ for any finite $a$, indicating the absence of a stable equilibrium. Furthermore, Fig.~\ref{hbiuuvu} shows that $V^{\prime\prime}(a_{\mathrm{eq}}) < 0$, meaning the potential is concave downward, which also does not support stability. Therefore, the potential function \eqref{wfwdfdwf} is  dynamically unstable.

Fig.~\eqref{xsdcdc} therefore indicates that, within the homogeneous minisuperspace dynamics considered here, the $w=1$ sector does not support a finite-radius stable equilibrium analogous to the dust or $w=1/2$ cases. This conclusion concerns the dynamics of the stellar surface and should be distinguished from the curvature behavior of the corresponding exterior metric, which is analyzed independently through the Misner-Sharp mass function in Sec.~\ref{dcwdkwdcd}.
\\

The above analysis shows that there are both dynamically stable and unstable potential functions for different values of $w$. We have chosen $w= \{-1/2, 0,1/2,1\}$, and it is also possible to find other potential functions for different EOS. In the following, we will attempt to study the exterior region of the gravitationally collapsing object and its properties as it evolves into a BH.  

%%%%%%%%%%%%%%%%%%%%%%%%%
%%%%%%%%%%%%%%%%%%%%%%%%%
%%%%%%%%%%%%%%%%%%%%%%%%%	
\section{Black hole solutions}
\label{dcwdkwdcd}
In this section, we derive BH solutions by matching the interior collapse
dynamics to an exterior static metric. 
It is well known that there is insufficient information about the EOS that governs the final state of gravitationally collapsed objects. The polytropic EOS is a possible choice for describing the gravitational collapse process that leads to a regular BH in the final stage of collapse \cite{Shojai:2022pdq}. Some known gravitational theories also predict complex EOS tending to $w = -1/2$  for collapsing stars at high energy densities \cite{Hassannejad:2023lrp}. In this section, we will consider the EOS in the range of $-1 \leq  w \leq 1$,  which is widely recognized in cosmology and encompasses both ordinary matter and quintessence.

In the previous sections, we extensively  studied the star's interior and surface dynamics, as well as the junction conditions connecting its interior and exterior. Our goal  now is to evaluate the physics governing the region outside the star. 
On the surface of the star, comparing Eqs.~\eqref{edbi4rbfr} and \eqref {dnjd}  yields the MS mass $m(R)$ as 
\begin{equation}\label{dhhdhh}
	m_{w}(R)=\frac{\zeta_{w}}{R^{3w}}~_2F_1\Big(1,1+w;2+w;-\frac{\varrho G_{0}}{R^3}\Big),
\end{equation}
where $\zeta_{w}={4\pi G_{0}\epsilon_{0}r_{b}^{3(1+w)}}/{3}$. 
\\
The small-$R$ behavior of $m_w(R)$ is particularly important because it determines whether the corresponding exterior geometry is regular at the center. For a static metric of the form $f(R) = 1 - 2m(R)/R$, a de Sitter-like regular core requires $m(R) \sim R^3$ as $R \to 0$. If $m(R)$ decreases more slowly than $R^3$, or contains logarithmic or inverse-power behavior that makes $2m(R)/R$ singular, curvature invariants generally diverge.

The mass function in Eq. \eqref{dhhdhh} is specific to the proper-distance cutoff prescription adopted above. For a general cutoff scheme $s$, the effective density and Misner-Sharp mass can be expressed as
	\begin{equation}
		\rho_e^{(s)}(\epsilon)=\frac{\Lambda_0}{8\pi}+\int_0^\epsilon G_s(\bar\epsilon)\,d\bar\epsilon,
	\end{equation}
	\begin{equation}
		m_s(R)=\frac{4\pi R^3}{3}\rho_e^{(s)}\!\left(\epsilon(R)\right).
	\end{equation}
	Calculating the mass-profile difference between two prescriptions shows that simple horizons shift continuously under moderate cutoff changes. Near a degenerate horizon, where $f'(R_h)\simeq0$, the sensitivity becomes much stronger, and a scheme change may alter the number of horizons.

\begin{figure}[h]
	\centering
	\includegraphics[scale=0.44]{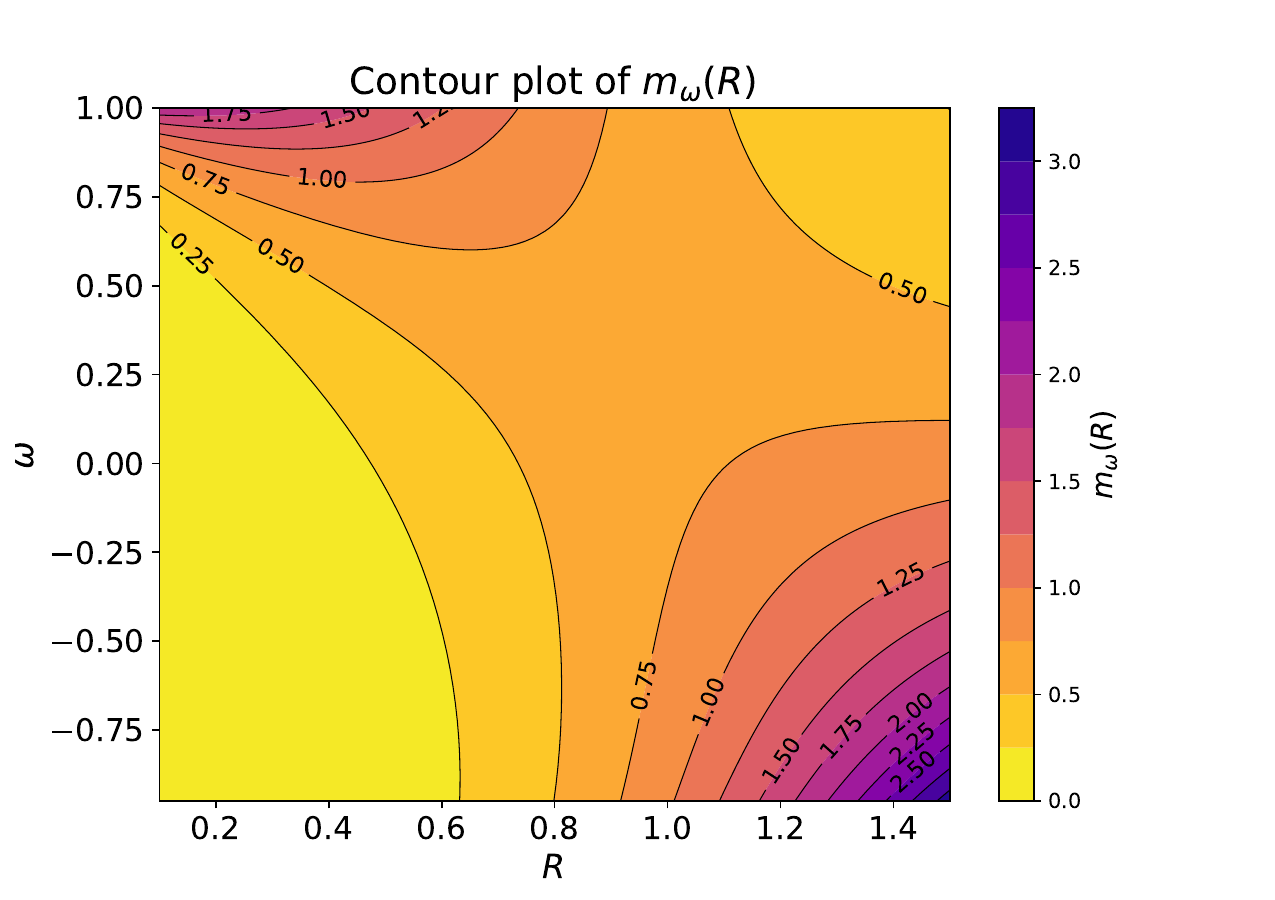}
	\caption{Contour plot of the MS mass function \eqref{dhhdhh} for $R \in [0.1, 1.5]$ and $w \in (-0.95, 1]$. The plot reveals how $m_{w}(R)$ varies with both parameters. The contours highlight regions of rapid variation and illustrate the behavior of the mass function across different scales. The parameter values are set to $\zeta_w=8\pi G_{0}=\varrho=1$ }
	\label{contour_4}
\end{figure}

\begin{figure}[h]
	\centering
	\centering
	\includegraphics[scale=0.37 ]{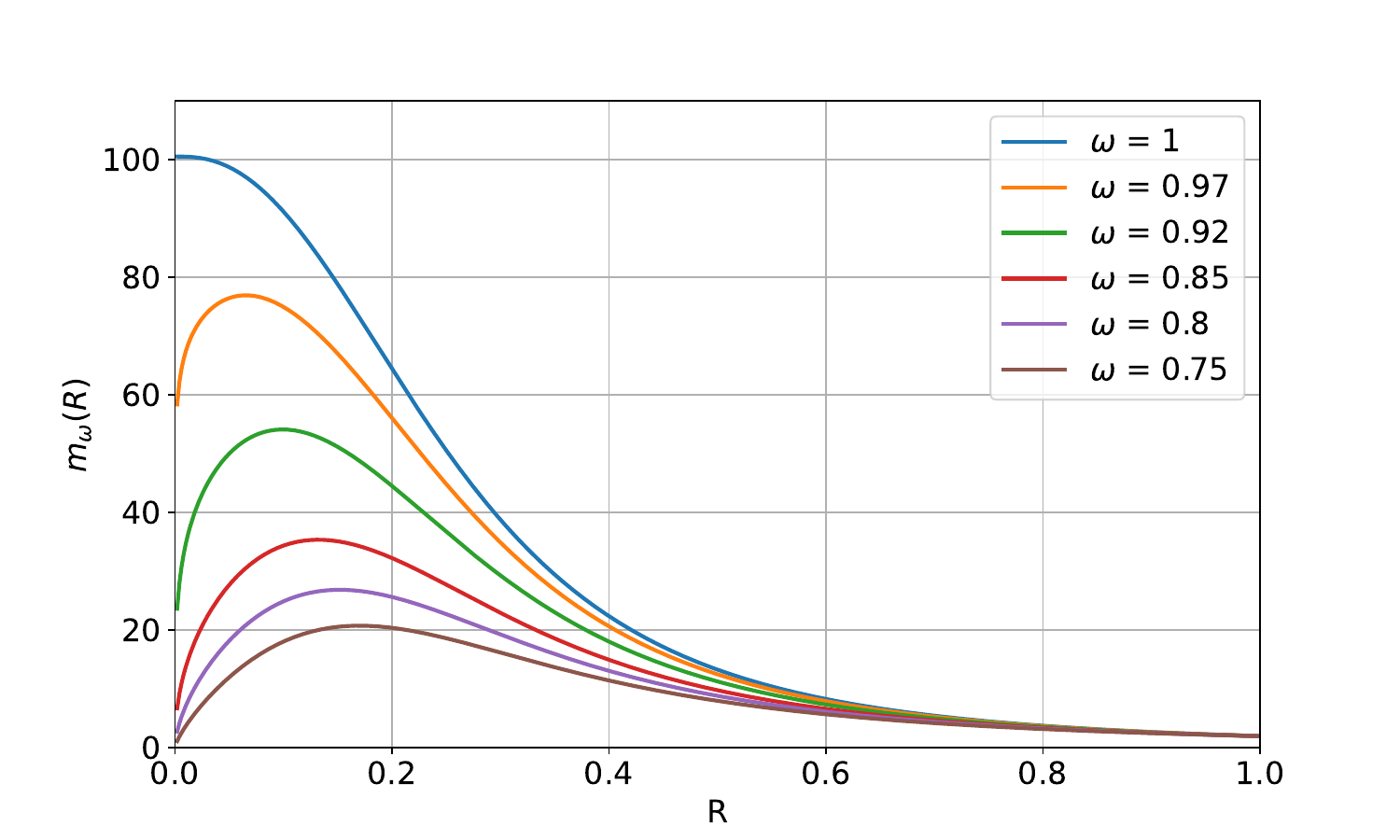}
	%	\caption{Figure 2}
	\label{fig:2}
	\caption{The MS mass function \eqref{dhhdhh} for some positive values of $w\in[0.75,1]$. At large distances, the MS mass approaches zero, independent of the values of $w$, while at small distances, its variation and peak value depend on $w$. The parameters are set as $ \epsilon_{0}=\bar{\alpha}=1,  G_{0}=1/8\pi$.}
	\label{fig:three_figures}
\end{figure}

\begin{figure}[h]
	\centering
	\includegraphics[scale=0.35]{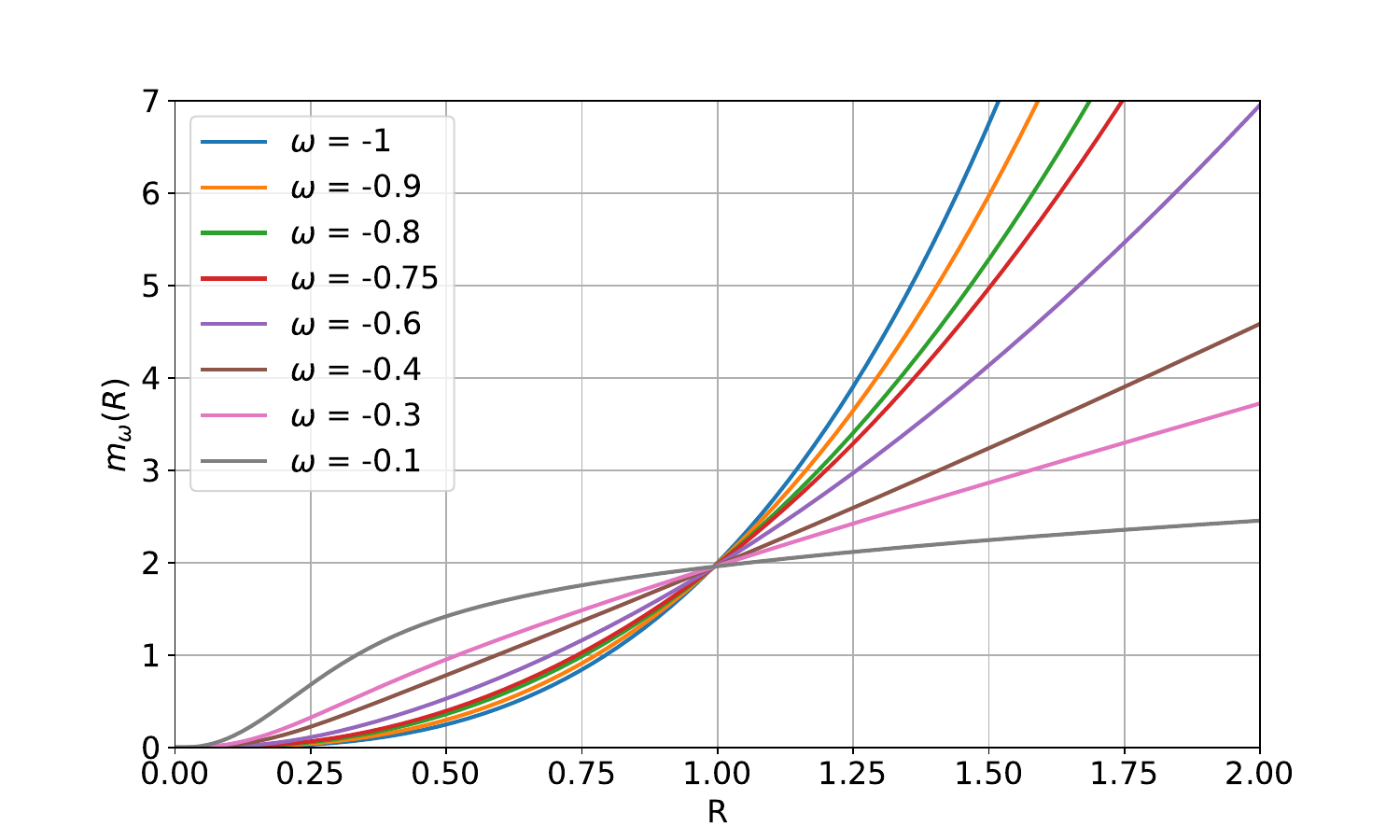}
	%	\caption{Figure 3}
	\label{fig:3}
	\caption{The MS mass function \eqref{dhhdhh}  for some negative values of $w\in[-1,-0.1]$. The MS mass function increases with radius and approaches zero at small distances. The parameters are set as $\varrho=\zeta_{w}=1$, $8\pi G_{0}=1$.}
	\label{3rf3rfr3f}
\end{figure}

\begin{figure}[h]
	\centering
	\includegraphics[scale=0.37]{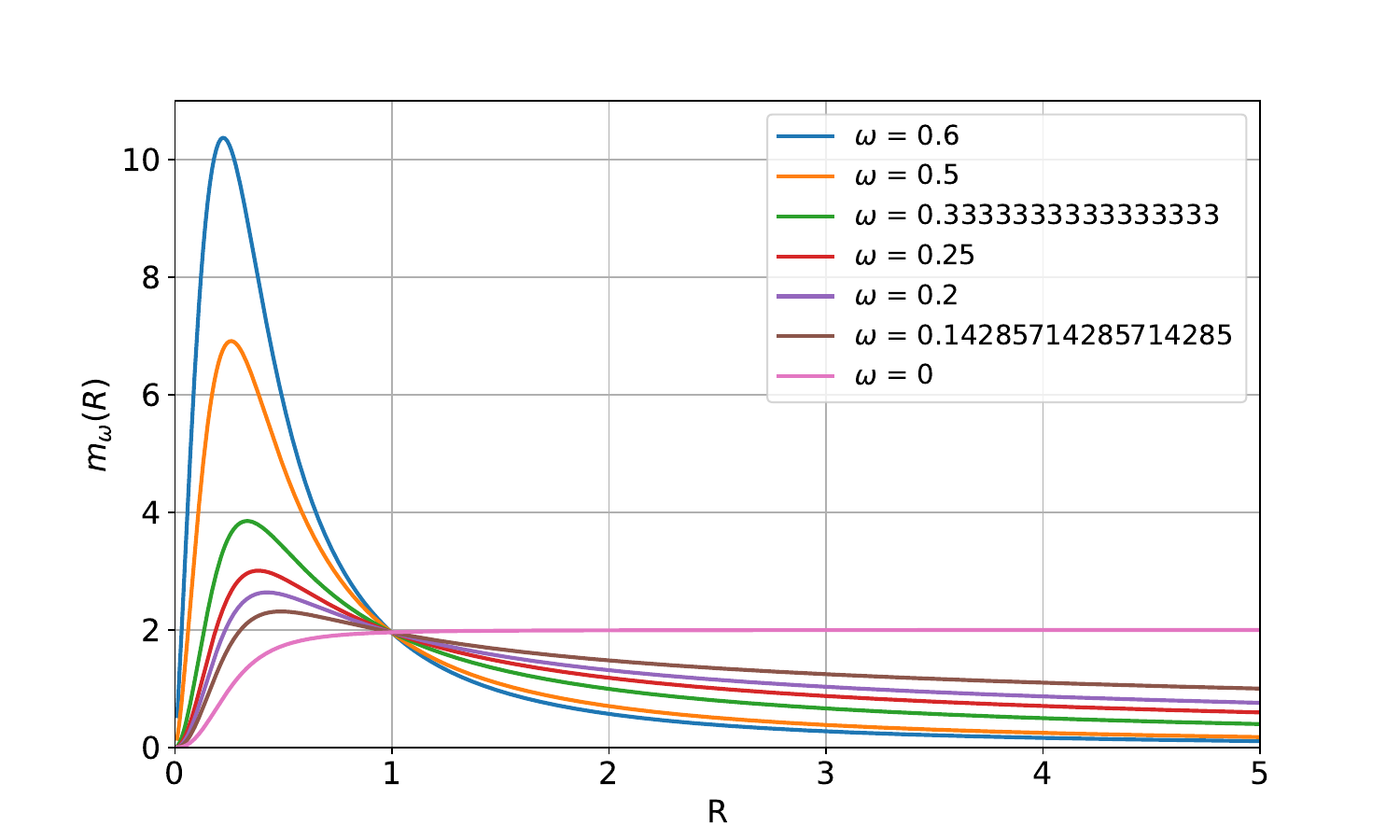}
	
	\caption{The MS mass function \eqref{dhhdhh} for selected non-negative values of $w \in [0,0.6]$. For $w>0$, the MS mass decreases toward zero at large distances, whereas the dust case $w=0$ approaches a constant asymptotic mass. At small distances, both the rate of variation and the maximum amplitude depend on $w$. The parameters are set as $\varrho=\zeta_{w}=1$, $8\pi G_{0}=1$. \label{fig:1}}
\end{figure}

Equation \eqref{dhhdhh} is the mass function selected by the EOS-dependent proper-distance prescription adopted here. The mass relation expressed in Ref.~\cite{Harada:2025cwd} can be reduced to Eq. \eqref{dhhdhh} by replacing $1/\gamma = 1+w$. With this identification, and after the corresponding normalization of the mass parameter, the mass relation discussed in Ref.~\cite{Harada:2025cwd}  reduces to the hypergeometric form in Eq.~\eqref{dhhdhh}.  Thus, Eq.~\eqref{dhhdhh} should be regarded as the EOS-selected representative
	of the broader mass-function class considered in Ref.~\cite{Harada:2025cwd}, rather than as an unrelated mass family. The mass function Eq. \eqref{dhhdhh} as a function of radial distance for different EOS parameters is shown in Figs.~\ref{fig:three_figures}, \ref{3rf3rfr3f}, and \ref{fig:1}. Furthermore, the behavior of $m_{w}(R)$ is depicted as a contour plot in Fig.~\ref{contour_4}. This motivates us to examine specific values of 
$w$ to assess the dynamics outside the star.
\subsubsection{ $w=0$}
\begin{figure}[h]
	\centering
	\includegraphics[scale=0.44]{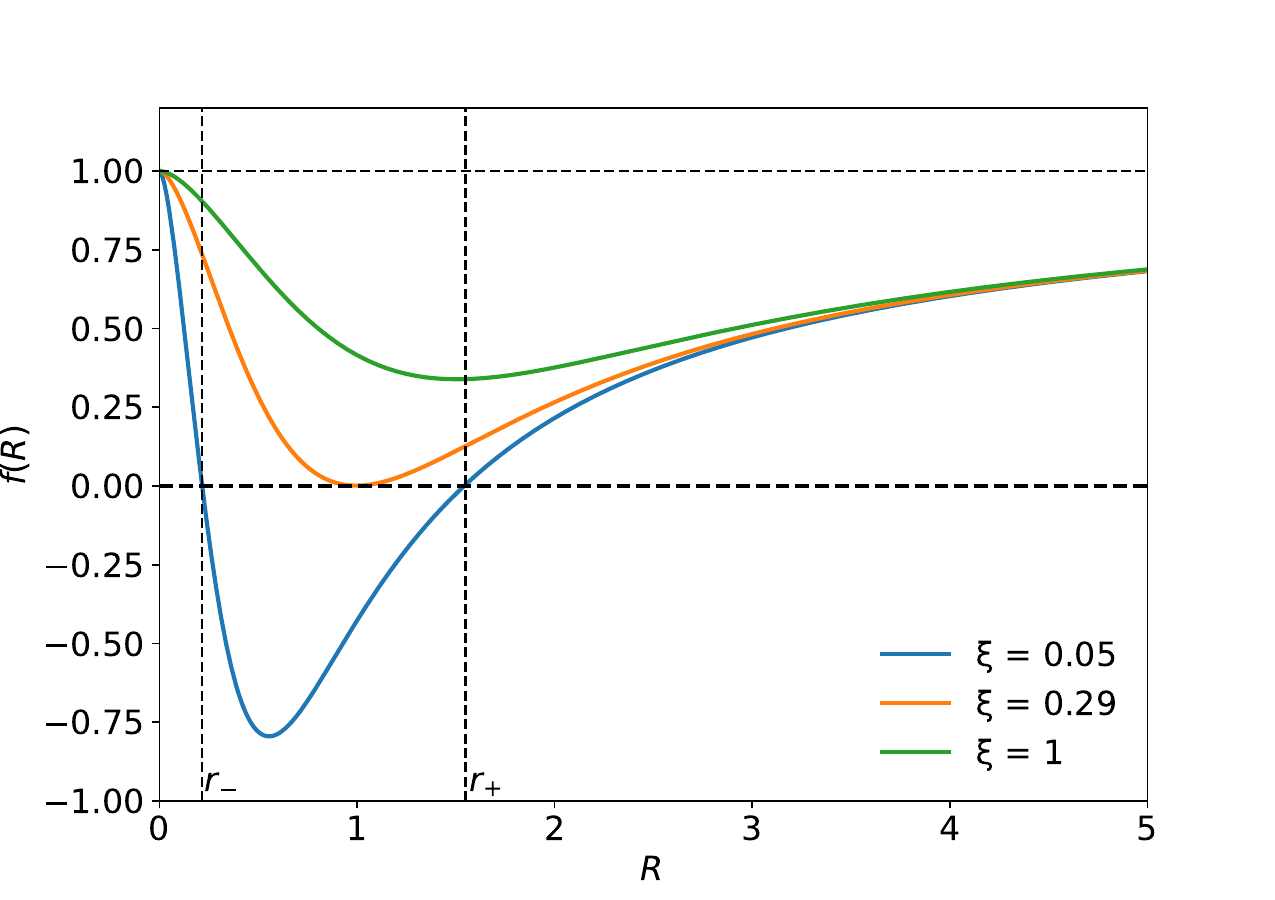}
	\caption{The behavior of $f(R)$ as a function of the radial distance $R$ according to Eq.~\eqref{djdjedjweoid}, for values $\xi=\{0.05, 0.29, 1\}$, $m_{0}=20$, and $8\pi G_{0} = 1$. The horizons are labeled as $r_{-}$ and $r_{+}$.}
	\label{eventhh}
\end{figure}

For dust gravitational collapse, $w=0$, the MS mass \eqref{dhhdhh} simplifies as follows 
\begin{equation}\label{fnffbf}
	m_{0}(R)=\frac{R^3}{6\xi}\ln\Big(1+\frac{6m_{0}G_{0}\xi}{R^3}\Big),
\end{equation}
where $\xi={3\beta^2}/{8g_*^{\mathrm{eff}}m_{0}}$, and the initial comoving energy density was assumed to be $\epsilon_{0}={3m_{0}}/{4\pi r^{3}_{b}}$. The MS mass \eqref{fnffbf} is recently derived for a dust fluid in Ref.~\cite{Bonanno:2023rzk}, where the interior matter is described as a dust fluid with $w=0$. The corresponding metric function is given by 
\begin{equation}\label{djdjedjweoid}
	f(R)=1-	\frac{R^2}{3\xi}\ln\Big(1+\frac{6m_{0}G_{0}\xi}{R^3}\Big).
\end{equation}	
At large distances, where $R^{3}\gg 6m_{0}G_{0}\xi$, Eq. \eqref{fnffbf} reduces to $f(R)\simeq 1-{2m_{0}G_{0}}/{R}$, which coincides with the Schwarzschild metric component. To determine the event horizon, we analyze Eq.  \eqref{djdjedjweoid} for $f(R)=0$. While it is not possible to obtain an exact analytical solution, numerical methods offer an efficient approach to find the solutions. The behavior of $f(R)$ as a function of radial distance is shown in Fig.~\ref{eventhh}. This figure illustrates that the roots of $ f(R)$ are highly sensitive to the values of $\xi$. For $ \xi = 1$, there is no horizon; for $\xi = 0.29$, there is a single event horizon; and for $\xi = 0.05$, there are two horizons.
To investigate the divergence of the metric at the center of a BH, where $R \to 0 $, we evaluate the Kretschmann scalar for the specified metric defined by Eq.~\eqref{djdjedjweoid}.
It is expressed as
\begin{align}\label{dkwbjdbweed}
	\mathcal{K}(R)=&\frac{4}{\xi^2}\bigg(\ln\Big(1+\frac{b}{R^3}\Big)-\frac{b}{2}\frac{2R^3+5b}{(R^3+b)^2}\bigg)^2
	+\frac{8}{\xi^2}\bigg(\ln\Big(1+\frac{b}{R^3}\Big)-\frac{b}{R^3+b}\bigg)^2
	+\frac{4}{3\xi^2}\bigg(\ln\Big(1+\frac{b}{R^3}\Big)\bigg)^2 \notag\\
	&
	-\frac{16}{3\xi^2}\bigg(\ln\Big(1+\frac{b}{R^3}\Big)-\frac{3b}{2(R^3+b)}\bigg)
	\bigg(\ln\Big(1+\frac{b}{R^3}\Big)-\frac{5b^2+2bR^3}{2(R^3+b)^2}\bigg) \notag\\
	&-\frac{16}{3\xi^2}\bigg(\ln\Big(1+\frac{b}{R^3}\Big)-\frac{b}{R^3+b}\bigg)\ln\bigg(1+\frac{b}{R^3}\bigg),
\end{align}	

where $b=6m_{0}G_{0}\xi$. At large radial distances, where $R^3 \gg b$, the Kretschmann scalar reduces to 
$
\mathcal{K}(R) \simeq 48 G_{0} m_{0}/R^6.
$
This expression coincides with the Kretschmann scalar associated with the Schwarzschild metric. However, at small radial distances, where $R^3 \ll b$, the curvature behavior is modified, and the Kretschmann scalar can be written as
\begin{equation}\label{wdfwfn}
	\mathcal{K}(R)\simeq \frac{24}{\xi^2} (\ln R)^2,
\end{equation}
As $R \to 0$, the Kretschmann scalar diverges, and therefore the $w = 0$ exterior extension contains a genuine curvature singularity. Introducing a cutoff $R > R_{\min}$ may keep the curvature finite within the effective domain of validity of the RG-improved description, as discussed in Ref.~\cite{Bonanno:2023rzk}.

\subsubsection{ $w=-1$}
A physically relevant value of the EOS parameter is $w = -1$, which is widely recognized in cosmological physics and alternative BH models, such as gravastars (gravitational vacuum condensate stars) \cite{Mazur:2004fk}. In this case, Eq.~\eqref{dhhdhh} takes the simplified form
\begin{equation}\label{fhfhfhh} 
	m_{-1}(R) = \zeta_{-1} R^{3},  
\end{equation}  
This mass term is equivalent to de Sitter spacetime, where $f(R)\big|_{w=-1} = 1 - \Lambda R^2/3$, with the cosmological constant $\Lambda = 8\pi G_{0} \epsilon_{0}$. The properties and implications of this spacetime have been widely investigated and are well established in gravitational and high-energy physics.
\subsubsection{ $w=-1/2$}
For $w=-1/2$, the mass function given in Eq.~\eqref{dhhdhh} reduces to
\begin{equation}
	m_{-\frac{1}{2}}(R)=\eta  R^3 \arctan \bigg(\sqrt{\frac{G_{0} \varrho}{R^3}}\bigg),
	\label{sjshhsh}
\end{equation}
where the constant $\eta$ is defined as  
$
\eta = 8\pi \epsilon_{0} \sqrt{G_{0} g_*^{\mathrm{eff}} r_{b}^3}/9\beta.
$
To derive Eq.~\eqref{sjshhsh}, we  employ Eq.~\eqref{dndndnd}, which establishes the relation between the hypergeometric function and the inverse tangent function.  
The corresponding metric function associated with this mass term is expressed as 
\begin{equation}\label{dkjccdjcw}
	f(R) = 1 - 2\eta R^2 \arctan \bigg(\sqrt{\frac{G_{0} \varrho}{R^3}}\bigg).
\end{equation}
Notably, the presence of the $\arctan$ function in $f(R)$ indicates a deviation from a simple power-law correction, indicating a nontrivial gravitational modification in this regime. The behavior of the metric function for large and small $R$ offers insights into the asymptotic structure of the solution, potentially revealing distinct gravitational signatures in comparison to standard cases. 
At sufficiently large distances, the metric function asymptotically behaves as  
\begin{equation}
	f(R) \simeq 1 - 2\eta\sqrt{G_{0}\varrho} \sqrt{R}.
\end{equation}
On the other hand, in the small-distance regime where $R^3 \ll G_{0}\varrho$, the metric function reduces to
\begin{equation}
	f(R) \simeq 1 - \pi \eta R^2,
\end{equation}
This expression corresponds to a de Sitter-like spacetime, characterized by an effective cosmological constant. The emergence of a quadratic dependence on $R$ at short distances indicates a dominant repulsive effect, resembling an inflationary or dark energy-like behavior. 
\\
To determine the event horizon, we solve the equation $f(R) = 0$ for Eq.~\eqref{dkjccdjcw}. Although an exact analytical solution is not feasible, the solutions can be straightforwardly obtained using numerical methods.
\begin{figure}[h]
	\centering
	\includegraphics[scale=0.44]{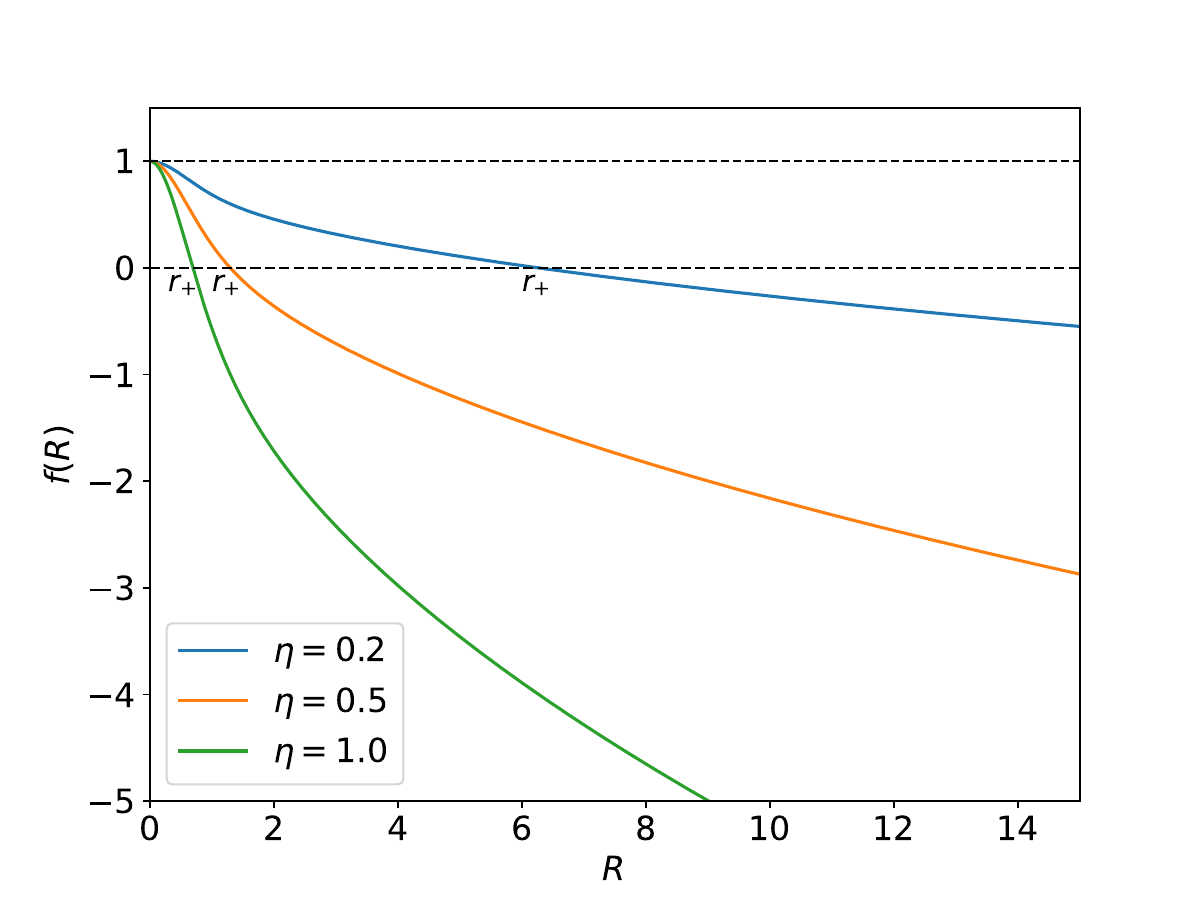}
	\caption{Metric function $f(R)$ from Eq.~\eqref{dkjccdjcw} for $\eta=\{0.2,0.5,1.0\}$ and $G_0\varrho=1$. Radius is expressed in reduced Planck units. Each curve possesses one positive zero, denoted by $r_+$.}
	\label{event}
\end{figure}
Fig.~\ref{event} shows that the spherically symmetric metric Eq.~\eqref{dkjccdjcw} possesses a single horizon. Since the spacetime is not asymptotically flat and the static region lies on the inner side of this horizon, its global causal
interpretation requires care; the horizon should not automatically be identified with an asymptotically flat BH event horizon. To explore the presence of a singularity in this spacetime, we evaluate curvature scalars, such as the Kretschmann scalar. 
As mentioned, at small distances the metric function becomes the de Sitter spacetime $f(R)\simeq 1-\pi\eta R^2$, and the Kretschmann scalar becomes $\simeq \pi^2\eta^2$. Therefore, spacetime described by Eq.~\eqref{dkjccdjcw} is regular at $R\to 0$. 
This regularity is consistent with the mass-function criterion stated above. In this case, the Misner-Sharp mass behaves as $m_{-1/2}(R) = \mathcal{O}(R^3)$ near the center, so the metric approaches a de Sitter-like core rather than a Schwarzschild-like singularity.
\\
\subsubsection{ $w=1$}
\begin{figure}[h]
	\centering
	\includegraphics[scale=0.44]{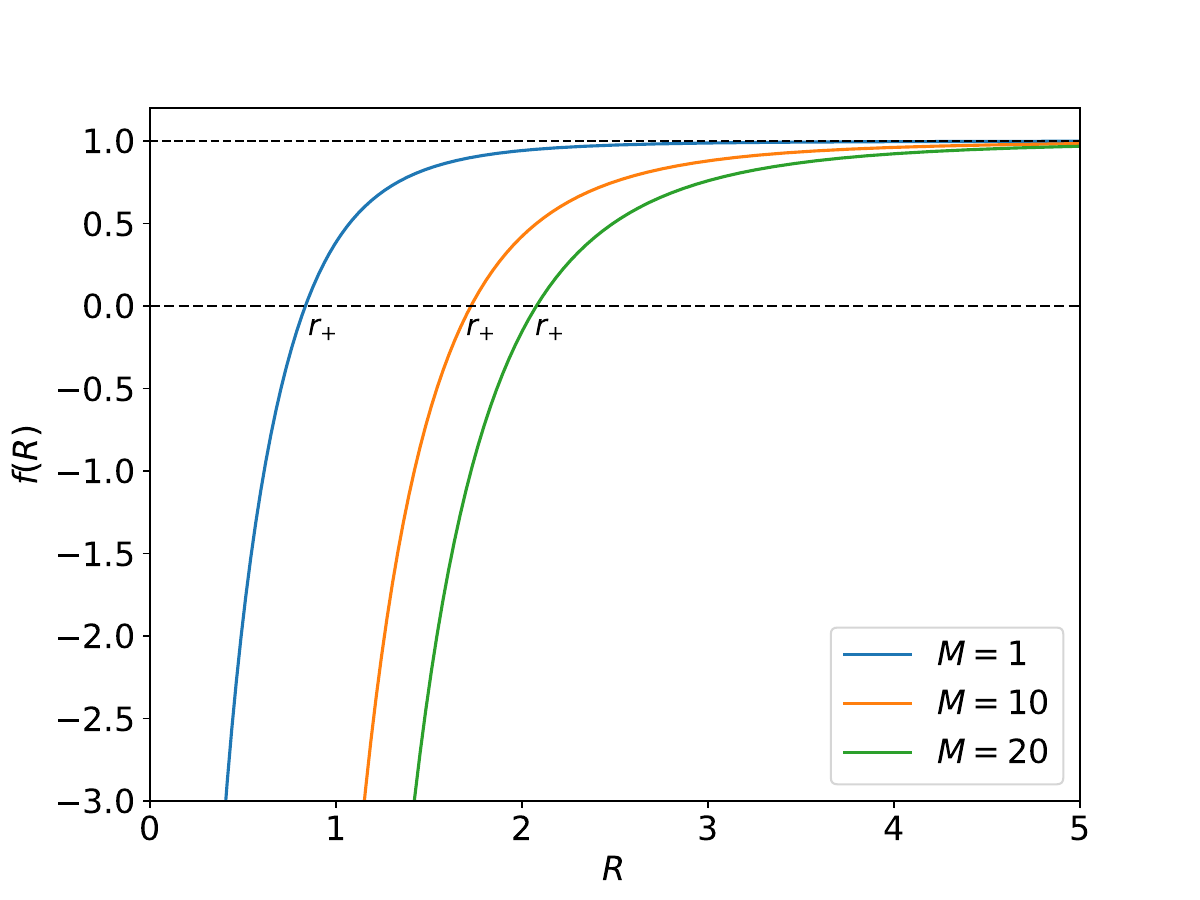}
	\caption{Metric function $f(R)$ from Eq.~\eqref{wbddwed} for $M=\{1,10,20\}$ and $G_0\varrho=1$, with $R$ and $M$ expressed in reduced Planck units. Each curve has a single positive horizon $r_+$.}
	\label{dfgfdvgfd}
\end{figure}

Now, let us consider another EOS with $w=1$. Substituting this into Eq.~\eqref{dhhdhh} yields the following expression for the MS mass equation as
\begin{align}\label{dkdjdj}
	m_{1}(R)=\frac{2\zeta_{1}}{\varrho G_{0}}-\frac{2\zeta_{1}R^3}{(\varrho G_{0})^2}\ln\bigg(1+\frac{\varrho G_{0}}{R^3}\bigg),
\end{align}
The metric function corresponding to this mass function is expressed as
\begin{equation}\label{wbddwed}
	f(R)=1-\frac{2M}{R}+\frac{2M}{\varrho G_{0}} R^2\ln\bigg(1+\frac{\varrho G_{0}}{R^3}\bigg),
\end{equation}
where $M=2\zeta_{1}/(\varrho G_{0})$. For large distances  $R\gg (G_{0}\varrho)^{{1}/{3}}$, the metric function \eqref{wbddwed} 
reduces to 
\begin{equation}
	f(R)\simeq 1-\frac{M\varrho G_{0}}{R^4}.
\end{equation}
The presence of a term proportional to $R^{-4}$ in the spherically symmetric spacetime metric has been investigated in several prominent theories of quantum gravity, including loop quantum gravity \cite{Lewandowski:2022zce, Kelly:2020uwj, Gambini:2020nsf} and quantum modified gravity theories \cite{Abedi:2015yga}.\\
At small distances $R\to0$, the metric function \eqref{wbddwed} takes the form
\begin{equation}
	f(R)\simeq 1-\frac{2M}{R}+\frac{2M\ln(\varrho G_{0})}{\varrho G_{0}}R^2-\frac{6M}{\varrho G_{0}} R^2\ln R,
\end{equation}
which corresponds to the Schwarzschild-de Sitter metric modified by the presence of a logarithmic term. \\
The computation of the Kretschmann scalar reveals the existence of a singularity as $R\to 0$ within this metric. This behavior is expected from the small-$R$ scaling of the mass function. The condition $m(R) = \mathcal{O}(R^3)$ is not satisfied in this case; instead, the metric contains a dominant Schwarzschild-like contribution together with logarithmic corrections. The $w=1$ exterior therefore does not develop a de Sitter-like regular core. Additionally, the behavior of the metric function $f(R)$ as a function of $R$ is depicted in Fig.~\ref{dfgfdvgfd}. The figure illustrates that there is a single finite radius at which the metric component $f(R)$ vanishes, indicating the presence of a single event horizon for this BH.

\subsubsection{ $w=1/2$}
\begin{figure}[h]
	\centering
	\includegraphics[scale=0.44]{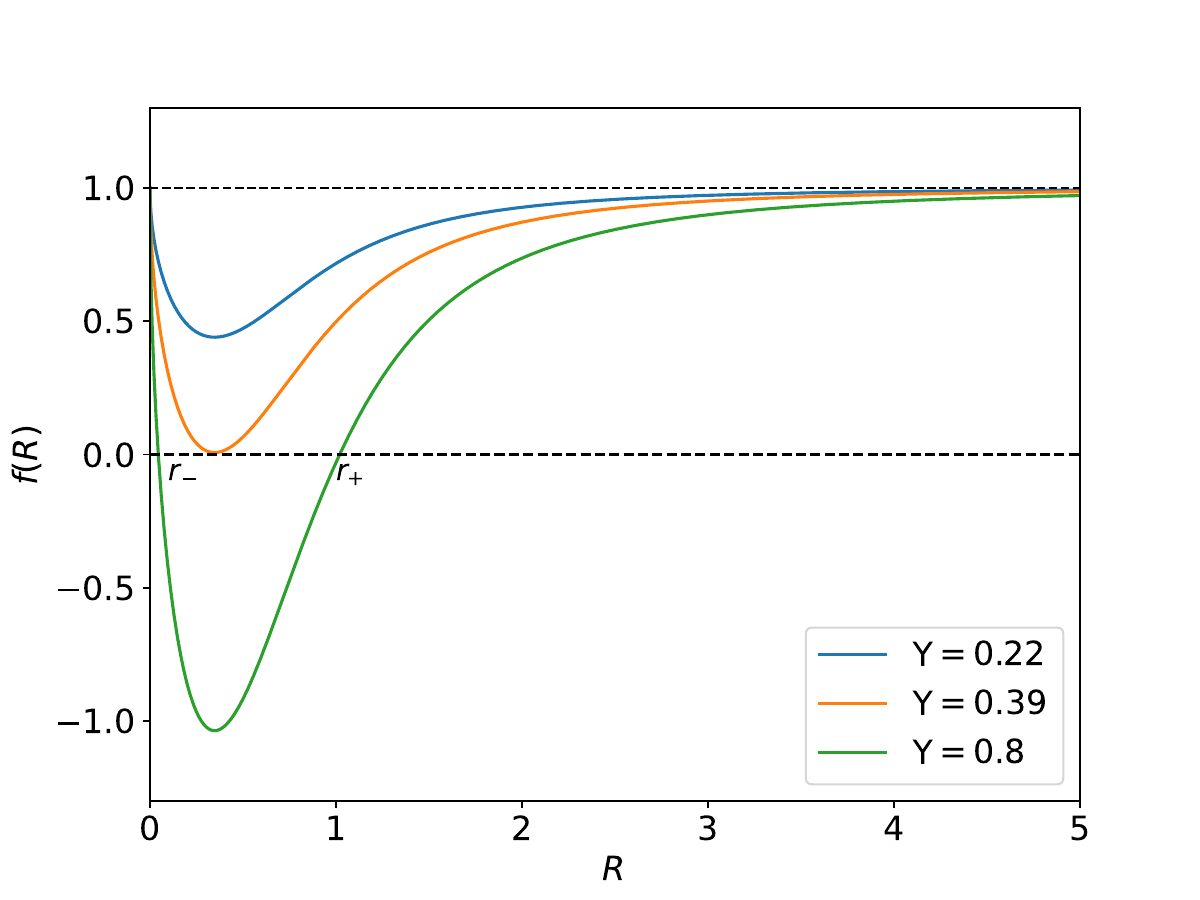}
	\caption{Metric function $f(R)$ from Eq.~\eqref{sdcfdev} for $\Upsilon=\{0.22,0.39,0.8\}$ and $G_0\varrho=1$. The inner and outer positive zeros are denoted by $r_-$ and $r_+$, respectively.}
	\label{rg3rgtg45tg}
\end{figure}

Choosing the EOS $w={1}/{2}$, the MS mass term Eq.~\eqref{dhhdhh} takes the form  
\begin{equation}\label{fhhfh}
	m_{1/2}(R)=3\Upsilon R^{3/2}\bigg(1-\frac{\arctan\big(\sqrt{\varrho G_{0}}R^{-3/2}\big)}{\sqrt{\varrho G_{0}}R^{-3/2}}\bigg),
\end{equation}
where $\Upsilon=\zeta_{1/2}/(\varrho G_{0})$. The metric function for the above mass function can be
written as 
\begin{equation}\label{sdcfdev}
	f(R)=1-6\Upsilon \sqrt{R}\bigg(1-\frac{\arctan\big(\sqrt{\varrho G_{0}}R^{-3/2}\big)}{\sqrt{\varrho G_{0}}R^{-3/2}}\bigg),
\end{equation}
As shown by the small-R behavior discussed below, the small-$R$ behavior of the metric function implies a curvature singularity in the exterior metric. However, this should be distinguished from the local stability of the homogeneous collapse potential at a finite value, $a_{\rm eq}>0$. If the collapse trajectory is trapped near the finite-radius minimum, the stellar surface does not reach the singular center. If the trajectory reaches $a=0$, however, the corresponding exterior geometry is singular. Thus, Fig.~\eqref{ffeve} and Fig.~\eqref{rg3rgtg45tg} address two different questions. Fig.~\eqref{ffeve} concerns the local stability of the homogeneous collapse potential, whereas Fig.~\eqref{rg3rgtg45tg} displays the horizon structure of a static exterior whose $R \to 0$ extension is singular. The coexistence of a locally stable finite-radius trajectory and a singular exterior extension is therefore not contradictory.
\\
In the small-distance regime, where $R^{3}\ll G_{0}\varrho$, the metric function ~\eqref{sdcfdev} reduces to $f(R)\simeq 1-6\Upsilon \sqrt{R}$.
In the large-distance limit, where $R\gg G_{0}\varrho$, the metric component behaves as $f(R)\big|_{w={1}/{2}}\simeq1-q^2/R^{5/2}$, where $q=\sqrt{2\zeta_{{1}/{2}}}$.

By calculating the Kretschmann scalar, we confirm that a singularity emerges as $R \to 0$ in this metric. Furthermore, Fig.~\ref{rg3rgtg45tg} illustrates how the metric function $f(R)$ varies with $R$. As shown in the figure, $f(R)$ vanishes at two finite radii, revealing that this BH has two horizons, namely the inner horizon $r_{-}$, and the outer horizon $r_{+}$.

\subsubsection{$|w| \gg 1$}
In this case, the MS mass term Eq.~\eqref{dhhdhh} takes the form
\begin{equation}\label{dnwddwd}
	m_{w}(R)\simeq \frac{\xi_{w}}{R^{3w}}\frac{1}{R^3+\varrho G_{0}},
\end{equation}	
The metric function corresponding to the given mass function can be written as
\begin{equation}\label{rgrtgrg}
	f(R)\simeq1- 2\xi_{w}\frac{R^{-3w}}{R^3+\varrho G_{0}}.
\end{equation}
For non-negative values of $w$, where $|w| = w$, the metric function \eqref{rgrtgrg} develops a singularity as $R \to 0$. On the other hand, when $|w| = -w$  (negative values), the small-$R$ behavior is controlled by the power of $R$ in the Misner-Sharp mass. Regularity of a metric of the form
$f(R)=1-2m(R)/R$ requires $m(R)=O(R^3)$, together with the corresponding regular behavior of its derivatives. Thus, the conclusion
about central regularity should be based on the central scaling of $m(R)$, not on the finiteness of $m(R)$ alone. In the negative-$w$
branch where this scaling is satisfied, the geometry can approach a regular core; otherwise, the $R \to 0$ extension remains
singular.
\subsubsection{Kiselev-Like Black Hole}
In general, according to the power series \eqref{jbbkk}, at large distances $R\gg(G_{0}\varrho)^{\frac{1}{3}}$, Eq.~\eqref{dhhdhh} reduces to $m_{w}(R)\simeq\zeta_{w}/R^{3w}$, and the spacetime metric takes the form
\begin{align}\label{dndjk}
	ds^2=&-\Big(1-\frac{2\zeta_{w}}{R^{1+3w}}\Big)dt^2+\Big(1-\frac{2\zeta_{w}}{R^{1+3w}}\Big)^{-1}dR^2\notag\\
	&+R^2d\Omega^2,
\end{align}
The above metric is the Kiselev BH \cite{Kiselev:2002dx}, with zero Schwarzschild mass, $m=0$, and $K=2\zeta_{m}$. It is well known that the Kiselev metric is the static spherically symmetric solution of Einstein's equations whose source is a non-perfect fluid with anisotropic pressures \cite{Visser:2019brz}. It has been demonstrated that the Kiselev metric reduces to the Reissner–Nordström metric when the averaged EOS satisfies $w=1/3$ \cite{Saadati:2020dje}\footnote{The metric of the Kiselev BH is expressed as
	\[
		ds^2 = - \left(1-{2m\over R} - {K\over R^{1+3w}} \right) dt^2 + {dR^2\over1-{2m\over R} - {K\over R^{1+3w}}} 
		+ R^2 \,d\Omega_2^2,
	\]
	where $m$ is the Schwarzschild mass. For $\omega=1/3$ and $K=-Q^2$, this metric reduces to the Reissner-Nordström metric, and the average pressure satisfies the radiation EOS $\bar{p}=(p^{R}+p^{\theta}+p^{\phi})/3=\rho/3$ (see Ref.~\cite{Saadati:2020dje}).
}. We have shown above that the Kiselev metric may also arise as a solution of AS gravity at large
distances whose source is a perfect fluid with a linear EOS (but in the presence of running gravitational coupling and running cosmological constant). 
From another point of view, if we take into account the equations of AS gravity  with the effective energy momentum tensor Eq.~\eqref{dndndnsosw},  then the source of the Kiselev metric is a fluid with anisotropic pressures, unless $m^{\prime\prime}(R)=2m^{\prime}(R)/R$ which gives the  Schwarzschild--de Sitter BH with the mass function $m(R)=A R^3/3+B$. By substituting the metric from Eq.~\eqref{dndjk} into Eq.~\eqref{ssbbjkdnwld}, and considering the average tangential and radial pressures, we can express the EOS as the ratio of the average pressure to the density. Consequently, the EOS acquires a linear form, as shown below
\begin{equation}\label{gcucvhbjjb}
	\bar{p}_{\mathrm{e}}=\frac{p^{R}_{\mathrm{e}}+p^{\theta}_{\mathrm{e}}+p^{\phi}_{\mathrm{e}}}{3}=-\frac{3w^2\xi_{w}}{4\pi R^{3(w+1)}};~~~\frac{\bar{p}_{\mathrm{e}}}{\rho_{\mathrm{e}}}=w,
\end{equation}
For the general mass function \eqref{dhhdhh}, the ratio of the average pressure to the density can be expressed as
\begin{align}\label{nddbddjk}
	w(R)=&\frac{9w^2(2+w)\alpha(R)-3(1+w)(2w+1)\beta(R)}{3\big(3w(2+w) \alpha(R)-(1+w)\beta(R)\big)}
	\notag\\
	&+\frac{2(2+w)(1+w)\gamma(R)}{3(3+w)\big(3w(2+w) \alpha(R)-(1+w)\beta(R)\big)}
\end{align}
where 
\begin{align}\label{dndndnkwowow}
	&\alpha(R)=~_2F_{1}\Big(1,1+w;2+w;-\frac{G_{0}\varrho}{R^3}\Big),
	\notag\\
	&\beta(R)=\frac{3\varrho G_{0}}{R^3}~_2F_{1}\Big(2,2+w;3+w;-\frac{G_{0}\varrho}{R^3}\Big),\notag\\
	&
	\gamma(R)=\frac{9\varrho^2 G_{0}^2}{R^6}~_2F_{1}\Big(3,3+w;4+w;-\frac{G_{0}\varrho}{R^3}\Big),
\end{align}
At large distances, $R^3\gg G_{0}\varrho$, the functions \eqref{dndndnkwowow} approach $\alpha(R)\simeq1$ and $\beta(R)\simeq\gamma(R)\simeq0$, for which the Eq.~\eqref{nddbddjk} simplifies to $w(R)\simeq w$.

%%%%%%%%%%%%%%%%%
%%%%%%%%%%%%%%%%%%%%%%
%%%%%%%%%%%%%%%%%%%%%%%%%%%%%%
\section{Conclusions}
In this study, we have investigated the gravitational collapse of a spherically symmetric star within the framework of AS gravity, taking into account a running gravitational coupling and a dynamical cosmological constant. The resulting equations, \eqref{sjshsh} and \eqref{nhhhh}, established a general framework that can be applied to a broad class of systems with different equations of state. Our findings indicate that AS gravity changes the mechanisms of BH formation compared to classical GR. This introduces novel possibilities that may contribute to resolving the longstanding issue of singularities. We have demonstrated that for negative values of the EOS parameter ($w < 0$), the spherically symmetric spacetimes remain regular at their  centers. For the negative-pressure cases $w = -1/2$ and $w = -1$, the corresponding Friedmann equations are given in Eqs.~\eqref{fdveferef} and \eqref{mhchgchgkcgh},  respectively. 
The associated exterior solutions are given by Eqs.~\eqref{fhfhfhh} and \eqref{dkjccdjcw}. Both geometries are curvature-regular at the
center. For the positive-pressure cases $w = 1/2$ and $w = 1$,  the corresponding Friedmann equations are given by Eqs.~\eqref{gfbeffbvfd} and \eqref{defbdfbdsffbg},  respectively. The associated exterior solutions are also given in Eqs.~\eqref{sdcfdev} and \eqref{wbddwed}, respectively. These metrics possess curvature singularities
in their $R\to 0$ extensions. The pressureless dust case $w=0$, described by Eq.~\eqref{djdjedjweoid}, is also curvature-singular at the center.  Therefore, within the effective model considered here, non-negative-pressure sectors do not generically yield central regularity.
\\
We emphasize that this result should be interpreted as a limitation of the effective RG-improved model considered here. A complete quantum theory of gravity is expected to resolve physical singularities; however, the present construction is based on an Einstein-Hilbert truncation, a phenomenological running coupling, and a specific scale-identification prescription. Within this restricted framework, the asymptotic weakening of the gravitational coupling is not by itself sufficient to guarantee central regularity. For a static spherically symmetric metric of the form
	\begin{equation}
		f(R) = 1 - \frac{2m(R)}{R},
	\end{equation}
	regularity at the center requires
	\begin{equation}
		m(R) = \mathcal{O}(R^3) \quad \text{as} \quad R \to 0,
	\end{equation}
	so that the metric approaches a de Sitter-like core. This condition is satisfied in the negative-pressure cases studied here, such as $w=-1$ and $w=-1/2$, but it is not satisfied for the non-negative-pressure cases $w=0$, $w=1/2$, and $w=1$. Consequently, the present model provides conditional singularity avoidance rather than universal singularity resolution. This distinction is important: the RG-improved dynamics identifies the matter sectors for which an effective regular core can emerge, but it also shows that additional high-density physics may be required to avoid singularity formation in realistic non-negative-pressure stellar collapse.
\\

Indeed, one of the main outcomes of our work is that the final fate of a collapsing star depends strongly on EOS of the matter inside it. Our results indicate that under certain conditions, AS gravity can prevent the appearance of singularity only for specific matter sectors. We have also analyzed the local stability of the reduced homogeneous dynamics through the effective potential $V(a)$.
The cases $w=-1/2$ and $w=1$, described by Eqs.~\eqref{wdkwddbowfd} and \eqref{wfwdfdwf}, do not possess locally stable minima in the potentials
considered here. By contrast, the cases $w=0$ and $w=1/2$, described by Eqs.~\eqref{dwodweif} and \eqref{fdvfegref}, admit local minima of
the homogeneous effective potential.
\\
	For the positive EOS parameters considered here, $w=1/2$ and $w=1$, the corresponding static exterior metrics possess curvature singularities in the limit $R\rightarrow0$. This geometric property does not imply that every collapse trajectory necessarily reaches the singular center. In particular, for $w=1/2$, the homogeneous effective potential admits a local minimum at a finite scale factor. When the Hamiltonian constraint and initial conditions confine the evolution near this minimum, the stellar surface may remain at a finite radius or undergo bounded oscillations, and the singular region of the exterior extension is not dynamically attained. Thus, the $w=1/2$ result represents a locally stable homogeneous potential together with a singular $R\rightarrow0$ exterior extension. For $w=1$, no such local minimum was found in the potential studied here.
\\
	These results should be interpreted in the context of the classical perfect-fluid collapse literature. Pressure and the EOS already affect the collapse rate, horizon formation, and possible final states in GR, and pressure alone does not guarantee singularity resolution. In the present model, the regularization mechanism is instead associated with the RG-induced weakening of gravity and the effective cosmological contribution at high density, whose influence is modulated by the EOS-dependent density scaling. Moreover, the FLRW ansatz restricts the analysis to the homogeneous sector, in which the compatibility difficulties of a constant linear EOS in general inhomogeneous spherical collapse do not arise. The present classification of central regularity and horizon structure should therefore not be extrapolated to inhomogeneous collapse without a separate analysis.
\\

Several directions remain open for future work. The shadow properties, thermodynamic behavior, Hawking radiation, and quasinormal-mode spectra of the exterior metrics derived here should be analyzed in detail and compared with EHT and
	gravitational-wave observations. Extending the present spherically symmetric analysis to rotating geometries is also essential
	for realistic astrophysical applications. In addition, anisotropic fluids, magnetic fields, and inhomogeneous density profiles
	should be incorporated to test whether the EOS-dependent regularity pattern found in the homogeneous model persists
	beyond the FLRW approximation.

	Finally, the RG improvement adopted in this work should be regarded as an effective and scheme-dependent description. The running coupling is motivated by the pure-gravity Einstein-Hilbert flow, while quantum matter backreaction is encoded only phenomenologically through an effective fixed-point and crossover parameter. Therefore, matter backreaction is expected primarily to shift crossover scales, horizon locations, critical parameters, and mass profiles.

 The scale identification $k \sim 1/d(R)$, evaluated with respect to the classical Schwarzschild background, is chosen to maintain analytical tractability while avoiding the potential horizon divergences associated with fully self-consistent proper-distance prescriptions. 	The cutoff identification introduces an unavoidable scheme dependence. Covariant prescriptions based on proper distance or curvature invariants may share the same leading ultraviolet weakening of the gravitational coupling, but they need not produce identical crossover functions. When evaluated on the Schwarzschild geometry, the inverse proper-distance and Kretschmann prescriptions both yield $k^2\propto R^{-3}$, so their leading difference can be absorbed into a scheme-dependent crossover parameter. In the FLRW interior, however, curvature-based prescriptions may also modify the functional dependence of $G$ on the density for $w\neq0$. Consequently, the precise Misner-Sharp mass profiles, horizon radii, critical parameter values, and, in some near-extremal cases, even the number of horizons are scheme dependent. The qualitative results should therefore be understood as conditional on the class of cutoff prescriptions that preserve ultraviolet weakening and the relevant fixed-point scaling, rather than as strictly scheme-independent predictions. In contrast, precise horizon radii, exact quantitative thresholds, and detailed EOS dependencies should be interpreted as scheme-sensitive predictions that may be refined by future coupled matter–gravity RG analyses and higher-order truncations. This distinction ensures that the physical insights derived herein are properly contextualized within the current effective approach to quantum gravity.
\\

In summary, the effective AS-inspired framework studied here provides a useful setting for exploring how RG-improved
	gravitational couplings may affect collapse and central regularity. By incorporating quantum-gravity-motivated running
	couplings, this approach offers a phenomenological route toward understanding possible departures from classical collapse
	at extreme densities.

\section*{Acknowledgement}
\noindent
The work of Kazuharu Bamba was supported in part by the JSPS KAKENHI Grants No.~21K03547, No.~24KF0100, No.~25KF0176, and a grant-in-aid of academic research of the Yamaguchi Scholarship Foundation.
F.Shojai gratefully acknowledges the support of the University of Tehran through a grant provided by the University Research Council.
%%%%%%%%%%%%%%%%%%
%%%%%%%%%%%%%%%%%%%%
\appendix
\section{Running gravitational constant}
\label{Appendix A}
Following \cite{PhysRevD.62.043008}, the averaged form of the Einstein-Hilbert action is written as
\begin{align}
	\label{ansk}
	\Gamma_{k}[g,\bar{g}]=\frac{1}{16\pi G(k)}\int d^{4}x\sqrt{g}(-R+2\Lambda(k))+S_{gf}[g,\bar{g}]
\end{align}
where $G(k)$ and $\Lambda(k)$ denote the scale-dependent (running) Newton constant and cosmological constant, respectively, and $S_{gf}$ represents the classical background gauge-fixing term. The flow equation governs the scale evolution of these couplings \cite{PhysRevD.62.043008,PhysRevD.57.971,wetterich1993exact} and can be expressed as follows
\begin{align}
	\label{ddcdnk}
	\partial_{t}\Gamma_{k}[g,\bar{g}] \ = \ &
	\frac{1}{2}{\rm Tr}\Bigg(\frac{\partial_{t}\mathcal{R}^{grav}_{k}[\bar{g}]}{\kappa^{-2}\Gamma^{(2)}_{k}[g,\bar{g}]+\mathcal{R}^{grav}_{k}[\bar{g}]}\Bigg)\nonumber\\
	&-{\rm Tr}\Bigg(\frac{\partial_{t}\mathcal{R}^{gh}_{k}[\bar{g}]}{-\mathcal{M}[g,\bar{g}]+\mathcal{R}^{gh}_{k}[\bar{g}]}\Bigg)
\end{align}
where $t=\ln k$, $\Gamma^{(2)}_{k}$ denotes the Hessian of $\Gamma_{k}$ with respect to $g_{\mu\nu}$, and $\mathcal{M}$ is the Faddeev-Popov ghost operator. The operators $\mathcal{R}^{\text{grav}}_{k}$ and $\mathcal{R}^{\text{gh}}_{k}$ are filtering functions implementing the infrared cutoff in the graviton and ghost sectors, respectively. They are defined as
\begin{equation}
	\mathcal{R}_{k}(p^{2}) \propto k^2 R^{0}(z),
	\qquad z = \frac{p^2}{k^2},
	\qquad R^0(z) = \frac{z}{e^{z}-1},
\end{equation}
see Refs.~\cite{PhysRevD.62.043008,PhysRevD.57.971}.
By inserting Eq.~\eqref{ansk} into Eq.~\eqref{ddcdnk} and projecting the flow onto the subspace spanned by the Einstein--Hilbert truncation, one obtains a coupled system of differential equations for the dimensionless Newton constant
\begin{equation}
	g(k) \equiv k^{2}G(k),
\end{equation}
and the dimensionless cosmological constant
\begin{equation}
	\lambda(k) \equiv \frac{\Lambda(k)}{k^2}.
\end{equation}
In deriving the approximate flow of $G(k)$ in this appendix, we neglect $\Lambda(k)$ in the vacuum Einstein-Hilbert RG system. This approximation is used only to obtain the analytical form of $G(k)$. The resulting differential equation for the dimensionless Newton constant reads \cite{PhysRevD.62.043008,PhysRevD.57.971}
\begin{align}
	\label{mdkkd}
	\frac{dg(t)}{dt}=\Big(2+\frac{B_{1}g(t)}{1-B_{2}g(t)}\Big)g(t)\,.
\end{align}
The constants $B_{1}$ and $B_{2}$ are determined as follows
\begin{align}
	\label{mxakxOS}
	&B_{1}=-\frac{1}{3\pi}\big(24\Phi^{2}_{2}(0)-\Phi^{1}_{1}(0)\big)\notag\\
	&B_{2}=\frac{1}{6\pi}\big(18\tilde\Phi^{2}_{2}(0)-5\tilde\Phi^{1}_{1}(0)\big)\,,
\end{align}
where the two auxiliary functions, $\Phi^{p}_{n}(w)$ and $\tilde\Phi^{p}_{n}(w)$,  are defined as follows
\begin{align}
	\label{x[mspdpw}
	&\Phi^{p}_{n}(w)\equiv\frac{1}{\Gamma(n)}\int^{\infty}_{0}dzz^{n-1}\frac{R^{(0)}(z)-zR^{(0)'}(z)}{(z+R^{(0)}(z)+w)^p}\notag\\
	&\tilde\Phi^{p}_{n}(w)\equiv\frac{1}{\Gamma(n)}\int^{\infty}_{0}dzz^{n-1}\frac{R^{(0)}(z)}{(z+R^{(0)}(z)+w)^p}\,.
\end{align}
Employing the exponential cutoff  $R^0(z)=z/(e^{z}-1)$, the explicit expressions are obtained as
\begin{align}
	\label{sclscmnls}
	&\Phi^{1}_{1}(0)=\frac{\pi^2}{6},\hspace{0.5cm}\Phi^{2}_{2}(0)=1\notag\\
	&\tilde\Phi^{1}_{1}(0)=1,\hspace{0.7cm}\tilde\Phi^{2}_{2}(0)=\frac{1}{2}\,.
\end{align}
Upon defining $\omega\equiv-\frac{1}{2}B_{1}$ and $\omega'=\omega+B_{2}$, the analytical solution to differential equation \eqref{mdkkd} can be expressed in the form \cite{PhysRevD.62.043008, PhysRevD.57.971}

\begin{align}
	\label{snddnoem}
	\frac{g}{(1-\omega' g)^{\frac{\omega}{\omega'}}}=\frac{g(k_{0})}{[1-\omega' g(k_{0})]^{\frac{\omega}{\omega'}}}\left(\frac{k}{k_{0}}\right)^2\,.
\end{align}
The ratio $\omega'/\omega$ is observed to be quantitatively close to unity, with a numerical value of $\omega'/\omega \approx 1.18$. Under the approximation $\omega \approx \omega'$, the solution presented in equation \eqref{snddnoem} can be recast in terms of the dimensionful Newton's constant, $G(k) \equiv g(k)/k^2$, yielding the following expression \cite{PhysRevD.62.043008,PhysRevD.57.971}.
\begin{align}
	\label{dwodnwdweo}
	G(k)=\frac{G_{0}}{1+\omega G_{0}k^2}
\end{align}
where $G_{0}=G(k=0)$ is the experimentally observed value of Newton's constant, as measured at the solar-system scales.
\\
Equation \eqref{dwodnwdweo} follows from the pure-gravity Einstein-Hilbert truncation. In the presence of quantum matter fields, the coefficients entering the anomalous dimension and the beta functions are modified, so that both the fixed-point value and the crossover parameter become matter dependent. For this reason, in the main text, we use the generalized notation
\begin{equation}
	G_{\mathrm{eff}}(k)=\frac{G_0}{1+\omega_{\mathrm{eff}}G_0k^2}, \quad
	\omega_{\mathrm{eff}}=\frac{1}{g_*^{\mathrm{eff}}},
\end{equation}
and regard $\omega_{\mathrm{eff}}$ as a phenomenological parameter encoding the leading matter-dressed correction.
\section{Potential Energy of the Fluid}
\label{dkfwfu}
We begin by considering a compressible elastic medium. In such a medium it is convenient to
introduce the quantity $\Pi$, representing the potential energy per unit mass, and to
express the stress tensor in terms of it. Let $a_1, a_2$, and $a_3$ denote a set of
Lagrangian coordinates, corresponding, for example, to the initial positions of the material
particles. At a later time $t$, the spatial coordinates of a given particle are
described by the functions
\begin{equation}
	x_i = x_i(a_1, a_2, a_3, t) \quad (i = 1, 2, 3)
\end{equation}
The deformation of the medium is then characterized by the quantities
\begin{equation}
	A_{mn} = \sum_{i=1}^3 
	\frac{\partial x_i}{\partial a_m}
	\frac{\partial x_i}{\partial a_n}
\end{equation}
The potential energy $\Pi$ depends on the deformation, and its differential can be expressed in the form
\begin{equation}
	d\Pi = \frac{1}{2\rho} \sum_{m,n=1}^3 P^{mn} \, dA_{mn},
	\qquad (P^{mn} = P^{nm})
\end{equation}
where the symmetric quantities $P^{mn}$ play the role of generalized stress
components. The physical stress tensor $p_{ik}$ can then be expressed in terms
of these coefficients as follows
\begin{equation}
	p_{ik} = 
	\sum_{m,n=1}^3 
	P^{mn} 
	\frac{\partial x_i}{\partial a_m}
	\frac{\partial x_k}{\partial a_n},
	\qquad (p_{ik} = p_{ki})
\end{equation}
Using the above relations, one readily finds that the rate of change of the
potential energy along the motion satisfies
\begin{equation}\label{dckdc}
	\rho \frac{d\Pi}{dt}
	= \rho \left(
	\frac{\partial \Pi}{\partial t}
	+ \sum_{i=1}^3 v_i \frac{\partial \Pi}{\partial x_i}
	\right)
	= \sum_{i,k=1}^3 p_{ik} \frac{\partial v_i}{\partial x_k}.
\end{equation}
For a fluid, the stress tensor takes the isotropic form
\begin{equation}
	p_{ik} = -p \delta_{ik}
\end{equation}
where $p$ denotes the pressure. Substituting this into \eqref{dckdc}, we obtain
\begin{equation}
	\rho \frac{d\Pi}{dt}
	= -p\, \mathrm{div}\,\mathbf{v}
	= \frac{p}{\rho} \frac{d\rho}{dt}
\end{equation}
Integrating this relation yields the well-known expression
\begin{equation}\label{dcedc}
	\Pi =
	\int \frac{p}{\rho^2} \, d\rho
	= \int \frac{dp}{\rho} - \frac{p}{\rho}.
\end{equation}

Let us assume that the energy--momentum tensor is given by
\begin{equation}
	T^{ik}
	= \left( \epsilon + p \right) u^i u^k
	- p\, e_k \delta_{ik}.
\end{equation}
Here $\epsilon$ and $p$ are scalar fields in four-dimensional spacetime, and they are related through the constitutive relation
\begin{equation}
	\epsilon = f(p).
\end{equation}
We now derive the equations of motion. For convenience, introduce the quantity
\begin{equation}
	Q = \sum_{k=0}^{3} \frac{\partial}{\partial x_k}
	\left[
	\left( \epsilon +p \right) u^k
	\right]
\end{equation}
Using the notation  $w^i$ for the acceleration, the equations of motion take the form
\begin{equation}
	\sum_{k=0}^{3} \frac{\partial T^{ik}}{\partial x_k}
	=
	Q u^i
	+ \left( \epsilon +p \right) w^i
	- e_i \frac{\partial p}{\partial x_i}
\end{equation}
In the absence of external forces, this expression must vanish.
Employing the relations $\sum_{i=0}^{3} u^{i} u_{i} = 1$ and $\sum_{i=0}^{3} u^{i} w_{i} = 0$, we obtain an alternative expression for $Q$, namely
\begin{equation}
	Q
	= \sum_{k=0}^{3} u^k \frac{\partial p}{\partial x_k}
	= \frac{1}{\sqrt{1 - v^2}} \frac{dp}{dt}
	=  \frac{dp}{d\tau}
\end{equation}
Here $dp/dt$ is the substantial (or convective) derivative of the pressure, and $d\tau$ is the differential of the particle’s proper time.  
Comparison with the non-relativistic limit confirms that $p$ represents the pressure.  
Equating the two forms of $Q$ yields
\begin{equation}\label{emejd}
	\sum_{k=0}^{3}
	\left[
	\left( \epsilon +p\right) \frac{\partial u^k}{\partial x_k}
	+ u^k \frac{\partial \epsilon}{\partial x_k}
	\right]
	= 0
\end{equation}
Assuming that $\epsilon$ depends only on $p$, we introduce a new variable $\rho^*$ defined through
\begin{equation}\label{fjefefr}
	\frac{d\rho^*}{\rho^*}
	=
	\frac{d\epsilon}{\epsilon + p}
\end{equation}
We may choose the constant of integration such that $\rho^{*} = \epsilon$ when $p = 0$.
Using Eq.~\eqref{emejd},  Eq.~\eqref{fjefefr} can then be written in the form
\begin{equation}\label{eq:3224}
	\sum_{k=0}^{3} \frac{\partial}{\partial x_{k}} \big( \rho^{*} u^{k} \big) = 0 
\end{equation}
Thus, $\rho^{*}$ can be interpreted as the invariant density of that portion of the rest mass which remains conserved during the motion.
\\
Let us define
\begin{equation}
	\epsilon = \rho^* \left( 1 +\Pi \right)
\end{equation}
Then, from the differential relation between $\epsilon$ and $\rho^*$, Eq.~\eqref{fjefefr}, we obtain 
\begin{equation}
	d\Pi = \frac{p \, d\rho^*}{\rho^{*2}} \Longrightarrow \Pi=\int \frac{dp}{\rho^{*}} - \frac{p}{\rho^{*}}
\end{equation}
The quantity $\Pi$ may be interpreted, in analogy with  Eq.~\eqref{dcedc}, as the potential energy per unit mass of the fluid.

\section{Variation of the Rest-Mass Density With Respect to the Metric}
\label{dmdjdwfw}
In relativistic fluid dynamics, the particle number current is given by
\begin{equation}
	J^\mu = n u^\mu ,
\end{equation}
where $n$ is the baryon number density in the fluid rest frame and $u^\mu$ is the four-velocity, normalized as
\begin{equation}
	u^\mu u_\mu = -1.
\end{equation}
The conservation of particles implies that $J^\mu$ is held fixed under variations of the metric
\begin{equation}
	\delta (\sqrt{-g}\, J^\mu) = 0
\end{equation}
Substituting $J^\mu = n u^\mu$ gives
\begin{equation}
	\delta (\sqrt{-g}\, n u^\mu) = 0
\end{equation}
Expanding the variation yields
\begin{equation}
	\delta(\sqrt{-g}\, n u^\mu)
	= \delta(\sqrt{-g}\, n)\, u^\mu
	+ \sqrt{-g}\, n\, \delta u^\mu 
\end{equation}
Contracting with $u_\mu$ gives
\begin{equation}
	\delta(\sqrt{-g}\, n) + \sqrt{-g}\, n\, u_\mu \delta u^\mu = 0
	\label{eq:contracted}
\end{equation}
Since $u^\mu u_\mu = -1$, its variation must satisfy
\begin{equation}
	u_\mu \delta u^\mu = -\frac{1}{2} u^\mu u^\nu \delta g_{\mu\nu}
\end{equation}
The variation of $\sqrt{-g}$ is
\begin{equation}
	\delta \sqrt{-g}
	= -\frac{1}{2} \sqrt{-g}\, g_{\mu\nu}\, \delta g^{\mu\nu}
\end{equation}
Thus,
\begin{equation}
	\delta(\sqrt{-g}\, n)
	= \sqrt{-g}\, \delta n
	- \frac{1}{2} \sqrt{-g}\, n\, g_{\mu\nu}\, \delta g^{\mu\nu}
\end{equation}
Putting everything together
\begin{equation}
	\sqrt{-g}\, \delta n
	- \frac{1}{2} \sqrt{-g}\, n\, g_{\mu\nu}\, \delta g^{\mu\nu}
	- \frac{1}{2} \sqrt{-g}\, n\, u^\mu u^\nu \delta g_{\mu\nu}
	= 0
\end{equation}
Use $\delta g_{\mu\nu} = - g_{\mu\alpha} g_{\nu\beta}\, \delta g^{\alpha\beta}$ to rewrite the last term, and simplify.  
The result is
\begin{equation}
	\delta n
	= \frac{1}{2} n (g_{\mu\nu} + u_\mu u_\nu)\, \delta g^{\mu\nu}
\end{equation}
Since the rest-mass density is $\rho = m n$, its variation is
\begin{equation}
	\frac{\delta \rho}{\delta g^{\mu\nu}}
	= \frac{1}{2}\rho\,(g_{\mu\nu} + u_\mu u_\nu).
\end{equation}
\section{The hypergeometric function}
\label{dkdkjdkjx}
The hypergeometric function $~_pF_{q}$ is defined in Ref.~\cite{koepf2014hypergeometric} as follows for $a_{a},...,a_{p}$, $b_{1},...,b_{q}$, $z\in\mathbb{C}$
\begin{equation}\label{vvchfuiyg;ou}
	~_pF_{q}(a_{1},...,a_{p};b_{1},...,b_{q};z)=\sum^{\infty}_{n=0}\frac{(a_{1})_{n}...(a_{p})_{n}}{(b_{1})_{n}...(b_{q})_{n}}\frac{z^{n}}{n!},
\end{equation}
where, for some parameter $\mu$, the Pochhammer symbol $(\mu)_{n}$ is defined as
\begin{equation}\label{dndedned}
	(\mu)_{0}=1,~~(\mu)_{n}=\mu(\mu+1)...(\mu+n-1),~~n=1,2,...
\end{equation}	
Another definition is that, for $c>b>0$ the hypergeometric function is given
by the Euler integral \cite{heckman2015tsinghua}, 
\begin{align}\label{dmwdwjd}
	~_2F_{1}(a,b&,c;z)=\notag\\
	&+\frac{\Gamma(c)}{\Gamma(b)\Gamma(c-b)}\int_{0}^{1}t^{b-1}(1-t)^{c-b-1}(1-zt)^{-a}dt,
\end{align}	
Moreover, the $n$-th derivative is given by
\begin{align}\label{xsxsxo}
	&\frac{d^{n}}{dz^{n}}\big[~_pF_{q}(a_{1},...,a_{p};b_{1},...,b_{q};z)\big]=\frac{(a_{1})_n...(a_{p})_{n}}{(b_{1})_n...(b_{q})_{n}}\notag\\
	&\times ~_pF_{q}(a_{1}+n,...,a_{p}+n;b_{1}+n,...,b_{q}+n;z).
\end{align}	

Importantly, in  specific case $(p,q)=(2,1)$, which is called the \textit{Gauss hypergeometric function}, the hypergeometric function is defined for $|z| < 1$ by the power series
\begin{align}\label{jbbkk}
	~_2F_{1}\big(a,b;c;z\big)&=\sum_{n=0}^{\infty}\frac{(a)_{n}(b)_{n}}{(c)_{n}}\frac{z^n}{n!},~~~~~~~~z\in \mathbb{C} \\&=1+\frac{ab}{c}\frac{z}{1!}+\frac{a(a+1)b(b+1)}{c(c+1)}\frac{z^2}{2}+...
\end{align}
and the first order time derivative is written as
\begin{equation}\label{dndnn}
	\frac{d}{dz}	~_2F_{1}(a,b;c;z)=\frac{ab}{c}	~_2F_{1}(a+1,b+1;c+1;z).
\end{equation}

Apart from that, there are some useful definitions for hypergeometric function which are frequently used in this paper, the first one is  written as
\begin{equation}
	{}_2F_1(1,w;w;z) = \sum_{n=0}^{\infty} \frac{(1)_n}{n!} z^n,
\end{equation}
since $(1)_n = n!$, we obtain
\begin{equation}\label{dckjwdwk}
	{}_2F_1(1,w;w;z) = \sum_{n=0}^{\infty} z^n = \frac{1}{1-z}, \quad \text{for } |z| < 1.
\end{equation}

We now focus on the case where $a = 1$, $b = 1$, and $c = 2$. By substituting these values into Eq.~\eqref{jbbkk}, we obtain
\begin{equation}
	{}_2F_1(1, 1; 2; -z) = \sum_{n=0}^\infty \frac{(1)_n (1)_n}{(2)_n} \frac{(-z)^n}{n!}=\sum_{n=0}^\infty \frac{(-z)^n}{n+1},
\end{equation}
Let $n+1 = m \implies n = m-1$. When $n = 0$, $m = 1$, the series becomes
\begin{equation}
	{}_2F_1(1, 1; 2; -z) =-\frac{1}{z} \sum_{m=1}^\infty \frac{(-z)^m}{m},
\end{equation}
using the Taylor expansion of $\ln(1+z)$
\begin{equation}
	\ln(1+z) = -\sum_{m=1}^\infty \frac{(-z)^m}{m},
\end{equation}
Therefore, for $a=1,b=1,c=2$, the Eq.~\eqref{jbbkk} is written as   
\begin{equation}\label{dnbkwdn}
	~_2F_{1}(1,1;2;-z)=\frac{\ln(1+z)}{z}.
\end{equation}
\\

We now substitute the values $a = 1$, $b = \frac{1}{2}$, and $c = \frac{3}{2}$ into the series \eqref{jbbkk}
\begin{equation}
	_2F_1\left(1, \frac{1}{2}; \frac{3}{2}; -z\right) = \sum_{n=0}^{\infty} \frac{(1)_n \left(\frac{1}{2}\right)_n}{\left(\frac{3}{2}\right)_n n!} (-z)^n=\sum_{n=0}^{\infty} \frac{ (-z)^n}{2n+1}, 
\end{equation}
Using 
\begin{equation}
	\arctan(\sqrt{z}) = \sum_{n=0}^{\infty} (-1)^n \frac{z^{n+1/2}}{2n+1},
\end{equation}
in the case $a=1,b=\frac{1}{2},c=\frac{3}{2}$,
the Gauss hypergeometric function can be written as
\begin{equation}\label{dndndnd}
	~_2F_{1}\left(1, \frac{1}{2}; \frac{3}{2}; -z\right) = \frac{\arctan\sqrt{z}}{\sqrt{z}}.
\end{equation}
\\

Another expression for the case $a=1, b=2, c=3$  is given as
\begin{equation}\label{dndjdjdkj}
	{}_2F_1(1,2;3;-z) = \sum_{n=0}^\infty \frac{(1)_n (2)_n}{(3)_n} \frac{(-z)^n}{n!},
\end{equation}
where 
\begin{equation}
	\frac{(1)_n (2)_n}{n!(3)_n} = \frac{2}{n+2},
\end{equation}
Inserting this equation into Eq.~\eqref{dndjdjdkj} yields the following equation
\begin{equation}\label{dwdfwdfwdfw}
	{}_2F_1(1,2;3;-z) =\frac{2}{z^2} \sum_{n=2}^\infty  \frac{(-z)^n}{n},
\end{equation}
This series resembles the Taylor series for the natural logarithm. Recall that 
\begin{equation}
	-\ln(1 + z) = \sum_{n=1}^{\infty} \frac{(-z)^{n}}{n} \quad \text{for } |z| < 1,
\end{equation}
To relate it to our series, we can write
\begin{equation}\label{ddjdhwedkju}
	\sum_{n=2}^{\infty} \frac{(-z)^{n}}{n} = -\ln(1 + z) + z,
\end{equation}
which is valid for $|z| < 1$. Substituting Eq.~\eqref{ddjdhwedkju} into Eq.~\eqref{dndjdjdkj} gives 
\begin{equation}\label{eferfrefr}
	{}_2F_1(1,2;3;-z) = \frac{2}{z}-\frac{2\ln(1+z)}{z^2}.
\end{equation}

Another significant case discussed in this paper corresponds to $a = 1$, $b = \frac{3}{2}$, and $c = \frac{5}{2}$, is presented as
\begin{equation}\label{fefevcfevcfev}
	{}_2F_1\Big(1,\frac{3}{2};\frac{5}{2};-z\Big) =-\frac{3}{z} \sum_{n=1}^\infty  \frac{(-z)^n}{1+2n},
\end{equation}
to analyze this series, we utilize the following integral
\begin{equation}\label{dndhdehweiu}
	\frac{1}{1+2n}=\int_{0}^{1}x^{2n}dx,
\end{equation}	
Substituting this integral into the series in Eq.~\eqref{fefevcfevcfev} leads to 
\begin{equation}\label{djdijwou}
	\sum_{n=1}^\infty  \frac{(-z)^n}{1+2n}=\int_{0}^{1}\sum_{n=1}^\infty (-zx^2)^{n}dx=-1+\frac{\arctan(\sqrt{z})}{\sqrt{z}},
\end{equation}
to derive the above equation, we employed the series $\sum_{n=0}^\infty y^n = \frac{1}{1-y}$. Substituting this series into Eq.~\eqref{fefevcfevcfev} yields
\begin{equation}\label{rtgtrgtr}
	{}_2F_1\Big(1,\frac{3}{2};\frac{5}{2};-z\Big) =\frac{3}{z}\Big(1-\frac{\arctan(\sqrt{z})}{\sqrt{z}}\Big).
\end{equation}


%\begin{thebibliography}{99}

% Please avoid comments such as "For a review'', "For some examples",
% "and references therein" or move them in the text. In general,
% please leave only references in the bibliography and move all
% accessory text in footnotes.

% Also, please have only one work for each \bibitem.

\nocite{*} 
\bibliographystyle{apsrev4-2}
\bibliography{Ramin}% Produces the bibliography via BibTeX.

@PREAMBLE{
 "\providecommand{\noopsort}[1]{}" 
 # "\providecommand{\singleletter}[1]{#1}%" 
}

@article{Hawking:1970zqf,
    author = "Hawking, S. W. and Penrose, R.",
    doi = "10.1098/rspa.1970.0021",
    journal = "Proc. Roy. Soc. Lond. A",
    volume = "314",
    pages = "529--548",
    year = "1970"
}

@article{Penrose:1964wq,
	author = "Penrose, Roger",
	doi = "10.1103/PhysRevLett.14.57",
	journal = "Phys. Rev. Lett.",
	volume = "14",
	pages = "57--59",
	year = "1965"
}

@article{Markov:1985py,
	author = "Markov, M. A. and Mukhanov, Viatcheslav F.",
	doi = "10.1007/BF02732276",
	journal = "Nuovo Cim. B",
	volume = "86",
	pages = "97--102",
	year = "1985"
}

@article{Bonanno:2023rzk,
	author = "Bonanno, Alfio and Malafarina, Daniele and Panassiti, Antonio",
	doi = "10.1103/PhysRevLett.132.031401",
	journal = "Phys. Rev. Lett.",
	volume = "132",
	number = "3",
	pages = "031401",
	year = "2024"
}

@inbook{Weinberg:1980gg,
	author = "Weinberg, Steven",
	booktitle = "{General Relativity}: {An Einstein Centenary Survey}",
	pages = "790--831",
	year = "1980",
    publisher = {Cambridge University Press},
}

@article{Kiselev:2002dx,
	author = "Kiselev, V. V.",
	doi = "10.1088/0264-9381/20/6/310",
	journal = "Class. Quant. Grav.",
	volume = "20",
	pages = "1187--1198",
	year = "2003"
}

@article{Visser:2019brz,
	author = "Visser, Matt",
	doi = "10.1088/1361-6382/ab60b8",
	journal = "Class. Quant. Grav.",
	volume = "37",
	number = "4",
	pages = "045001",
	year = "2020"
}

@article{Hassannejad:2023lrp,
	author = "Hassannejad, R. and Sadeghi, A. and Shojai, F.",
	doi = "10.1088/1361-6382/acbd81",
	journal = "Class. Quant. Grav.",
	volume = "40",
	number = "7",
	pages = "075002",
	year = "2023"
}

@article{Abedi:2015yga,
	author = "Abedi, Jahed and Arfaei, Hessamaddin",
	doi = "10.1007/JHEP03(2016)135",
	journal = "JHEP.",
	volume = "03",
	pages = "135",
	year = "2016"
}

@article{Lewandowski:2022zce,
	author = "Lewandowski, Jerzy and Ma, Yongge and Yang, Jinsong and Zhang, Cong",
	doi = "10.1103/PhysRevLett.130.101501",
	journal = "Phys. Rev. Lett.",
	volume = "130",
	number = "10",
	pages = "101501",
	year = "2023"
}

@article{Kelly:2020uwj,
	author = "Kelly, Jarod George and Santacruz, Robert and Wilson-Ewing, Edward",
	doi = "10.1103/PhysRevD.102.106024",
	journal = "Phys. Rev. D",
	volume = "102",
	number = "10",
	pages = "106024",
	year = "2020"
}

@article{Gambini:2020nsf,
	author = "Gambini, R. and Olmedo, J. and Pullin, J.",
	doi = "10.1088/1361-6382/aba842",
	journal = "Class. Quant. Grav.",
	volume = "37",
	number = "20",
	pages = "205012",
	year = "2020"
}

@article{Caldwell:1999ew,
	author = "Caldwell, R. R.",
	doi = "10.1016/S0370-2693(02)02589-3",
	journal = "Phys. Lett. B",
	volume = "545",
	pages = "23--29",
	year = "2002"
}

@article{Caldwell:2003vq,
	author = "Caldwell, Robert R. and Kamionkowski, Marc and Weinberg, Nevin N.",
	doi = "10.1103/PhysRevLett.91.071301",
	journal = "Phys. Rev. Lett.",
	volume = "91",
	pages = "071301",
	year = "2003"
}

@article{Carroll:2003st,
	author = "Carroll, Sean M. and Hoffman, Mark and Trodden, Mark",
	doi = "10.1103/PhysRevD.68.023509",
	journal = "Phys. Rev. D",
	volume = "68",
	pages = "023509",
	year = "2003"
}

@article{Shojai:2022pdq,
	author = "Shojai, F. and Sadeghi, A. and Hassannejad, R.",
	doi = "10.1088/1361-6382/ac5924",
	journal = "Class. Quant. Grav.",
	volume = "39",
	number = "8",
	pages = "085003",
	year = "2022"
}

@article{Ashtekar:2008zu,
	author = "Ashtekar, Abhay",
	doi = "10.1007/s10714-009-0763-4",
	journal = "Gen. Rel. Grav.",
	volume = "41",
	pages = "707--741",
	year = "2009"
}

@book{hawking1973,
  author={Hawking, S. W. and Ellis, G. F. R.},
  publisher={Cambridge University Press},
  address={Cambridge},
  year={1973},
    title = {The Large Scale Structure of Space-Time},
}

@incollection{weinberg1979,
  author={Weinberg, Steven},
  booktitle={General Relativity: An Einstein Centenary Survey},
  editor={Hawking, S. W. and Israel, W.},
  pages={790--831},
  publisher={Cambridge University Press},
  address={Cambridge},
  year={1979}
}

@article{bonanno2001,
  author={Bonanno, A. and Reuter, M.},
  journal={Physical Review D},
  volume={65},
  number={4},
  pages={043508},
  publisher={American Physical Society},
  doi={10.1103/PhysRevD.65.043508},
  year={2001}
}

@article{wetterich1993exact,
  author = {Wetterich, Christof},
  journal = {Physics Letters B},
  volume = {301},
  number = {1},
  pages = {90--94},
  doi = {10.1016/0370-2693(93)90726-X},
  year = {1993}
}

@article{goroff1986ultraviolet,
  author = {Goroff, Marc H. and Sagnotti, Augusto},
  journal = {Nuclear Physics B},
  volume = {266},
  number = {3–4},
  pages = {709--736},
  doi = {10.1016/0550-3213(86)90193-8},
  year = {1986}
}

@article{thooft1974one,
  
  author = {'t Hooft, Gerard and Veltman, Martinus J. G.},
  journal = {Annales de l'Institut Henri Poincaré},
  volume = {20},
  number = {1},
  pages = {69--94},
  year = {1974}
}

@article{reuter1998nonperturbative,
 
  author = {Reuter, Martin},
  journal = {Physical Review D},
  volume = {57},
  number = {2},
  pages = {971--985},
  doi = {10.1103/PhysRevD.57.971},
  year = {1998}
}

@article{Dou:1997fg,
    author = "Dou, Djamel and Percacci, Roberto",
    doi = "10.1088/0264-9381/15/11/011",
    journal = "Class. Quant. Grav.",
    volume = "15",
    pages = "3449--3468",
    year = "1998"
}

@article{oppenheimer1939continued,
  author={Oppenheimer, J. Robert and Snyder, Hartland},
  journal={Physical Review},
  volume={56},
  pages={455--459},
  year={1939},
  publisher={American Physical Society},
  doi={10.1103/PhysRev.56.455}
}

@article{hawking1976breakdown,
  
  author={Hawking, S. W.},
  journal={Physical Review D},
  volume={13},
  pages={191--197},
  year={1976},
  publisher={American Physical Society},
  doi={10.1103/PhysRevD.13.191}
}

@article{hawking1974black,
 
  author={Hawking, S. W.},
  journal={Nature},
  volume={248},
  pages={30--31},
  year={1974},
  publisher={Nature Publishing Group},
  doi={10.1038/248030a0}
}

@article{Hawking1972,
  author    = {Stephen W. Hawking},
  journal   = {Communications in Mathematical Physics},
  volume    = {25},
  pages     = {152},
  year      = {1972},
  doi       = {10.1007/BF01645966}
}

@book{becker2006string,
  
  author={Becker, K. and Becker, M. and Schwarz, J.H.},
  isbn={9781139460484},
  url={https://books.google.de/books?id=WgUkSTJWQacC},
  year={2006},
  publisher={Cambridge University Press},
    title = {String Theory and M-Theory: A Modern Introduction},
}

@book{Gambini:2011,
  author    = {Rodolfo Gambini and Jorge Pullin},
  publisher = {Oxford University Press},
  year      = {2011},
  edition   = {1},
  isbn      = {978-0199692888},
    title = {A First Course in Loop Quantum Gravity},
}

@book{Platania:2018,
    author = {Platania, Alessia Benedetta},
    publisher = {Springer},
    year = {2018},
    series = {Springer Theses},
    isbn = {978-3-319-98793-4},
    doi = {10.1007/978-3-319-98794-1},
    pages = {142},
    title = {Asymptotically Safe Gravity: From Spacetime Foliation to Cosmology},
}

@article{Datt1938,
  author = {S. Datt},
  journal = {Journal of Physics},
  volume = {108},
  pages = {314--321},
  year = {1938},
  doi = {10.1007/BF01374951}
}

@article{Bojowald:2005ah,
    author = "Bojowald, Martin",
    doi = "10.1103/PhysRevLett.95.061301",
    journal = "Phys. Rev. Lett.",
    volume = "95",
    pages = "061301",
    year = "2005"
}

@article{Ashtekar:2018lag,
    author = "Ashtekar, Abhay and Olmedo, Javier and Singh, Parampreet",
    doi = "10.1103/PhysRevLett.121.241301",
    journal = "Phys. Rev. Lett.",
    volume = "121",
    number = "24",
    pages = "241301",
    year = "2018"
}

@phdthesis{Maldacena:1996ky,
    author = "Maldacena, Juan Martin",
    eprint = "hep-th/9607235",
    archivePrefix = "arXiv",
    reportNumber = "UMI-96-27605",
    school = "Princeton U.",
    year = "1996"
}

@article{Litim:2003vp,
    author = "Litim, Daniel F.",
    doi = "10.1103/PhysRevLett.92.201301",
    journal = "Phys. Rev. Lett.",
    volume = "92",
    pages = "201301",
    year = "2004"
}

@article{tHooft:1984kcu,
    author = "'t Hooft, Gerard",
    doi = "10.1016/0550-3213(85)90418-3",
    journal = "Nucl. Phys. B",
    volume = "256",
    pages = "727--745",
    year = "1985"
}

@book{fock1959theory,
  author={Fock, Vladimir},
  year={1959},
  publisher={Pergamon Press},
  edition={2nd edition},
  address={Oxford, UK},
    title = {Theory of Space, Time and Gravitation},
}

@book{koepf2014hypergeometric,
  author={Wolfram Koepf},
  series={Universitext},
  year={2014},
  publisher={Springer},
  address={London},
  edition={2nd},
  isbn={978-1-4471-6463-0},
  doi={10.1007/978-1-4471-6464-7},
    title = {Hypergeometric Summation: An Algorithmic Approach to Summation and Special Function Identities},
}

@misc{heckman2015tsinghua,
  author       = {Gert Heckman},
  title= {Tsinghua Lectures on Hypergeometric Functions},
  year         = {2015},
  institution  = {Radboud University Nijmegen},
  note         = {December 8, 2015, G.Heckman@math.ru.nl}
}

@article{Bonanno:2017pkg,
    author = "Bonanno, Alfio and Saueressig, Frank",
    doi = "10.1016/j.crhy.2017.02.002",
    journal = "Comptes Rendus Physique",
    volume = "18",
    pages = "254--264",
    year = "2017"
}

@article{PhysRevD.111.064069,
  title = {Gravitational collapse in scale-dependent gravity},
  author = {Hassannejad, Ramin and Lambiase, Gaetano and Scardigli, Fabio and Shojai, Fatimah},
  journal = {Phys. Rev. D},
  volume = {111},
  issue = {6},
  pages = {064069},
  numpages = {30},
  year = {2025},
  month = {Mar},
  publisher = {American Physical Society},
  doi = {10.1103/PhysRevD.111.064069},
  url = {https://link.aps.org/doi/10.1103/PhysRevD.111.064069}
}

@article{Israel:1966rt,
    author = "Israel, W.",
    doi = "10.1007/BF02710419",
    journal = "Nuovo Cim. B",
    volume = "44S10",
    pages = "1",
    year = "1966"
}

@article{Fernandes:2022zrq,
    author = "Fernandes, Pedro G. S. and Carrilho, Pedro and Clifton, Timothy and Mulryne, David J.",
    doi = "10.1088/1361-6382/ac500a",
    journal = "Class. Quant. Grav.",
    volume = "39",
    number = "6",
    pages = "063001",
    year = "2022"
}

@article{Glavan:2019inb,
    author = "Glavan, Dra\v{z}en and Lin, Chunshan",
    doi = "10.1103/PhysRevLett.124.081301",
    journal = "Phys. Rev. Lett.",
    volume = "124",
    number = "8",
    pages = "081301",
    year = "2020"
}

@article{Einstein:1916vd,
    author = "Einstein, Albert",
    editor = "Hsu, Jong-Ping and Fine, D.",
    doi = "10.1002/andp.19163540702",
    journal = "Annalen Phys.",
    volume = "49",
    number = "7",
    pages = "769--822",
    year = "1916"
}

@article{Zeldovich:1967lct,
    author = "Zel'dovich, Ya. B. and Novikov, I. D.",
    journal = "Sov. Astron.",
    volume = "10",
    pages = "602",
    year = "1967"
}

@article{Carr:1974nx,
    author = "Carr, Bernard J. and Hawking, S. W.",
    doi = "10.1093/mnras/168.2.399",
    journal = "Mon. Not. Roy. Astron. Soc.",
    volume = "168",
    pages = "399--415",
    year = "1974"
}

@article{Hawking:1971ei,
    author = "Hawking, Stephen",
    doi = "10.1093/mnras/152.1.75",
    journal = "Mon. Not. Roy. Astron. Soc.",
    volume = "152",
    pages = "75",
    year = "1971"
}

@book{weinberg2008cosmology,
  title        = {Cosmology},
  author       = {Weinberg, Steven},
  year         = 2008,
  publisher    = {Oxford University Press},
  location     = {Oxford, UK},
  isbn         = {978-0-19-852682-7}
}

@article{Fradkin:1978yf,
    author = "Fradkin, E. S. and Vilkovisky, G. A.",
    doi = "10.1016/0370-2693(78)90702-5",
    journal = "Phys. Lett. B",
    volume = "77",
    pages = "262--266",
    year = "1978"
}

@article{Mazur:2004fk,
    author = "Mazur, Pawel O. and Mottola, Emil",
    doi = "10.1073/pnas.0402717101",
    journal = "Proc. Nat. Acad. Sci.",
    volume = "101",
    pages = "9545--9550",
    year = "2004"
}

@article{Harada:2025cwd,
    author = "Harada, Tomohiro and Chen, Chiang-Mei and Mandal, Rituparna",
    title = "{Singularity resolution and regular black hole formation in gravitational collapse in asymptotically safe gravity}",
    eprint = "2502.16787",
    archivePrefix = "arXiv",
    primaryClass = "gr-qc",
    reportNumber = "RUP-25-3",
    doi = "10.1103/pt9s-jqjz",
    journal = "Phys. Rev. D",
    volume = "111",
    number = "12",
    pages = "126017",
    year = "2025"
}

@article{Bekenstein:1973ur,
    author = "Bekenstein, Jacob D.",
    doi = "10.1103/PhysRevD.7.2333",
    journal = "Phys. Rev. D",
    volume = "7",
    pages = "2333--2346",
    year = "1973"
}

@article{Bekenstein:1974ax,
    author = "Bekenstein, Jacob D.",
    doi = "10.1103/PhysRevD.9.3292",
    journal = "Phys. Rev. D",
    volume = "9",
    pages = "3292--3300",
    year = "1974"
}

@article{Bekenstein:1972tm,
    author = "Bekenstein, J. D.",
    doi = "10.1007/BF02757029",
    journal = "Lett. Nuovo Cim.",
    volume = "4",
    pages = "737--740",
    year = "1972"
}

@article{Hawking:1975vcx,
    author = "Hawking, S. W.",
    editor = "Gibbons, G. W. and Hawking, S. W.",
    doi = "10.1007/BF02345020",
    journal = "Commun. Math. Phys.",
    volume = "43",
    pages = "199--220",
    year = "1975"
}

@article{Bardeen:1973gs,
    author = "Bardeen, James M. and Carter, B. and Hawking, S. W.",
    doi = "10.1007/BF01645742",
    journal = "Commun. Math. Phys.",
    volume = "31",
    pages = "161--170",
    year = "1973"
}

@article{Hawking:1982dh,
    author = "Hawking, S. W. and Page, Don N.",
    doi = "10.1007/BF01208266",
    journal = "Commun. Math. Phys.",
    volume = "87",
    pages = "577",
    year = "1983"
}

@article{Hawking:1976ra,
    author = "Hawking, S. W.",
    doi = "10.1103/PhysRevD.14.2460",
    journal = "Phys. Rev. D",
    volume = "14",
    pages = "2460--2473",
    year = "1976"
}

@article{Almheiri:2020cfm,
    author = "Almheiri, Ahmed and Hartman, Thomas and Maldacena, Juan and Shaghoulian, Edgar and Tajdini, Amirhossein",
    doi = "10.1103/RevModPhys.93.035002",
    journal = "Rev. Mod. Phys.",
    volume = "93",
    number = "3",
    pages = "035002",
    year = "2021"
}

@article{Penrose:1969pc,
    author = "Penrose, R.",
    doi = "10.1023/A:1016578408204",
    journal = "Riv. Nuovo Cim.",
    volume = "1",
    pages = "252--276",
    year = "1969"
}

@article{Frolov:1981mz,
    author = "Frolov, Valeri P. and Vilkovisky, G. A.",
    doi = "10.1016/0370-2693(81)90542-6",
    journal = "Phys. Lett. B",
    volume = "106",
    pages = "307--313",
    year = "1981"
}

@inproceedings{Callan:1996ey,
    author = "Callan, Curtis G. and Maldacena, Juan Martin",
    booktitle = "{ICTP Summer School in High-energy Physics and Cosmology}",
    pages = "1--65",
    month = "6",
    year = "1996"
}

@article{Maldacena:1997de,
    author = "Maldacena, Juan Martin and Strominger, Andrew and Witten, Edward",
    doi = "10.1088/1126-6708/1997/12/002",
    journal = "JHEP.",
    volume = "12",
    pages = "002",
    year = "1997"
}

@article{Horowitz:2003he,
    author = "Horowitz, Gary T. and Maldacena, Juan Martin",
    doi = "10.1088/1126-6708/2004/02/008",
    journal = "JHEP.",
    volume = "02",
    pages = "008",
    year = "2004"
}

@article{Maldacena:2013xja,
    author = "Maldacena, Juan and Susskind, Leonard",
    doi = "10.1002/prop.201300020",
    journal = "Fortsch. Phys.",
    volume = "61",
    pages = "781--811",
    year = "2013"
}

@article{Ashtekar:2018cay,
    author = "Ashtekar, Abhay and Olmedo, Javier and Singh, Parampreet",
    doi = "10.1103/PhysRevD.98.126003",
    journal = "Phys. Rev. D",
    volume = "98",
    number = "12",
    pages = "126003",
    year = "2018"
}

@article{Ashtekar:1997yu,
    author = "Ashtekar, A. and Baez, J. and Corichi, A. and Krasnov, Kirill",
    doi = "10.1103/PhysRevLett.80.904",
    journal = "Phys. Rev. Lett.",
    volume = "80",
    pages = "904--907",
    year = "1998"
}

@article{Bonanno:2017zen,
    author = "Bonanno, Alfio and Koch, Benjamin and Platania, Alessia",
    doi = "10.1007/s10701-018-0195-7",
    journal = "Found. Phys.",
    volume = "48",
    number = "10",
    pages = "1393--1406",
    year = "2018"
}

@article{Codello:2006in,
    author = "Codello, Alessandro and Percacci, Roberto",
    doi = "10.1103/PhysRevLett.97.221301",
    journal = "Phys. Rev. Lett.",
    volume = "97",
    pages = "221301",
    year = "2006"
}

@article{Platania:2019kyx,
    author = "Platania, Alessia",
    doi = "10.1140/epjc/s10052-019-6990-2",
    journal = "Eur. Phys. J. C",
    volume = "79",
    number = "6",
    pages = "470",
    year = "2019"
}

@article{Machado:2007ea,
    author = "Machado, Pedro F. and Saueressig, Frank",
    doi = "10.1103/PhysRevD.77.124045",
    journal = "Phys. Rev. D",
    volume = "77",
    pages = "124045",
    year = "2008"
}

@article{Lauscher:2002sq,
    author = "Lauscher, O. and Reuter, M.",
    doi = "10.1103/PhysRevD.66.025026",
    journal = "Phys. Rev. D",
    volume = "66",
    pages = "025026",
    year = "2002"
}

@article{Reuter:2001ag,
    author = "Reuter, M. and Saueressig, Frank",
    doi = "10.1103/PhysRevD.65.065016",
    journal = "Phys. Rev. D",
    volume = "65",
    pages = "065016",
    year = "2002"
}

@article{Lauscher:2001ya,
    author = "Lauscher, O. and Reuter, M.",
    doi = "10.1103/PhysRevD.65.025013",
    journal = "Phys. Rev. D",
    volume = "65",
    pages = "025013",
    year = "2002"
}

@article{Souma:1999at,
    author = "Souma, Wataru",
    doi = "10.1143/PTP.102.181",
    journal = "Prog. Theor. Phys.",
    volume = "102",
    pages = "181--195",
    year = "1999"
}

@article{Knorr:2018kog,
    author = "Knorr, Benjamin and Saueressig, Frank",
    doi = "10.1103/PhysRevLett.121.161304",
    journal = "Phys. Rev. Lett.",
    volume = "121",
    number = "16",
    pages = "161304",
    year = "2018"
}

@article{Polonyi:2001se,
    author = "Polonyi, Janos",
    doi = "10.2478/BF02475552",
    journal = "Central Eur. J. Phys.",
    volume = "1",
    pages = "1--71",
    year = "2003"
}

@article{Christiansen:2017cxa,
    author = "Christiansen, Nicolai and Litim, Daniel F. and Pawlowski, Jan M. and Reichert, Manuel",
    doi = "10.1103/PhysRevD.97.106012",
    journal = "Phys. Rev. D",
    volume = "97",
    number = "10",
    pages = "106012",
    year = "2018"
}

@article{Dona:2015tnf,
    author = "Don\`a, Pietro and Eichhorn, Astrid and Labus, Peter and Percacci, Roberto",
    doi = "10.1103/PhysRevD.93.129904",
    journal = "Phys. Rev. D",
    volume = "93",
    number = "4",
    pages = "044049",
    year = "2016"
}

@article{Meibohm:2015twa,
    author = "Meibohm, Jan and Pawlowski, Jan M. and Reichert, Manuel",
    doi = "10.1103/PhysRevD.93.084035",
    journal = "Phys. Rev. D",
    volume = "93",
    number = "8",
    pages = "084035",
    year = "2016"
}

@article{Dona:2013qba,
    author = "Don\`a, Pietro and Eichhorn, Astrid and Percacci, Roberto",
    doi = "10.1103/PhysRevD.89.084035",
    journal = "Phys. Rev. D",
    volume = "89",
    number = "8",
    pages = "084035",
    year = "2014"
}

@article{Vacca:2010mj,
    author = "Vacca, G. P. and Zanusso, O.",
    doi = "10.1103/PhysRevLett.105.231601",
    journal = "Phys. Rev. Lett.",
    volume = "105",
    pages = "231601",
    year = "2010"
}

@article{Platania:2025imw,
    author = "Platania, Alessia",
    doi = "10.1007/s10714-025-03390-5",
    journal = "Gen. Rel. Grav.",
    volume = "57",
    number = "3",
    pages = "58",
    year = "2025",
}

@article{Weinberg:2009wa,
    author = "Weinberg, Steven",
    doi = "10.1103/PhysRevD.81.083535",
    journal = "Phys. Rev. D",
    volume = "81",
    pages = "083535",
    year = "2010"
}

@article{Bonanno:2015fga,
    author = "Bonanno, Alfio and Platania, Alessia",
    doi = "10.1016/j.physletb.2015.10.005",
    journal = "Phys. Lett. B",
    volume = "750",
    pages = "638--642",
    year = "2015"
}

@article{Liu:2018hno,
    author = "Liu, Lei-Hua and Prokopec, Tomislav and Starobinsky, Alexei A.",
    doi = "10.1103/PhysRevD.98.043505",
    journal = "Phys. Rev. D",
    volume = "98",
    number = "4",
    pages = "043505",
    year = "2018"
}

@article{Silva:2024wit,
    author = "Silva, Agust\'\i{}n",
    doi = "10.1016/j.physletb.2024.139154",
    journal = "Phys. Lett. B",
    volume = "860",
    pages = "139154",
    year = "2025"
}

@article{Platania:2019qvo,
    author = "Platania, Alessia",
    editor = "Eichhorn, Astrid and Percacci, Roberto and Saueressig, Frank",
    doi = "10.3390/universe5080189",
    journal = "Universe",
    volume = "5",
    number = "8",
    pages = "189",
    year = "2019"
}

@article{Reuter:2005kb,
    author = "Reuter, M. and Saueressig, Frank",
    doi = "10.1088/1475-7516/2005/09/012",
    journal = "JCAP.",
    volume = "09",
    pages = "012",
    year = "2005"
}

@article{Koch:2010nn,
    author = "Koch, Benjamin and Ramirez, Israel",
    doi = "10.1088/0264-9381/28/5/055008",
    journal = "Class. Quant. Grav.",
    volume = "28",
    pages = "055008",
    year = "2011"
}

@article{Bonanno:2007wg,
    author = "Bonanno, Alfio and Reuter, Martin",
    doi = "10.1088/1475-7516/2007/08/024",
    journal = "JCAP.",
    year = "2007",
    volume = "08",
    pages = "024"
}

@article{Bonanno:2011yx,
    author = "Bonanno, Alfio and Carloni, Sante",
    doi = "10.1088/1367-2630/14/2/025008",
    journal = "New J. Phys.",
    volume = "14",
    pages = "025008",
    year = "2012"
}

@article{Kofinas:2016lcz,
    author = "Kofinas, Georgios and Zarikas, Vasilios",
    doi = "10.1103/PhysRevD.94.103514",
    journal = "Phys. Rev. D",
    volume = "94",
    number = "10",
    pages = "103514",
    year = "2016"
}

@article{Bonanno:2022jjp,
    author = "Bonanno, Alfio and Khosravi, Amir-Pouyan and Saueressig, Frank",
    doi = "10.1103/PhysRevD.107.024005",
    journal = "Phys. Rev. D",
    volume = "107",
    number = "2",
    pages = "024005",
    year = "2023"
}

@article{Laporte:2021kyp,
    author = "Laporte, Cristobal and Pereira, Antonio D. and Saueressig, Frank and Wang, Jian",
    doi = "10.1007/JHEP12(2021)001",
    journal = "JHEP",
    volume = "12",
    pages = "001",
    year = "2021",
    number = {12},
}

@article{Bosma:2019aiu,
    author = "Bosma, Lando and Knorr, Benjamin and Saueressig, Frank",
    doi = "10.1103/PhysRevLett.123.101301",
    journal = "Phys. Rev. Lett.",
    volume = "123",
    number = "10",
    pages = "101301",
    year = "2019"
}

@article{Saadati:2020dje,
    author = "Saadati, R. and Shojai, F.",
    doi = "10.1088/1361-6382/abfed5",
    journal = "Class. Quant. Grav.",
    volume = "38",
    number = "13",
    pages = "135025",
    year = "2021"
}

@misc{Zhumabek:2024tvp,
    author = "Zhumabek, Tilek and Mukhamediya, Azamat and Chakrabarty, Hrishikesh and Malafarina, Daniele",
    title = "{Running gravitational constant induced dark energy as a solution to $\sigma_8$ tension}",
    eprint = "2411.05965",
    archivePrefix = "arXiv",
    primaryClass = "astro-ph.CO",
    month = "11",
    year = "2024"
}

@misc{Zholdasbek:2024pxi,
    author = "Zholdasbek, Aknur and Chakrabarty, Hrishikesh and Malafarina, Daniele and Bonanno, Alfio",
    title = "{An emergent cosmological model from running Newton constant}",
    eprint = "2405.02636",
    archivePrefix = "arXiv",
    primaryClass = "gr-qc",
    month = "5",
    year = "2024"
}

@article{Casadio:2010fw,
    author = "Casadio, Roberto and Hsu, Stephen D. H. and Mirza, Behrouz",
    title = "{Asymptotic Safety, Singularities, and Gravitational Collapse}",
    doi = "10.1016/j.physletb.2010.10.060",
    journal = "Phys. Lett. B",
    volume = "695",
    pages = "317--319",
    year = "2011"
}

@article{Falkenberg:1996bq,
    author = "Falkenberg, Sven and Odintsov, Sergei D.",
    title = "{Gauge dependence of the effective average action in Einstein gravity}",
    doi = "10.1142/S0217751X98000263",
    journal = "Int. J. Mod. Phys. A",
    volume = "13",
    pages = "607--623",
    year = "1998"
}

@article{PhysRevD.62.043008,
  author = {Bonanno, Alfio and Reuter, Martin},
  journal = {Phys. Rev. D},
  volume = {62},
  issue = {4},
  pages = {043008},
  numpages = {21},
  year = {2000},
  month = {Jul},
  publisher = {American Physical Society},
  doi = {10.1103/PhysRevD.62.043008},
  url = {https://link.aps.org/doi/10.1103/PhysRevD.62.043008}
}

@article{PhysRevD.57.971,  
  author = {Reuter, M.},
  journal = {Phys. Rev. D},
  volume = {57},
  issue = {2},
  pages = {971--985},
  numpages = {0},
  year = {1998},
  month = {Jan},
  publisher = {American Physical Society},  
  url = {https://link.aps.org/doi/10.1103/PhysRevD.57.971}
}

@misc{Platania:2023srt,
    author = "Platania, Alessia",
    title = "{Black Holes in Asymptotically Safe Gravity}",
    eprint = "2302.04272",
    archivePrefix = "arXiv",
    primaryClass = "gr-qc",
    reportNumber = "NORDITA 2022-085",
    doi = "10.1007/978-981-19-3079-9_24-1",
    year = "2023"
}

@misc{Held:2021vwd,
    author = "Held, Aaron",
    title = "{Invariant Renormalization-Group improvement}",
    eprint = "2105.11458",
    archivePrefix = "arXiv",
    primaryClass = "gr-qc",
    reportNumber = "Imperial/TP/2021/AH/04",
    month = "5",
    year = "2021"
}

@article{Held:2019xde,
    author = "Held, Aaron and Gold, Roman and Eichhorn, Astrid",
    primaryClass = "gr-qc",
    doi = "10.1088/1475-7516/2019/06/029",
    journal = "JCAP",
    volume = "06",
    pages = "029",
    year = "2019",
    number = {06},
}

@article{Liu:2012ee,
    author = "Liu, Dao-Jun and Yang, Bin and Zhai, Yong-Jia and Li, Xin-Zhou",
    primaryClass = "gr-qc",
    doi = "10.1088/0264-9381/29/14/145009",
    journal = "Class. Quant. Grav.",
    volume = "29",
    pages = "145009",
    year = "2012"
}

@article{Kumar:2019ohr,
    author = "Kumar, Rahul and Singh, Balendra Pratap and Ghosh, Sushant G.",
    primaryClass = "gr-qc",
    doi = "10.1016/j.aop.2020.168252",
    journal = "Annals Phys.",
    volume = "420",
    pages = "168252",
    year = "2020"
}

@article{Bonanno:2025dry,
    author = "Bonanno, Alfio M. and Konoplya, Roman A. and Oglialoro, Giovanni and Spina, Andrea",
    primaryClass = "gr-qc",
    doi = "10.1088/1475-7516/2025/12/042",
    journal = "JCAP",
    volume = "12",
    pages = "042",
    year = "2025",
    number = {12},
}

@article{Semiz:2020lxj,
    author = "Semiz, {\.I}brahim",
    archivePrefix = "arXiv",
    primaryClass = "gr-qc",
    doi = "10.1088/1361-6382/ac8cca",
    journal = "Class. Quant. Grav.",
    volume = "39",
    number = "21",
    pages = "215002",
    year = "2022"
}

@article{Semiz:2008rd,
    author = "Semiz, Ibrahim",
    archivePrefix = "arXiv",
    primaryClass = "gr-qc",
    doi = "10.1142/S0218271810016208",
    journal = "Int. J. Mod. Phys. D",
    volume = "19",
    pages = "1--20",
    year = "2010"
}

@article{PhysRevD.109.024045,
  title = {Quantum black hole physics from the event horizon},
  author = {Del Piano, Manuel and Hohenegger, Stefan and Sannino, Francesco},
  journal = {Phys. Rev. D},
  volume = {109},
  issue = {2},
  pages = {024045},
  numpages = {32},
  year = {2024},
  month = {Jan},
  publisher = {American Physical Society},
  doi = {10.1103/PhysRevD.109.024045},
  url = {https://link.aps.org/doi/10.1103/PhysRevD.109.024045}
}

@article{Donoghue2020,
  author        = {Donoghue, John F.},
  title         = {A Critique of the Asymptotic Safety Program},
  journal       = {Frontiers in Physics},
  volume        = {8},
  pages         = {56},
  year          = {2020},
  doi           = {10.3389/fphy.2020.00056},
  url           = {https://doi.org/10.3389/fphy.2020.00056},
  eprint        = {1911.02967},
  archivePrefix = {arXiv},
  primaryClass  = {hep-th}
}

@misc{Mosani:2019wfu,
    author = "Mosani, Karim and Dey, Dipanjan and Joshi, Pankaj S.",
    title = "{On viability of isentropic perfect fluid collapse with a linear equation of state}",
    eprint = "1910.13678",
    archivePrefix = "arXiv",
    primaryClass = "gr-qc",
    month = "10",
    year = "2019"
}

@article{Joshi:2011rlc,
    author = "Joshi, Pankaj S. and Malafarina, Daniele",
    title = "{Recent developments in gravitational collapse and spacetime singularities}",
    eprint = "1201.3660",
    archivePrefix = "arXiv",
    primaryClass = "gr-qc",
    doi = "10.1142/S0218271811020792",
    journal = "Int. J. Mod. Phys. D",
    volume = "20",
    pages = "2641--2729",
    year = "2011"
}

\end{document}